\documentclass{article}
\usepackage[preprint]{neurips_2021}
\usepackage[utf8]{inputenc} % allow utf-8 input
\usepackage[T1]{fontenc}    % use 8-bit T1 fonts
\usepackage{hyperref}       % hyperlinks.
\usepackage{url}            % simple URL typesetting
\usepackage{booktabs}       % professional-quality tables
\usepackage{amsfonts}       % blackboard math symbols
\usepackage{nicefrac}       % compact symbols for 1/2, etc.
\usepackage{microtype}      % microtypography
\usepackage{xcolor}         % colors

\usepackage{dsfont}
\usepackage{easy-todo}
\usepackage{algorithm}
\usepackage{algpseudocode}
\usepackage{amsthm}
\usepackage{amsmath}
\usepackage{amssymb}
\usepackage{graphicx}
\usepackage{caption}
\usepackage{subcaption}

\newcommand{\muzero}{MuZero}
\newcommand{\muzerorc}{MuZero-RC}
\newcommand{\libvpx}{libvpx}
\newcommand{\sgn}{\text{sgn}}
\newcommand{\psnr}{\text{PSNR}}
\newcommand{\overshoot}{\text{overshoot}}
\newcommand{\bitrate}{\text{Bitrate}}
\newcommand{\cliptozero}[1]{\lfloor #1 \rfloor_+}

\title{MuZero with Self-competition for Rate Control in VP9 Video Compression}

\author{%
  Amol Mandhane\thanks{Equal contributions}\hspace{4pt}\thanks{DeepMind, London, UK}
  \And
  Anton Zhernov\footnotemark[1]\hspace{4pt}\footnotemark[2]
  \And
  Maribeth Rauh\footnotemark[1]\hspace{4pt}\footnotemark[2]
  \And
  Chenjie Gu\footnotemark[1]\hspace{4pt}\footnotemark[2]
  \And
  Miaosen Wang\footnotemark[2]
  \And
  Flora Xue\footnotemark[2]
  \And
  Wendy Shang\footnotemark[2]
  \And
  Derek Pang\thanks{Google, Mountain View, CA, USA}
  \And
  Rene Claus\footnotemark[3]
  \And
  Ching-Han Chiang\footnotemark[3]
  \And
  Cheng Chen\footnotemark[3]
  \And
  Jingning Han\footnotemark[3]
  \And
  Angie Chen\footnotemark[2]
  \And
  Daniel J.\ Mankowitz\footnotemark[2]
  \And
  Jackson Broshear\footnotemark[2]
  \And
  Julian Schrittwieser\footnotemark[2]
  \And
  Thomas Hubert\footnotemark[2]
  \And
  Oriol Vinyals\footnotemark[2]
  \And
  Timothy Mann\footnotemark[2]
}

\begin{document}

\maketitle

\newcommand\blfootnote[1]{%
  \begingroup
  \renewcommand\thefootnote{}\footnote{#1}%
  \addtocounter{footnote}{-1}%
  \endgroup
}
\blfootnote{Correspondence to: Amol Mandhane \texttt{<mandhane@deepmind.com>}, Anton Zhernov \texttt{<typedef@deepmind\\.com>}, Chenjie Gu \texttt{<gcj@deepmind.com>}}

\begin{abstract}

Video streaming usage has seen a significant rise as entertainment, education, and business increasingly rely on online video. Optimizing video compression has the potential to increase access and quality of content to users, and reduce energy use and costs overall. In this paper, we present an application of the \muzero{} algorithm to the challenge of video compression. Specifically, we target the problem of learning a rate control policy to select the quantization parameters (QP) in the encoding process of \libvpx{}, an open source VP9 video compression library widely used by popular video-on-demand (VOD) services. We treat this as a sequential decision making problem to maximize the video quality with an episodic constraint imposed by the target bitrate. Notably, we introduce a novel self-competition based reward mechanism to solve constrained RL with variable constraint satisfaction difficulty, which is challenging for existing constrained RL methods. We demonstrate that the \muzero{}-based rate control achieves an average 6.28\% reduction in size of the compressed videos for the same delivered video quality level (measured as PSNR BD-rate) compared to \libvpx{}'s two-pass VBR rate control policy, while having better constraint satisfaction behavior.

\end{abstract}

\newcommand{\expect}[2]{\mathds{E}_{{#1}} \left[ {#2} \right]}
\newcommand{\myvec}[1]{\boldsymbol{#1}}
\newcommand{\myvecsym}[1]{\boldsymbol{#1}}
\newcommand{\vx}{\myvec{x}}
\newcommand{\vy}{\myvec{y}}
\newcommand{\vz}{\myvec{z}}
\newcommand{\vtheta}{\myvecsym{\theta}}

\section{Introduction}

In recent years, video traffic has dominated global internet traffic and is expected to grow further \citep{ciscoreport}. \emph{Efficient video encoding algorithms} are key to reducing bandwidth and storage costs, as well as improving user Quality of Experience (QoE) in scenarios such as video-on-demand (VOD) and live streaming.

In video compression, rate control is a critical component that determines the trade-off between rate (video size) and distortion (video quality) by assigning a quantization parameter (QP) for each frame in the video. A lower QP results in lower distortion (higher quality), but uses more bits.  In this paper, we focus on the rate control problem in \libvpx{} \citep{libvpx}, an open source VP9 codec library \citep{vp9}. The objective in this setting is to select a QP for each of the video frames to maximize the quality of the encoded video subject to a bitrate constraint as seen in \eqref{eqn:highlevel-objective}. 

\begin{equation}
    \max_{\text{QPs}} \,\,\text{Encoded Video Quality} \,\,\,\,\,\text{s.t.}\, \text{Bitrate} \leq \text{Target}
    \label{eqn:highlevel-objective}
\end{equation}

\begin{figure}[tb]
	\centering
	\begin{subfigure}[c]{0.65\columnwidth}
	   \centering
    	\includegraphics[width=\textwidth]{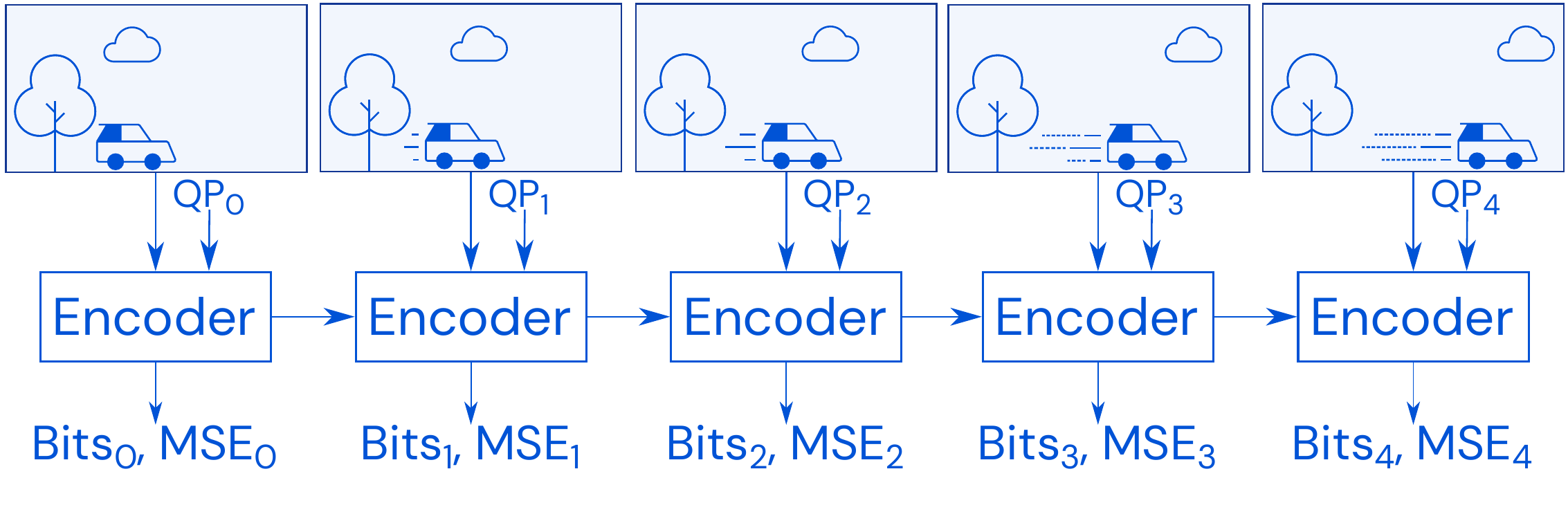}
    	\caption{Overview of the Rate Control process. For illustration purpose, we only draw the show frames. In practice, the encoder also assigns a QP for hidden alternate reference frames (ARF).}
    	\label{fig:qp-problem}
    \end{subfigure}
    \hfill
    \begin{subfigure}[c]{0.3\columnwidth}
	   \centering
    	\includegraphics[width=\textwidth]{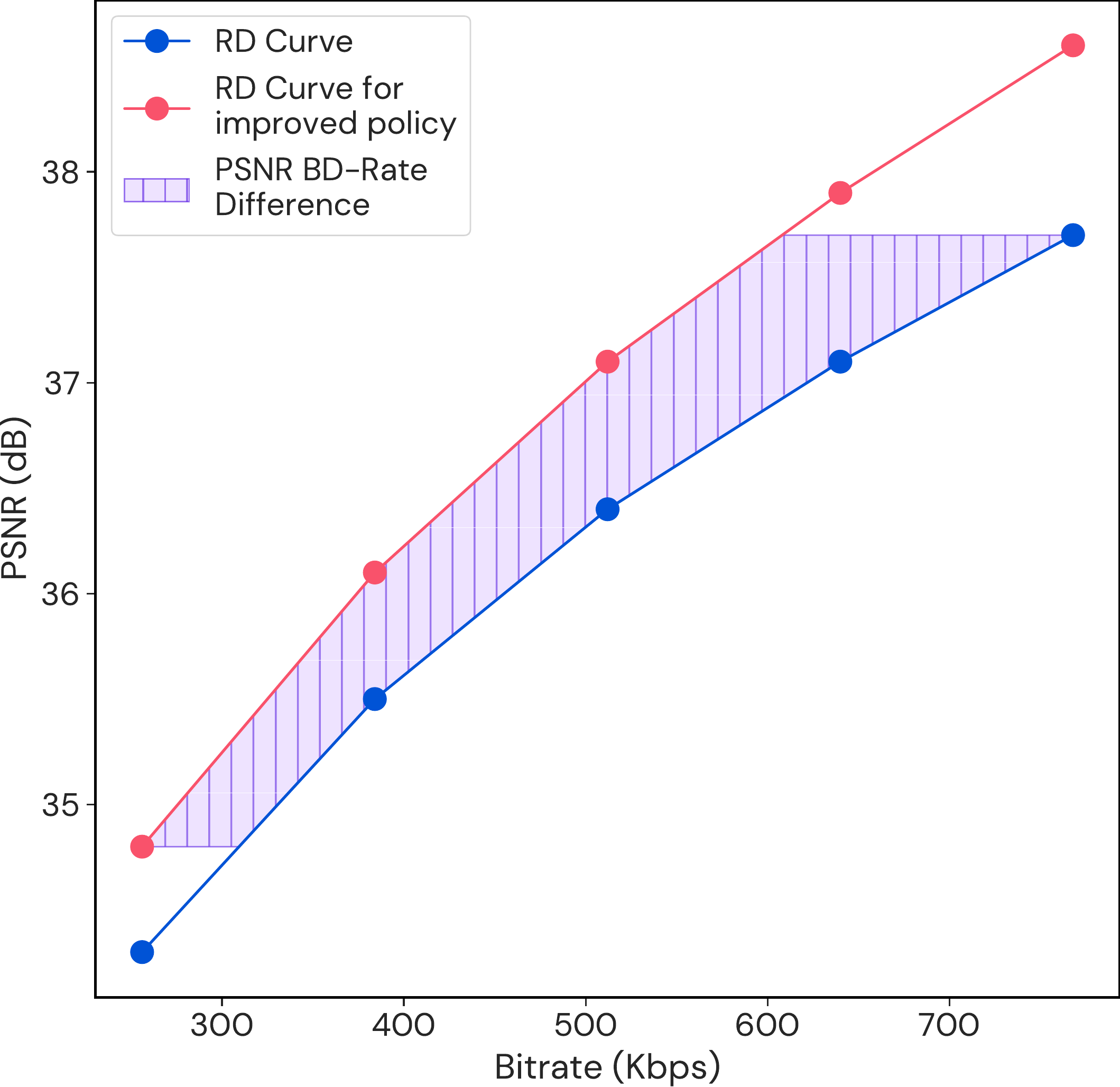}
     	\caption{Example of Rate-Distortion (RD) curve.}
    	\label{fig:rd-curve-example}
    \end{subfigure}
    \caption{Illustrative example of the rate control process, Rate-Distortion curve, and the BD-rate metric for encoding efficiency.}
\end{figure}

The rate control problem can be seen as a constrained planning problem. In particular, the QPs are chosen sequentially for each video frame as seen in Figure \ref{fig:qp-problem}, with the goal of optimizing for video quality under a bitrate constraint.  Classical optimization algorithms such as mixed integer linear programming cannot be used for rate control as the codec operates as a black box without an explicit equation and the interactions between the video frames cannot easily be modeled by a linear objective.  To provide some intuition for the planning element, imagine a video that has little motion for the first half of the video (e.g., a person talking or a static scene), and then has a lot of motion for the latter half of the video (e.g., a sports highlight). In this case, using up more of the bitrate budget on the first half of the video will cause the second half of the video to have lower quality. Intuitively, the rate control algorithm needs to allocate more bits to complex/dynamic frames and less bits to simple/static frames. This requires the algorithm to understand the rate-distortion tradeoff for each frame and their relationship to the QP values.  However, the inter-dependency of QP decisions and the rate-distortion metrics are often non-trivial, making it challenging to engineer optimal rate control algorithms that work for a diverse set of videos.

Reinforcement Learning (RL) \citep{sutton2018reinforcement} is a sequential decision making framework that has had recent success in solving numerous planning problems (in many cases achieving super-human performance) ranging from  games \citep{dqn,tessler2017deep, vinyals2019grandmaster,alphago} to chip design \citep{mirhoseini2020chip} and robotics \citep{levine2016end}. RL algorithms are now solving problems at scale and are potentially well-suited for solving rate control.

In this work, we apply the \muzero{} RL algorithm \citep{muzero} along with a novel self-competition based constrained RL method to rate control in VP9.  Specifically, our main contributions are as follows:

\begin{itemize}
    \item We propose a self-competition based reward mechanism that optimizes for the constrained rate control objective.
    \item We combine this self-competition based reward mechanism with the \muzero{} algorithm to create `\muzerorc{}',  an RL based rate control agent.
    \item We evaluate this agent against \libvpx{}'s two-pass VBR rate control implementation \citep{vp9modes} on 3062 video clips, each of which are 5-seconds in length, obtained from the YouTube UGC dataset \citep{youtube_ugc} which has been created for evaluating video compression algorithms. We show that \muzerorc{} achieves an average 6.28\% reduction in bitrate (measured as PSNR BD-rate) compared to the \libvpx{} baseline which is canonically used for VP9 encoding, while having better bitrate constraint satisfaction.
\end{itemize}

\section{Related Work}

\textbf{Rate control in video codecs: } Traditionally, rate control algorithms are based on empirical models using relevant features, such as mean absolute differences (MAD) and sum of absolute transformed differences (SATD) from past encoded frames. Common empirical models include quadratic models used in MPEG-4's VM8~\citep{derek2} and H.264/AVC~\citep{derek3,derek5,derek6}, $\rho$-domain models that map between rate and the proportion of non-zero quantized coefficients~\citep{derek7}, and dynamic programming based models that exploit interframe dependency~\citep{derek8}.  In modern codecs, such as HEVC and VVC, \citet{rdmodel2} introduced $\lambda$-domain model to control both rate and mode selections through rate-distortion optimization instead of relying on QP decisions. Despite recent advances in rate control models, the empirical rate control models require complex hand-designed heuristics to effectively adapt to different dynamic video sequences and satisfy different requirements for different applications. 

\textbf{ML for rate control: }
Machine learning enables more complex and nonlinear models to be formulated. \citet{derek10} use a radial basis function network to predict QP offset and ensure quality consistency. \citet{derek11} utilize a game theory method to allocate CTU-level bit allocation and optimize for SSIM in HEVC.  More recently, Reinforcement learning approaches in rate control have also been proposed to HEVC  ~\citep{derek12,derek13,rl-rate-control-2020-ieee,ho_dual_critic,chen_hevc_rl}.  \citet{vp9imitiation} use an imitation learning approach on evolutionary search based policy \citep{es} with a feedback-based correction for rate control in VP9.

Unlike the previously proposed ML based approaches, our proposed solution can work with a wide range of target bitrates specified for longer video durations with multiple groups-of-pictures (GOPs) instead of specifying bitrates for each GOP, which is close to the practical setting.  The solution does not rely on demonstrations from prior experience, which allows discovery of rate control strategies from scratch.

\section{Background}
\subsection{VP9 Video Compression}
In this section, we briefly describe the two-pass variable bitrate (VBR) mode, which is a popular libvpx-vp9 rate control mode for Videos-On Demand (VOD). In this work, we use the statistics collected in the first-pass and the second-pass encoding as the input features to a neural network model that predicts the QP for every frame.

\textbf{First-pass encoding:} The encoder computes statistics for every frame in the video, by dividing the frame into non-overlapping $16\times 16$ blocks followed by per-block intra and inter-frame prediction (of each block) and calculation of the residual error. The statistics contain information such as average motion prediction error, average intra-frame prediction error, average motion vector magnitude, percentage of zero motion blocks, noise energy, etc.\ \citep{firstpass}.

\textbf{Second-pass encoding:} The encoder uses the first-pass statistics to decide key-frame locations and insert hidden alternate reference frames.  The key-frames and alternate reference frames are used as references for encoding other frames, so their encoding quality affects other frames in the video as well.  With those decisions made, the encoder starts to encode video frames \emph{sequentially} as seen in Figure \ref{fig:qp-problem}. The rate controller regulates the trade-off between rate and distortion by specifying a QP to each frame in order to maximize the quality and reduce the bits as explained in section \ref{section:rc_crl}. The QP is an integer in range [0, 255] that can be monotonically mapped to a quantization step size which is used for quantizing the frequency transform of the prediction residue for entropy coding. Smaller quantization step sizes lead to smaller reconstruction error (measured by mean squared error), but also higher bits usage for the frame.

\subsection{Constrained Markov Decision Process}
\label{sec:cmdp}
A Markov Decision Process (MDP) \citep{sutton2018reinforcement} $\mathcal{M}$ is defined by the tuple $\langle S, A, P, R, \gamma \rangle$ where  $S$ is the set of states; $A$ is the set of available state-dependent actions; $R:S\times A \rightarrow \mathbb{R}$ is a bounded reward function;  $P:S\times A\times S \rightarrow \left[0, 1\right]$ is the transition function, where $P\left(s'|s, a\right)$ is the probability of transitioning to state $s'$ from $s$ when action $a$ was taken; $\gamma \in [0, 1]$ is the discount factor.  A policy $\pi : S \times A \rightarrow \left[0, 1\right]$ is a mapping from a state to a distribution over actions where $\pi\left(a|s\right)$ denotes the probability of taking action $a$ in state $s$.  The goal is to obtain a policy which maximizes the expected sum of discounted rewards ($J_{\pi}^{R}$).

\begin{equation*}
    \max_{\pi}J_{\pi}^{R}(s) := \underset{a \sim \pi, s \sim P}{\mathds{E}}\left[\sum_{t}\gamma^t R(s_t, a_t)\right]
\end{equation*}

\noindent A Constrained Markov Decision Process (CMDP) \citep{cmdp_book} extends the MDP formulation by introducing $k$ constraint functions $c_k:S\times A \rightarrow \mathbb{R}$ and corresponding thresholds $\beta_k$.  The goal is to obtain a policy which maximizes the expected sum of discounted rewards subject to the discounted constraints ($J_{\pi}^{c_k}$) being under the target thresholds $\beta_k$.

\begin{equation}
    \max_{\pi}J_{\pi}^{R} \;\;\; \text{s.t.} \;\;\; J_{\pi}^{c_k} \leq \beta_k \; \forall k
\end{equation}

There are a number of approaches to solving this constrained RL objective \citep{cmdp_book,rcpo,efroni2020explorationexploitation,AchiamHTA17,bohez2019value,chow2018lyapunovbased,Paternain2019,zhang2020reward}. One approach involves solving the Lagrangian relaxation of the original problem as $\max_{\pi} \min_{\lambda \geq 0} J_{\pi}^{R} - \lambda ( J_{\pi}^{c_k} - \beta_k)$. Here, the typical approach is to alternate the optimization between the policy $\pi$ and the Lagrangian parameter $\lambda$ \citep{rcpo,metal_rcpo}. Another powerful approach is to fix the Lagrangian parameter and perform reward shaping \citep{AchiamHTA17,rcpo}.

\subsection{\muzero{}}
\muzero{} \citep{muzero} is a Reinforcement Learning algorithm based on deep neural networks and Monte-Carlo tree search (MCTS) planning algorithm \citep{mcts} which has achieved state-of-the-art performance in various domains like Chess, Go, Shogi, and Atari.  \muzero{} uses  a learned model of the environment dynamics with MCTS to look ahead and plan from a given state in order to identify good actions.  It uses a policy neural network as a prior for the MCTS, and a value network to truncate the MCTS.  Unlike its predecessors, \muzero{} does not need to interact with the environment for planning; instead it uses a learned model of environment dynamics to look ahead.

\section{Rate Control with \muzero}
The following section is divided into three parts. First, we formulate the rate control problem as a CMDP and define the rate control CMDP objective that we wish to optimize. Next, we introduce the self-competition based reward mechanism to address this constrained objective that can be incorporated into any RL agent. Finally, we provide an overview of the architecture and training setup of our \muzero{} agent combined with the self-competition based reward for rate control (\muzerorc{}). 
\subsection{Rate Control as a Constrained RL Problem}
\label{section:rc_crl}
In this paper, we focus on the ‘two-pass, variable bitrate (VBR)’ mode in the \libvpx{} implementation of VP9. However, our methods do not focus on the specifics of this mode and can be generalized to other settings.  The objective of rate control in this mode is to maximize the quality of the encoded video while keeping the size under a user-specified target.  Quality of the encoding is commonly measured as Peak Signal-to-Noise Ratio (PSNR) $:= 10 \log_{10}{\left(\nicefrac{255^2}{\text{MSE(original, encoded)}}\right)}$, although any other quality measure such as SSIM \citep{ssim} or VMAF \citep{vmaf} can be used.

The rate control problem can be formulated as the Constrained MDP (CMDP) tuple $\langle S, A, P, R_\psnr, $ $ \gamma=1, c_\bitrate, \beta_\text{Target} \rangle$.  The state space $S$ consists of the first-pass statistics and the frame information generated by the encoder (described in section \ref{section:architecture}).  The action space $A$ consists of an integer QP in the range $[0, 255]$ which is applied to the frame being encoded. The transition function $P(s' | s, a)$ transitions to a new state $s' \in S$ (next frame to be encoded) given that action $a\in A$ (QP) was executed from state $s$ (current frame being encoded) as seen in Figure \ref{fig:qp-problem}. The bounded reward function $R_\psnr$ is defined as the PSNR of the encoded video at the final timestep of the episode and 0 otherwise. The constraint function $c_\bitrate$ is the bitrate of the encoded video at the final timestep of the episode and 0 otherwise.  $\beta_\text{Target}$ is the user-specified size target for the encoding.  We consider the size target range of [256, 768] kbps (kilobits-per-second) for 480p videos in this paper, although our methods can be extended to larger ranges.  The rate control constrained objective to optimize is defined as:

\begin{equation}
  \label{eqn:vp9_cmdp_objective}
    \max_{\pi\left(\text{QP}_t|s_t, \beta_\text{target}\right)} J_\pi^\psnr \;\;\;\;\text{s.t.}\, J_\pi^\bitrate \leq \beta_\text{Target}
\end{equation} 

As mentioned in Section \ref{sec:cmdp}, it is possible to solve the unconstrained Lagrangian relaxation of the CMDP objective by performing alternating optimization on the Lagrange parameter $\lambda$ and the policy $\pi$. However, as mentioned in previous works (e.g., \citep{rcpo}), multi-timescale stochastic approximation is very sensitive to learning rates and is therefore typically difficult to tune. Additionally, for learning a rate control policy, a large number of Lagrangian parameters would be needed to solve the constrained RL objective (per video and per target bitrate) as there is no single multiplier that is optimal across videos and bitrates. We validated this by combining \muzero{} with the Lagrangian relaxation method and observed that it was both difficult to tune during training and was unable to provide a reasonable solution in terms of quality and constraint satisfaction during evaluation. A more in-depth analysis can be found in the appendix \ref{appendix:lagrange}.  As such, we propose a different approach which we refer to as self-competition which solves a slightly different form of the above CMDP objective and removes the need to use Lagrangian parameters. 

\subsection{Constrained RL via Self-Competition}

In this section, we present the self-competition reward mechanism that enables solving the rate control constrained optimization problem defined in Equation \ref{eqn:vp9_cmdp_objective}. The high-level intuition of this reward mechanism is that the agent attempts to outperform its own historical performance on the constrained objective over the course of the training. We adapt \muzero{} to perform self-competition and refer to this agent as \emph{\muzero{}-Rate-Controller} (\muzerorc{}).  Note that this mechanism differs from \muzero{}'s self-play as self-play runs two-player games with the latest version of the agent, while self-competition agent is competing against its own historical performance in a single player setting.

To track the historical performance of the agent, we maintain exponential moving averages (EMA) of PSNR and overshoot (bitrate - target $\beta$) for each unique [video, target bitrate $\beta$] pair over the course of the training in a buffer.  When the agent newly encodes a video with a target bitrate $\beta$, the encoder computes the PSNR and overshoot of the episode ($P_\text{episode}$ and $O_\text{episode}$ respectively).  The agent looks up the historical PSNR and overshoot ($P_\text{EMA}$ and $O_\text{EMA}$ respectively) for that [video, target bitrate $\beta$] pair from the buffer. We first compare $O_\text{episode}$ with the $O_\text{EMA}$, and then compare $P_\text{episode}$ with $P_\text{EMA}$.  The return for the episode is set to $\pm 1$ depending on whether the episode is better than the EMA-based historical performance according to equation \ref{eqn:self-compete}.  We also update the EMAs in the buffer for the [video, target bitrate $\beta$] pair with the $P_\text{episode}$ and $O_\text{episode}$.  Algorithm \ref{alg:binary_reward_algo}  provides a detailed pseudocode of this self-competition algorithm.

\begin{equation}
\label{eqn:self-compete}
    \text{Return} =
    \underbrace{\sgn(P_\text{episode} - P_\text{EMA})}_{\text{Improve PSNR}}
    \underbrace{\mathds{1}_{O_\text{EMA} \leq 0, O_\text{episode} \leq 0}}_{\text{if constraints are satisfied}}
    - \underbrace{\sgn( \cliptozero{O_\text{episode}} - \cliptozero{O_\text{EMA}})}_\text{otherwise, ensure constraints are satisfied}
\end{equation}

where $\sgn$ is the sign function, $\mathds{1}$ is the indicator function, and $\cliptozero{x} = \max(x, 0)$.
 
To understand why this solution is suitable for variable difficulty constrained RL, we can interpret the EMA-based historical performance as a baseline for the agent to beat.  As the agent learns to beat this baseline, the baseline itself improves via EMA updates. This leads the agent to improve its performance further to beat the newer baseline, which creates a cycle of performance improvement.  Since the performance is measured by comparing bitrate constraint satisfaction first, the agent learns to satisfy the constraint first, and then learns to improve the PSNR.  Unlike Lagrangian relaxation methods, this mechanism doesn’t contain any direct parameters which represent the difficulty of constraint satisfaction.  This allows the agent to infer the difficulty of satisfying the constraints from the observations.

This formulation can be related to applying Lagrangian relaxation methods on an augmented CMDP.  The state of the CMDP can be augmented with EMAs, and the constrained objective becomes

\begin{align*}
  \max_{\pi(\text{QP}_t|s_t, P_\text{EMA}, O_\text{EMA}, \beta_\text{Target})}\;\;\; &
  \sgn(P_\text{episode} - P_\text{EMA})\mathds{1}_{O_\text{EMA} \leq 0, O_\text{episode} \leq 0}
  \\
  \text{s.t. }\;\;\;&\sgn( \cliptozero{O_\text{episode}} - \cliptozero{O_\text{EMA}}) \leq 0
\end{align*}

Using a Lagrangian relaxation method with $\lambda=1$ on this objective results in the same return as equation \ref{eqn:self-compete}.  Reward mechanisms with historical performance based self-competition have been proposed in the past \citep{rankedreward, selfplay_reward}, although they haven’t been applied to constrained RL to the best of our knowledge.

\subsection{Agent Architecture}\label{section:architecture}
Our training setup closely follows \citet{muzero}. We train this agent in the canonical asynchronous distributed actor-learner setting with experience replay \citep{replay, actorlearner}.  We maintain a shared buffer across all actors to track agent's EMA-based historical performance for each [video, target bitrate] pair.  Actor processes sample [video, target bitrate] pairs randomly from the training dataset, and generate QPs for encoding them using the latest network parameters and 200 simulations of \muzero{}'s MCTS algorithm.  The self-competition based reward for the episode is computed using equation \ref{eqn:self-compete} and the episode is added to the experience replay buffer.  The learner process samples transitions from the experience replay buffer to train the networks, and sends the updated parameters to the actor processes at regular intervals.

\textbf{Features:} At every step in the encoding process, the agent receives an observation from the encoder containing first-pass statistics for the video, PSNR and bits for the frames in the video encoded so far, index and type of the next frame to be encoded, and the target bitrate for the encoding.  Appendix \ref{appendix:architecture} lists the features and preprocessing methods in detail.

\textbf{Network Architecture: } Similar to \muzero{}, the agent has three subnetworks. The representation network maps the features to an embedding of the state using Transformer-XL \citep{transformerxl} encoders and ResNet-V2 style blocks \citep{resnet}.  The dynamics network generates the embedding of the next state given the embedding of the previous state and an action using ResNet-V2 style blocks. Given an embedding, the prediction network computes a policy using a softmax over the QPs, the value of the state, and several auxiliary predictions of the metrics related to the encoding using independent layer-normalised \citep{layernorm} feedforward networks.  Appendix \ref{appendix:architecture} lists these auxiliary predictions and describes the architecture of the subnetworks in detail.

\textbf{Training: } At every step in the training, the learner uniformly samples a batch of states, five subsequent actions, and the necessary labels from the experience replay buffer.  The representation network generates the embedding of the state, and the dynamics network is unrolled five times to generate the subsequent embeddings.  The prediction network outputs the predictions for policy, value, and auxiliary metrics for the current and the five subsequent states.  The policy prediction $\hat{\pi}_t$ is trained to match the MCTS policy generated at acting time $\pi_t$ using cross-entropy loss.  The value prediction $\hat{v}_t$ is trained to match the self-competition reward $v_t$ for the episode using IQN loss \citep{iqn}, and the auxiliary predictions $\hat{y}_t$ are trained to match the corresponding metrics $y_t$ using a quantile regression loss \citep{qr}.  We use equation \ref{eq:loss} to combine the losses.

\begin{equation}
    \label{eq:loss}
    \text{Loss} = \frac{1}{6}\sum_{t=0}^{5}
    \left[
     \underbrace{L_\text{CE}(\pi_t, \hat{\pi}_t)}_\text{Policy}
      +
     \underbrace{0.5\cdot L_\text{IQN}(v_t, \hat{v}_t)}_\text{Value}
      + 0.1\cdot\sum_\text{Auxiliary}L_\text{QR}(y_t, \hat{y}_t)
    \right]
    +
    \underbrace{10^{-3}\cdot\lVert\theta\rVert_2^2}_\text{L2-Reg}
\end{equation}

In our experiments, we found the auxiliary losses crucial for learning the dynamics of the encoding in order to perform the MCTS.  The hyperparameters of the training are described in appendix \ref{appendix:training}.

\subsection{Augmented \muzerorc{}} We also propose a small variant of the \muzerorc{} agent. We augment the self-competition mechanism in equation \ref{eqn:self-compete} by replacing the $\psnr$ term with `$\psnr - 0.005\times\overshoot$' in order to get the agent to reduce bitrate if it can't improve PSNR.   We label this agent `Augmented \muzerorc{}' and report the results alongside \muzerorc{} in section \ref{section:expt}.  The factor of 0.005 was selected as it is roughly the average slope of VP9's RD curve across the training dataset.  In our experiments, we didn't find significant performance difference with small adjustments to this factor.

\section{Experiments}
\label{section:expt}

In this section, we first describe the experiment setup.  This includes an overview of the environment, the dataset of videos used for training and evaluation, and the evaluation metrics.  We then describe the performance of \muzerorc{} agents in this setup compared to the baselines.

\subsection{Setup}
\label{sec:setup}
\textbf{Environment setup and baseline: } We use the \libvpx{} implementation of VP9 in our experiments.
In particular, we implemented an RL environment based on the SimpleEncode API \citep{simpleencode} in \libvpx{}. This allows us to override the \libvpx{}'s QP decision with an external QP computed by our agent. We train and evaluate the learned rate control policy using the environment.
In our experiments, we use \libvpx{}'s two-pass variable bitrate (VBR) mode \citep{vp9modes}
at encode speed 1 (i.e., \verb|--cpu-used=1|), which is a commonly used setting for video-on-demand (VOD). We use \libvpx{}'s default rate control implementation as our baseline as it is canonically used in VP9.

\textbf{Dataset: } We limit the focus of our experiments to 480p videos with target bitrate in [256, 768] kbps range, although our methods can be generalised to other resolutions and bitrates. 
For training, we use 20,000 randomly selected 480p videos that have varying duration (3 to 7 seconds), varying aspect ratios, varying frame rates, and diverse content types. The target bitrate for training is uniformly sampled from \{256, 384, 512, 640, 768\} kbps.
For evaluation, we use videos with resolution 480p and higher from the YouTube UGC dataset \citep{youtube_ugc}, designed for benchmarking video compression and quality assessment research available under CC-BY 3.0 license. We resize the videos to 480p, and chunk them into 5-second clips, resulting in 3062 clips.

\textbf{Metrics: }
To evaluate the coding efficiency and overall bitrate reduction, we measure the Bjontegaard-delta rate (BD-rate) \citep{bdrate} using \libvpx{}'s default VBR rate control policy as the reference. Given the bitrate v.s.\ PSNR curves of two policies, BD-rate computes the average bitrate difference for the same PSNR across the overlapped PSNR range, and therefore, measures the average bitrate reduction for encoding videos at the same quality.  See figure \ref{fig:rd-curve-example} for an example.  We report the mean PSNR BD-rate difference compared to \libvpx{} on the evaluation set, as well as BD-rate difference with respect to perceptual quality measures SSIM and VMAF. To evaluate the bitrate constraint satisfaction, we report the fraction of evaluation cases on which the policies violate the bitrate constraint.  Since a small amount of constraint violation is acceptable in practice, we also report the fraction of cases on which the policies violate the constraint by >5\% of the target bitrate.  We also note that since PSNR and bitrate are competing objectives, the optimal behavior is to use the target bitrate budget fully in order to maximize the PSNR.  To evaluate this, we report the fraction of cases on which the policies achieve bitrate within 5\% of the target.  We report the histograms of BD-rate differences and bitrates in appendices \ref{appendix:bd_rate_hists} and \ref{appendix:overshoots}.

\textbf{Evaluation: }  We use the final network checkpoint generated by the training process for evaluation.  Practically, the rate control step should not take a large amount of time.  So, instead of performing \muzero{}-style MCTS, we select the QPs with the highest probability assigned by the policy output of the prediction network.  We iterate over the evaluation dataset and encode each video with \libvpx{} rate control and both the \muzerorc{} agents at nine target bitrate values uniformly spaced between [256, 768] kbps.  We evaluate the constraint satisfaction behavior of the agent for all nine target bitrates, and use the RD-curve generated by these encodings for each video to compute the BD-rate difference.

\subsection{Results}
\textbf{PSNR BD-Rate improvement: }First, we evaluate the PSNR BD-rate of \muzerorc{} agents compared to \libvpx{}.  Table \ref{table:bdrate} shows the difference of PSNR BD-rate achieved by \muzerorc{} over the evaluation set videos compared to \libvpx{}'s rate control.  It shows that \muzerorc{} can achieve better PSNR v.s.\ bitrate tradeoff with 4.72\% average bitrate savings for encoding videos at same quality as \libvpx{}, and Augmented \muzerorc{} increases it further to 6.28\%.  The agent reduces the BD-rate with respect to perceptual quality measures SSIM and VMAF as well.  We present the histogram of BD-rate differences over the evaluation video set in appendix \ref{appendix:bd_rate_hists}.

\begin{table}[htb]
  \caption{BD-rate difference of \muzerorc{} agents compared to \libvpx{} with respect to various quality measures.  We report $\pm 1$ standard error interval computed over 5 random seeds. As seen in the table, both \muzerorc{} and Augmented \muzerorc{} lead to BD-rate reductions compared to the \libvpx{} baseline with respect to PSNR, SSIM and VMAF metrics.}
  \label{table:bdrate}
  \centering
  \begin{tabular}{cccc}
    \toprule
    & Mean PSNR  & Mean SSIM  & Mean VMAF \\
    Agent &   BD-rate difference & BD-rate difference & BD-rate difference \\
    \midrule
    \muzerorc{} & -4.72\% $\pm$ 0.32\% &  -3.68\% $\pm$ 0.33\% & -0.53\% $\pm$ 0.21\% \\
    Augmented \muzerorc{}& \textbf{-6.28\%} $\pm$ 0.19\% & \textbf{-5.11\%} $\pm$ 0.24\% & \textbf{-1.88\%} $\pm$ 0.12\% \\
    \bottomrule
  \end{tabular}
\end{table}

\textbf{Constraint satisfaction: } Then, we evaluate the bitrate constraint satisfaction behaviour.  Table \ref{table:overshoot} lists the constraint satisfaction metrics achieved by \libvpx{} baseline policy and the \muzerorc{} agents.  The \muzerorc{} agent violates the constraints on substantially fewer videos compared to \libvpx{}.  Even when the constraints are violated, large magnitude violations by \muzerorc{} are less frequent compared to \libvpx{}.   Also, \muzerorc{} agent encodes videos with bitrates within 5\% margin around the target bitrate more frequently than \libvpx{}.  This indicates that \muzerorc{} not only overshoots less, but it also tends to fully use the target bitrate budget more frequently compared to \libvpx{}.  We also see that Augmented \muzerorc{} satisfies constraints just as well as \muzerorc{}, and its bitrate accuracy is comparable to \libvpx{}.  We present the histogram of overshoots over the evaluation video set for different target bitrates in appendix \ref{appendix:overshoots}.  Figure \ref{fig:overshoot_ex1} demonstrates a typical behavior of \muzerorc{} regarding bitrate control. As shown in Figure \ref{fig:overshoot_ex1_cumu_bits}, the \libvpx{} baseline overshoots the target bitrate by about 5\%, while \muzerorc{} is accurate in allocating the target bitrate (marked by black line). \muzerorc{} avoids overshooting by using fewer bits (and thus sacrificing the PSNR) in early frames of the video as shown in Figure \ref{fig:overshoot_ex1_frame_psnr}. This is a good decision because the PSNR of the first part of the video is relatively high. In fact, if a policy sacrifices the PSNR in the last 40 frames, it would hurt the overall PSNR. This example highlights agent's ability to reason about frame complexity and bitrate-PSNR tradeoff, and plan accordingly. 

\begin{table}[htb]
  \caption{Bitrate constraint satisfaction behavior over the evaluation set.  We report $\pm 1$ standard error interval computed over 5 different random seeds for \muzerorc{} agents.  As \libvpx{} rate control is a  deterministic process, we do not report the standard error intervals for it.  As seen in the table, both \muzerorc{} and Augmented \muzerorc{} overshoot the target bitrate for substantially fewer test cases compared to \libvpx{}.  \muzerorc{} agent achieves higher bitrate accuracy compared to \libvpx{} while Augmented \muzerorc{} achieves a comparable accuracy.}
  \label{table:overshoot}
  \centering
  \begin{tabular}{cccc}
    \toprule
    & \multicolumn{3}{c}{Fraction of videos with} \\
    \cmidrule(r){2-4}
    &  & overshoot > 5\% & bitrate within 5\% \\
    Agent & overshoot > 0 & of target & of target \\
    \midrule
    libvpx & 64.00\%  & 6.13\%  &  71.34\%  \\
    \muzerorc{} & 20.34\% $\pm$ 2.16\% & \textbf{2.04\%} $\pm$ 0.20\% & \textbf{84.14\%} $\pm$ 0.68\% \\
    Augmented \muzerorc{} & \textbf{16.10\%} $\pm$ 2.06\% & 2.25\% $\pm$ 0.15\% & 70.12\% $\pm$ 1.15\%  \\
    \bottomrule
  \end{tabular}
\end{table}

\begin{figure}[htb]
	\centering
	\begin{subfigure}[c]{0.49\columnwidth}
	   \centering
    	\includegraphics[width=\textwidth]{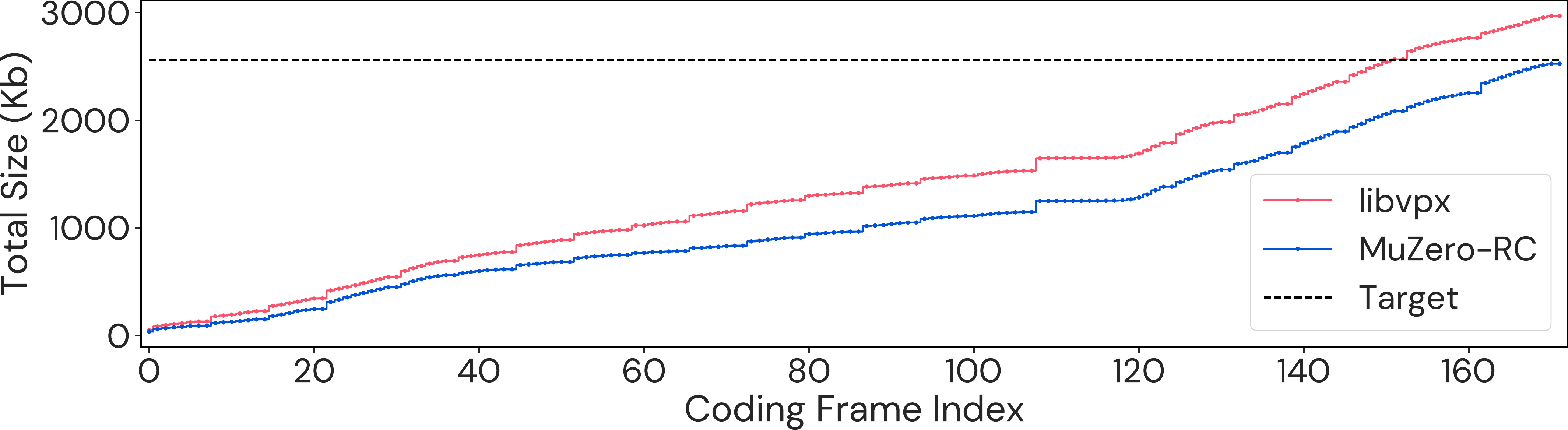}
    	\caption{Cumulative frame bits trajectories.}
    	\label{fig:overshoot_ex1_cumu_bits}
    \end{subfigure}
    \begin{subfigure}[c]{0.49\columnwidth}
	   \centering
    	\includegraphics[width=\textwidth]{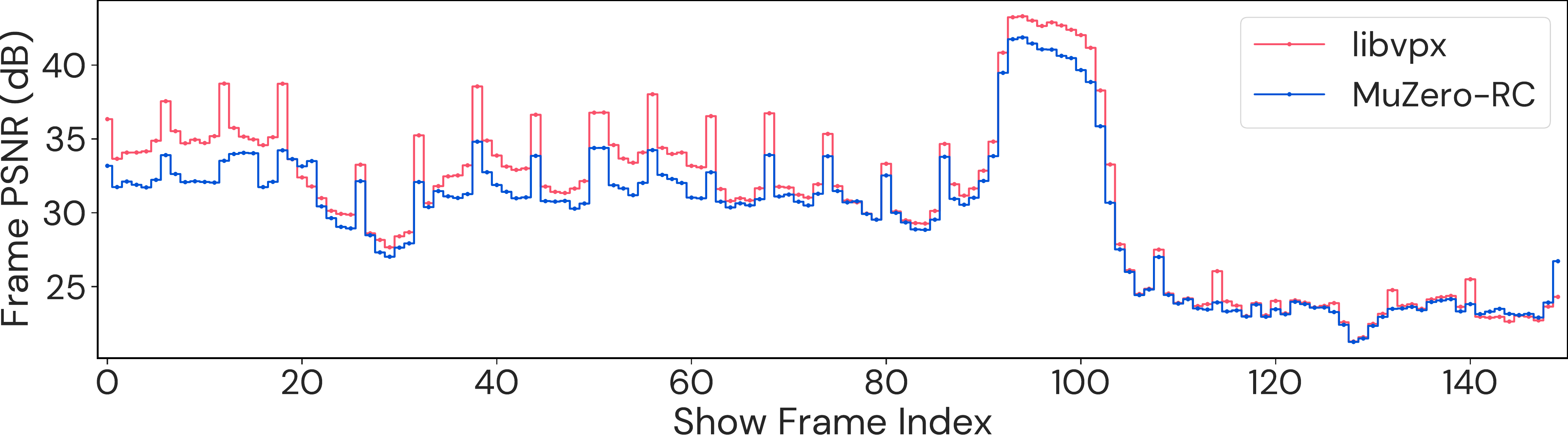}
     	\caption{Show frame PSNR trajectories.}
    	\label{fig:overshoot_ex1_frame_psnr}
    \end{subfigure}
    \caption{Encoding trajectories of video {\texttt{LyricVideo\_480P-1484}}. The learned policy sacrifices PSNR in early frames in order to save bits to avoid bitrate overshooting in the end.}
    \label{fig:overshoot_ex1}
\end{figure}

\begin{figure}[ht]
	\centering
	\begin{subfigure}[c]{0.49\columnwidth}
	   \centering
    	\includegraphics[width=\textwidth]{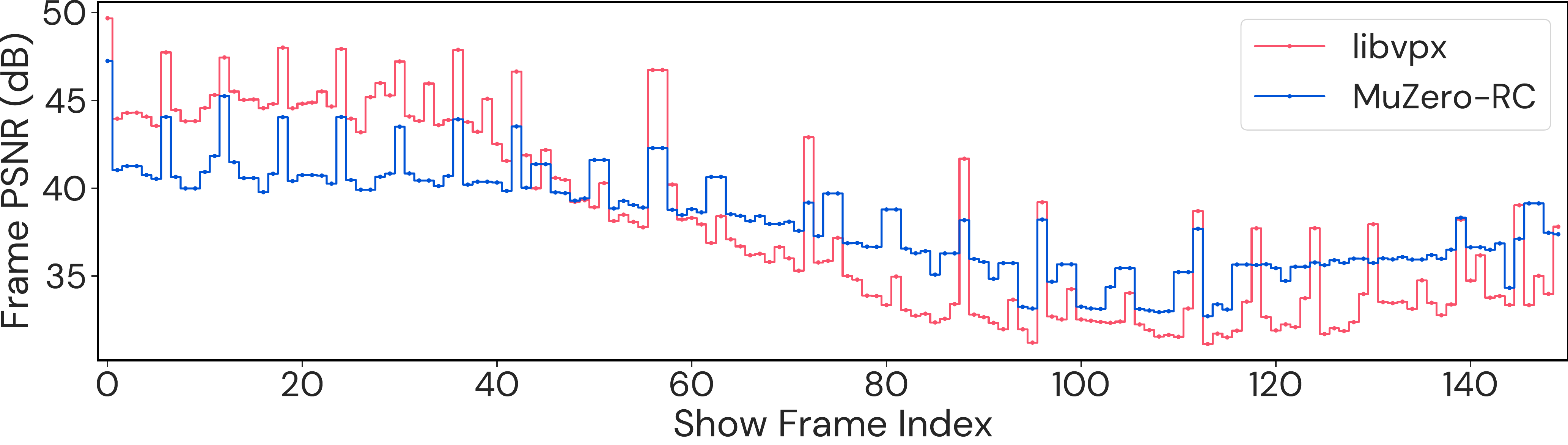}
    	\caption{\texttt{Animation\_480P-073c}.}
    	\label{fig:better_bdrate_ex1}
    \end{subfigure}
    \begin{subfigure}[c]{0.49\columnwidth}
	   \centering
    	\includegraphics[width=\textwidth]{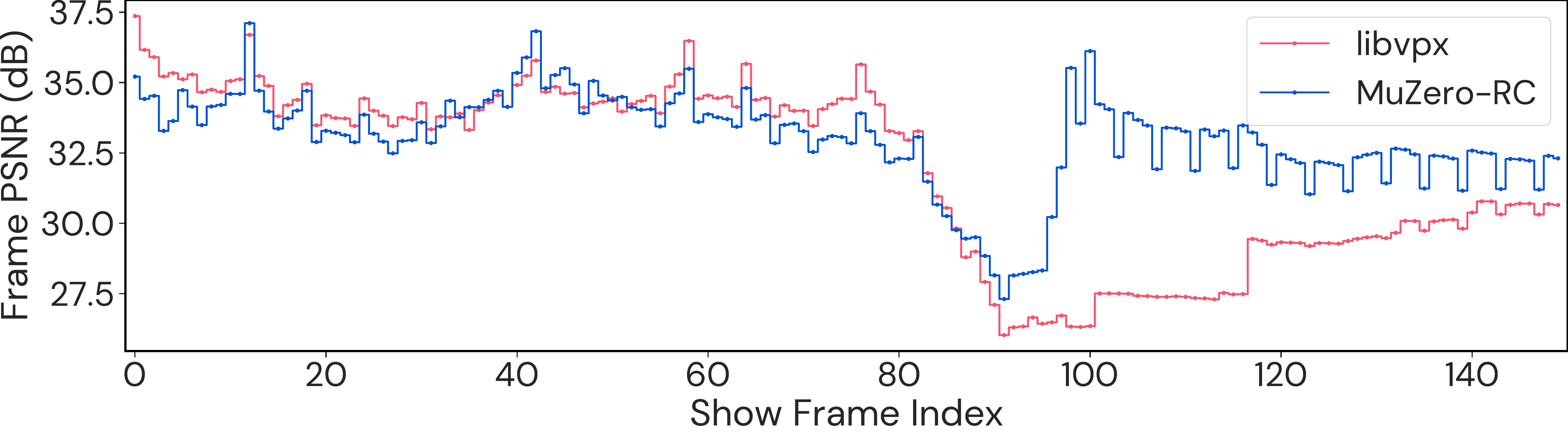}
     	\caption{\texttt{Gaming\_1080P-26dc}.}
    	\label{fig:better_bdrate_ex2}
    \end{subfigure}
    \begin{subfigure}[c]{0.49\columnwidth}
	   \centering
    	\includegraphics[width=\textwidth]{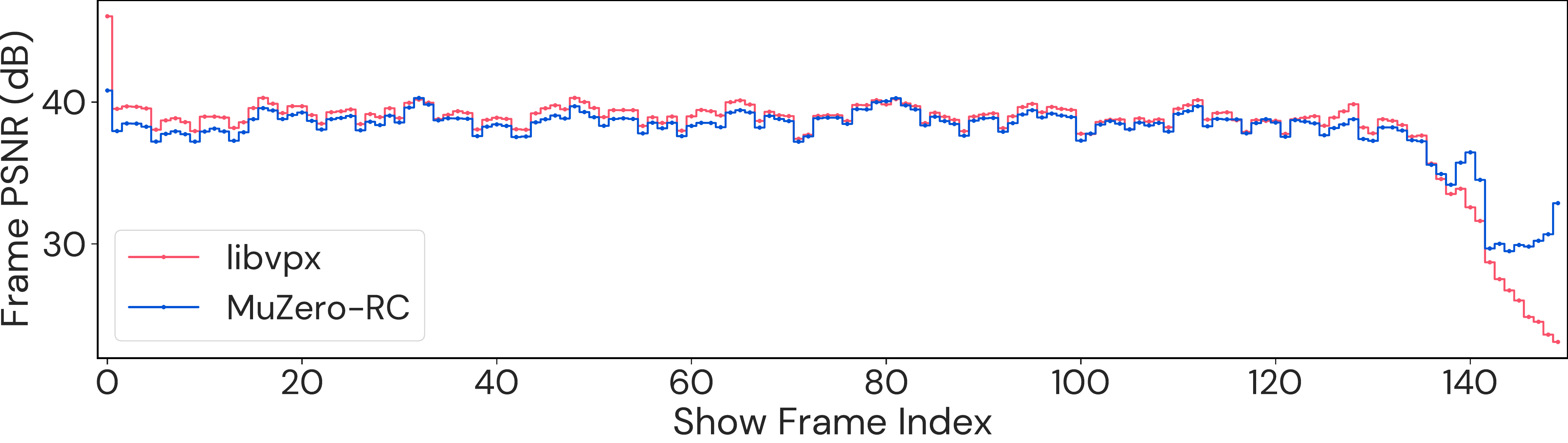}
     	\caption{\texttt{CoverSong\_720P-05d6}.}
    	\label{fig:better_bdrate_ex3}
    \end{subfigure}
    \begin{subfigure}[c]{0.49\columnwidth}
	   \centering
    	\includegraphics[width=\textwidth]{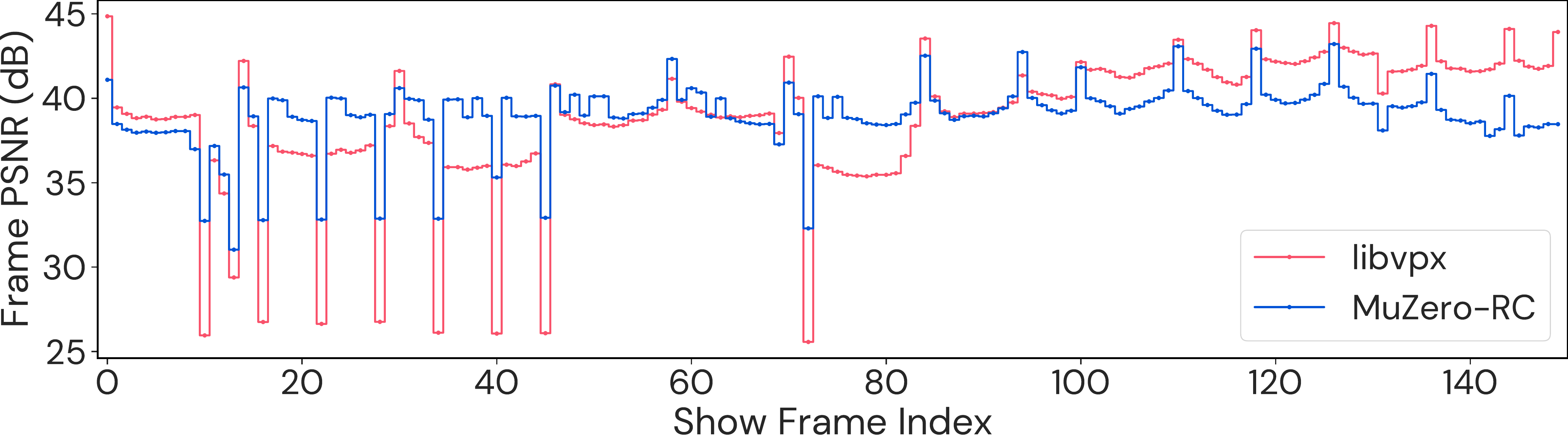}
     	\caption{\texttt{LiveMusic\_480P-58fb}.}
    	\label{fig:better_bdrate_ex4}
    \end{subfigure}
    \caption{Show frame PSNR trajectories of 4 video clips.}
    \label{fig:better_bdrate_examples}
\end{figure}

\textbf{Planning performance: } We investigate the encoding trajectories of videos and observe a few recurring patterns where \muzerorc{} demonstrates better planning performance than \libvpx.
Figure \ref{fig:better_bdrate_examples} shows examples where \muzerorc{} achieves better BD-rate. 
In Figure \ref{fig:better_bdrate_ex1}, \muzerorc{} sacrifices the PSNR of the first 60 frames in order to improve the PSNR of the rest of the frames. Because the latter part of the video is more complex (indicated by lower PSNR), improving the PSNR of the latter part of the video improves the overall PSNR. 
In Figure \ref{fig:better_bdrate_ex2}, \muzerorc{} boosts PSNR of a reference frame, leading to significant PSNR improvement for all the frames following that frame.
In Figure \ref{fig:better_bdrate_ex3}, the video has a scene change towards the very end. Because \libvpx{} runs out of bitrate budget, it decides to encode the last few frames with low quality. In contrast, \muzerorc{} correctly anticipates the scene change, and saves some bitrate budget from the first 140 frames so that it can obtain a reasonable PSNR for the last few frames.
Figure \ref{fig:better_bdrate_ex4} shows another pattern where \muzerorc{} has overall less frame PSNR fluctuation. In particular, it avoids having extremely low PSNR on some frames, which in turn improves the overall PSNR.

\section{Conclusions}
\label{section:conclusions}
In this paper, we have demonstrated that the \muzero{} reinforcement learning algorithm can be used for rate control in VP9.  Our formulation of the self-competition based reward mechanism allows the agent to tackle the complex constrained optimization task and achieve better quality-bitrate tradeoff and better bitrate constraint satisfaction than \libvpx{}'s VBR rate control algorithm.  The final agent results in 6.28\% average reduction in bitrate (measured as PSNR BD-rate) on videos from the evaluation set, and can be readily deployed in \libvpx{} via the SimpleEncode API.

\textbf{Limitations: } The self-competition based reward mechanism requires that every unique [video, target bitrate] pair be encoded a few times so that the historical performance converges and provides a reasonable baseline for reward computation.  Because of this, the amount of data the actors need to generate increases linearly with the number of videos in the training dataset and the number of target bitrate samples.  For very large training datasets, this method might not scale well.  However, in future work, it may be possible to learn these baseline values based on observations using a neural network which can generalise to unseen videos in a large dataset.  

\textbf{Future Work: } Our proposed methods are agnostic to the specifics of VP9/\libvpx{}, and they can potentially be generalized not only to other coding formats and implementations, but also to other components within video encoders such as block partitioning and reference frame selection. Our method also opens the possibility of allowing codec developers and users to develop new rate control modes. For example, we can replace PSNR with other video quality metrics such as VMAF. We can also modify the reward to minimize bitrate given a minimum PSNR constraint --  which is similar to the constrained quality (CQ) mode in \libvpx{}, but reinforcement learning is likely to learn a policy that has more precise control of the PSNR.

\begin{ack}
We would like to thank Balu Adsumilli, Ross Wolf, Yaowu Xu, James Bankoski, Nishant Patil, Marisabel Hechtman, Dan A.\ Calian, Luis C.\ Cobo, Chris Gamble, and David Silver for the insightful discussions about this work and for their support throughout the project.
\end{ack}

\bibliographystyle{abbrvnat}
\bibliography{refs}

\appendix

\section{Appendix}

\subsection{Network Architecture}
\label{appendix:architecture}

As required by the \muzero{} algorithm, the \muzerorc{} agent has three subnetworks.

\textbf{Representation Network: } This network takes the features provided by the environment as the input and produces an embedding of the current state of the environment as the output.  For any state, the environment generates following observations.

\begin{enumerate}
    \item A sequence of first-pass statistics for all the show frames in the video.
    \item A sequence of PSNR, number of used bits, and applied QPs for all the previously encoded frames in the video so far, along with indices of those frames.
    \item The index and the type of the frame to be encoded next.  The type can be one of five frame types from the SimpleEncode API.
    \item The duration of the video.
    \item Target bitrate for the encoding.
\end{enumerate}

We use the same first pass statistics and features normalization methods used by \citet{vp9imitiation}.  Additionally, we generate the fraction of the target bitrate used so far in the encoding using the bits used by previously encoded frames and video duration.  We use this fraction as an additional scalar feature.

\begin{figure}[ht]
    \centering
    \includegraphics[width=0.98\textwidth]{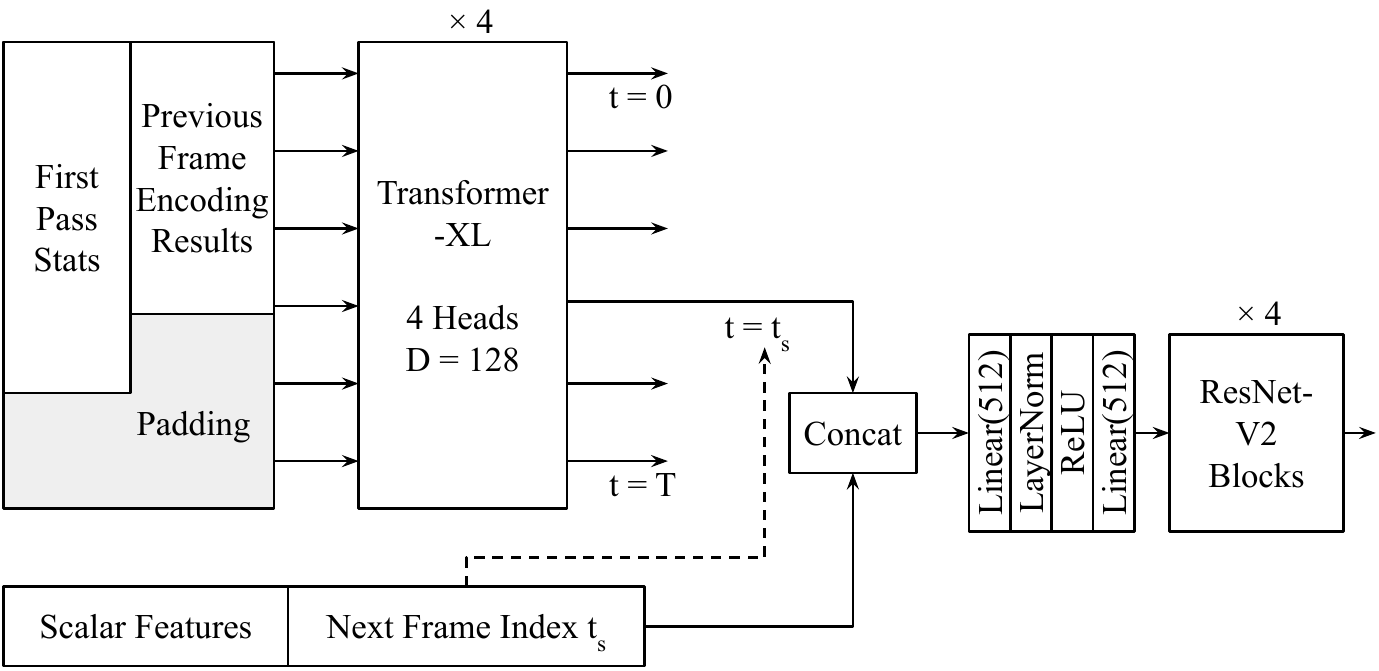}
    \caption{\muzerorc{} Representation Network}
    \label{fig:repr_net}
\end{figure}

The representation network aligns the first two sequential features along the indices of the frames and concatenates the aligned sequences along the feature dimension.  This concatenated sequence is then processed using a series of 4 Transformer-XL encoder blocks \citep{transformerxl}.  From this sequence, the entry at index of the frame to be encoded next is extracted.  This entry is concatenated with the remaining scalar features and processed using two feedforward layers with intermediate layer normalization \citep{layernorm}.  The network processes the output of these layers with a series of 4 layer normalized ResNet-V2 blocks \citep{resnet}.  The output of these blocks is the embedding of the state.  We use an embedding of 512 units in all our experiments.  All the layers use ReLU as the activation function.  Figure \ref{fig:repr_net} shows a diagram of this network.

\begin{figure}[ht]
    \centering
    \includegraphics[width=0.49\textwidth]{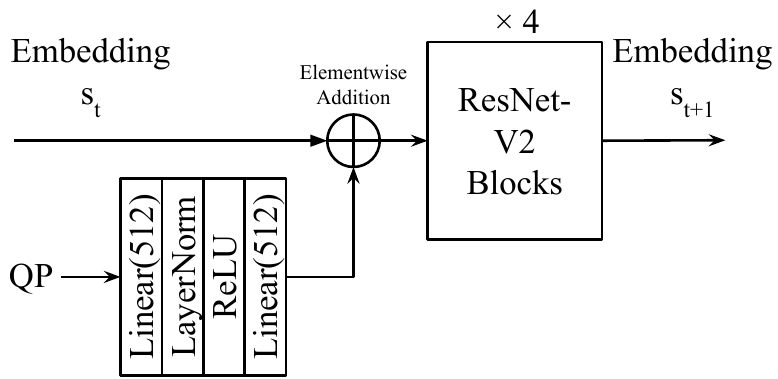}
    \caption{\muzerorc{} Dynamics Network}
    \label{fig:dyn_net}
\end{figure}

\textbf{Dynamics Network: } This network takes an embedding of the state and the QP to be applied in that state as the input.  It produces an embedding of the next state reached after applying the QP as output.  This network processes the QP using two feedforward layers with intermediate layer normalization to output a vector with the same dimension as the embedding of the previous state.  It performs elementwise addition of this vector and the embedding of the previous state, and processes the result with a series of 4 layer normalized ResNet-V2 blocks.   The output of these blocks is the embedding of the next state reached after applying the QP.   All the layers use ReLU as the activation function.  Figure \ref{fig:dyn_net} shows a diagram of this network.

\textbf{Prediction Network: } This network takes the embedding of a state and produces the policy, value, and several auxiliary predictions as the output.  For the policy prediction, the network processes the state embedding with two feedforward layers with 256 hidden units and layer normalization followed by a linear layer of 256 units representing the logits for each QP value.  A softmax function is applied to these logits to produce the policy.  For the value prediction, the network processes the state embedding with two feedforward layers with 256 hidden units and layer normalization followed by a linear layer of 64 units.  The output of this layer is used as an embedding for the IQN layer which produces samples of the value prediction.  We apply the tanh function to these samples to limit them in range (-1, 1) as the value in the self-competition based reward mechanism is limited to [-1, 1].  At training time, we draw 8 samples from the IQN layer to match the self-competition reward.  At inference time, we use the expected value instead of sampling.

For each of the auxiliary predictions, the network processes the state embedding with two feedforward layers with 256 hidden units and layer normalization followed by a linear layer of 64 units.  The output of this layer represents the uniformly spaced quantiles of the corresponding auxiliary prediction.  In our experiments, we predict the following metrics as auxiliary predictions.
\begin{enumerate}
    \item The PSNR of the last encoded frame (0 when no frames are encoded).
    \item The log of the number of bits used by the last encoded frame (0 when no frames are encoded).
    \item The expected PSNR of the video being encoded.
    \item The expected bitrate of the video being encoded.
\end{enumerate}

\begin{figure}[ht]
    \centering
    \includegraphics[width=0.8\textwidth]{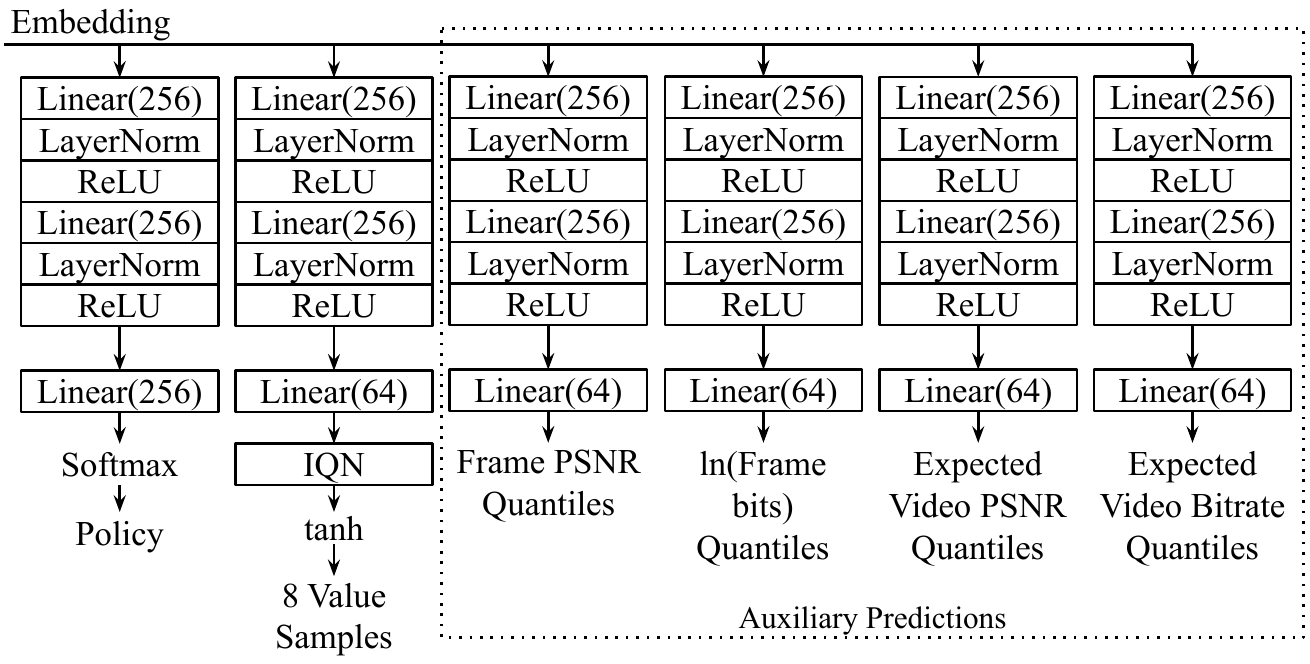}
    \caption{\muzerorc{} Prediction Network}
    \label{fig:pred_net}
\end{figure}

These auxiliary predictions help the agent understand the dynamics of the video encoding process, which we found to be crucial in our experiments.  All the layers use ReLU as the activation function unless specified otherwise.  Figure \ref{fig:pred_net} shows a diagram of this network.

All experiments in this paper are implemented using the JAX framework \citep{jax}.  We use the Haiku library to implement the subnetworks \citep{haiku}.

\subsection{Training}
\label{appendix:training}

We train all the subnetworks jointly similar to \citet{muzero} in order to minimize the loss in equation \ref{eq:loss}.  We use the SGD algorithm with momentum factor of 0.9 for optimization.  The learning rate for the optimization reduces over the course of training as follows

\begin{equation*}
    \text{lr}(t) = \text{lr}_\text{init} \cdot \text{decay}^{\frac{t}{\text{interval}}}
\end{equation*}

where t is the training step, $\text{lr}_\text{init}=0.05$, decay is 0.1 and interval is 300,000.  We use the Optax library \citep{optax} for optimization.  We train the agent for 1 million steps.  400 million frames are generated by the actor processes over the course of training.  For experience replay, we keep a buffer of 50,000 latest episodes generated by the actor processes, and draw samples with batch size 512 from the replay buffer.

\subsection{Computational Resources}
\label{appendix:compute}

All experiments in this paper are run using Google Cloud TPUs \citep{tpu}.  The learner processes use two 3rd generation TPUs, and the actor processes use four 2nd generation TPUs for 15 hours.  Additionally, we use 3000 CPU cores from a shared heterogeneous compute cluster for running video encoding with \libvpx{} for actors.

\subsection{Lagrangian-relaxation Method for Rate Control}
\label{appendix:lagrange}

To evaluate the hypothesis that Lagrangian-relaxation methods for constrained RL are not suitable for rate control, we train a \muzero{} agent with Lagrangian-relaxation based value function similar to \citet{rcpo}.  We maintain all other hyperparameters and architecture same as \muzerorc{} apart from the tanh activation on value network, which is removed in this agent as the value is no longer in [-1, 1].  We observe that this agent was difficult to tune, and we had to run multiple hyperparameter sweeps to get a stable instance.  We evaluated this instance using the protocol described in section \ref{section:expt} and report the metrics in tables \ref{table:lagrange_bdrate} and \ref{table:lagrange_overshoot} respectively.  We notice that the agent exhibits poor performance compared to \libvpx{} in terms of the PSNR BD-rate reduction.  We also note that while the agent satisfies the constraint for a reasonable number of test videos, it tends to use very little of the bitrate target budget as seen in the last column of table \ref{table:lagrange_overshoot}.  Also, the number of videos on which agent overshoots by >5\% of the target is significantly higher compared to the other policies.

\begin{table}[htb]
  \caption{BD-rate difference of Lagragnian-relaxation method compared to \libvpx{}.  Columns in this table are same as those in table \ref{table:bdrate}.}
  \label{table:lagrange_bdrate}
  \centering
  \begin{tabular}{cccc}
    \toprule
    & Mean PSNR  & Mean SSIM  & Mean VMAF \\
    Agent &   BD-rate difference & BD-rate difference & BD-rate difference \\
    \midrule
    Lagrangian & 256.37\% & 41.44\% & 195.12\%\\
    \muzerorc{} & -4.72\% $\pm$ 0.32\% &  -3.68\% $\pm$ 0.33\% & -0.53\% $\pm$ 0.21\% \\
    \bottomrule
  \end{tabular}
\end{table}

\begin{table}[htb]
  \caption{Bitrate constraint satisfaction behavior of Lagrangian-method over the evaluation set.  Columns in this table are same as those in table \ref{table:overshoot}.}
  \label{table:lagrange_overshoot}
  \centering
  \begin{tabular}{cccc}
    \toprule
    & \multicolumn{3}{c}{Fraction of videos with} \\
    \cmidrule(r){2-4}
    &  & overshoot > 5\% & bitrate within 5\% \\
    Agent & overshoot > 0 & of target & of target \\
    \midrule
    libvpx & 64.00\%  & 6.13\%  &  71.34\%  \\
    Lagrangian & 25.83\% & 25.49\% & 0.68\% \\
    \muzerorc{} & 20.34\% $\pm$ 2.16\% & 2.04\% $\pm$ 0.20\% & 84.14\% $\pm$ 0.68\% \\
    \bottomrule
  \end{tabular}
\end{table}

\newpage
\subsection{Pseudocode for Self-competition Reward}
\label{section:pseudocode}

Algorithm \ref{alg:binary_reward_algo} illustrates the pseudocode for the self-competition reward mechanism.  In our experiments, we set the parameters $\text{PSNR}_0=30\text{ dB}$ and $\alpha=0.9$.  

\begin{algorithm}[ht]
\caption{Self-competition reward mechanism for \muzero{}-RC}\label{alg:binary_reward_algo}

\begin{algorithmic}[1]

\State Dataset $\mathcal{D}$ of videos with assigned target bitrates

\State Buffer $\mathcal{B} \leftarrow \left[[\text{Video}_i, \beta_i] \rightarrow \left(\text{PSNR}_0, 0\right) \forall\,\text{videos} \right] $

\State $\alpha:\,$ Buffer update factor

\Repeat \Comment{Actor Process}
\State video$_i$, target\_bitrate $\beta_i \sim \mathcal{D}$\Comment{Sample video and target bitrate from dataset}

\State $\psnr_\text{Episode}$, $\bitrate_\text{Episode} \gets$ encode\_video(video$_i\vert\beta_i$, agent policy)
\State\Comment{Encode video using agent policy and compute PSNR and bitrate}

\State

\State\Call{update\_buffer}{$\mathcal{B}$, video$_i$, $\psnr_\text{Episode}$, $\overshoot_\text{Episode}$,  $\beta_i$, $\alpha$}

\State Return $\gets$ \Call{self\_competition\_reward}{$\mathcal{B}$, video$_i$, $\psnr_\text{Episode}$, $\overshoot_\text{Episode}$, $\beta_i$}

\Until{end}

\State
\Procedure{update\_buffer}{$\mathcal{B}$, video$_i$, $\psnr_\text{Episode}$, $\overshoot_\text{Episode}$, $\beta_i$, $\alpha$}

\State $\mathcal{B}\left[\text{video}_i, \beta_i\right][0] \gets (1-\alpha)\cdot\mathcal{B}\left[\text{video}_i, \beta_i\right][0] + \alpha\cdot\psnr_\text{Episode}$
\State   $\mathcal{B}\left[\text{video}_i, \beta_i\right][1] \gets (1-\alpha)\cdot\mathcal{B}\left[\text{video}_i, \beta_i\right][1] + \alpha\cdot\overshoot_\text{Episode}$
\EndProcedure

\State

\Procedure{self\_competition\_reward}{$\mathcal{B}$, video$_i$, $\psnr_\text{Episode}$, $\overshoot_\text{Episode}$, $\beta_i$}
\State\Comment{Equivalent to equation \ref{eqn:self-compete}}
  \State $\psnr_\text{EMA}$, $\overshoot_\text{EMA} \gets \mathcal{B}\left[\text{video}_i, \beta_i\right]$
  \If{$\overshoot_\text{Episode}$ $>0$ \textbf{or} $\overshoot_\text{EMA}$ $>0$}
      \If{$\overshoot_\text{Episode}$ $\le$ $\overshoot_\text{EMA}$}
        \State \textbf{return} $1$
      \Else
        \State \textbf{return} $-1$
      \EndIf
    \Else
      \If{$\psnr_\text{Episode}$ $\geq$ $\psnr_\text{EMA}$}
        \State \textbf{return} $1$
      \Else
        \State \textbf{return} $-1$
      \EndIf
  \EndIf
\EndProcedure

\end{algorithmic}
\end{algorithm}

\newpage
\subsection{Histograms of BD-Rate differences compared to \libvpx{}}
\label{appendix:bd_rate_hists}

\begin{figure}[ht]
	\centering
	\begin{subfigure}[c]{0.48\columnwidth}
	   \centering
    	\includegraphics[width=\textwidth]{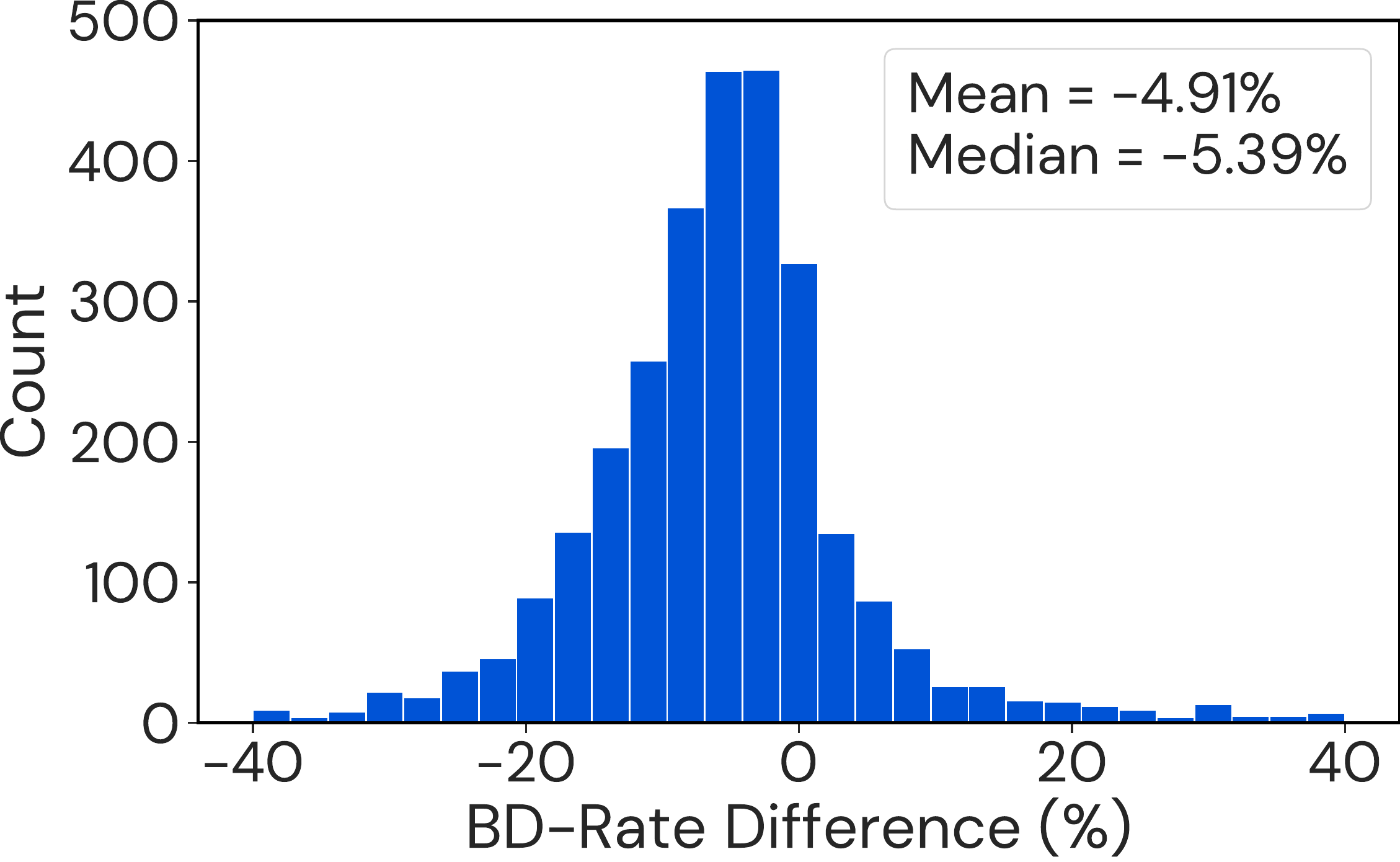}
    	\caption{\muzerorc{}}
    \end{subfigure}
    \begin{subfigure}[c]{0.48\columnwidth}
	   \centering
    	\includegraphics[width=\textwidth]{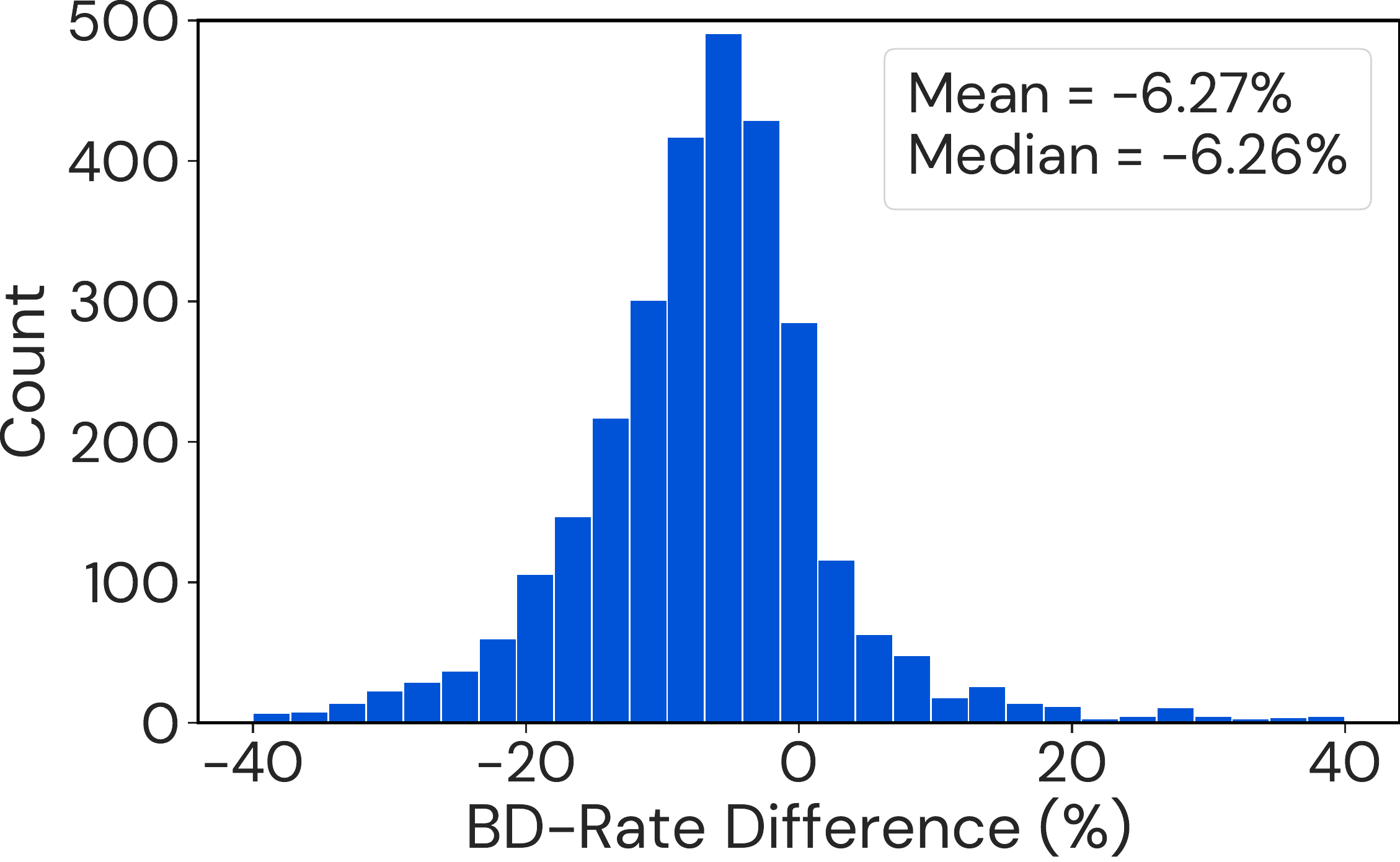}
     	\caption{Augmented \muzerorc{}}
    \end{subfigure}
    \caption{Histogram of PSNR BD-Rate difference of agent compared to \libvpx{} on evaluation set.}
    \label{figure:psnr_hist}
\end{figure}

\begin{figure}[ht]
	\centering
	\begin{subfigure}[c]{0.48\columnwidth}
	   \centering
    	\includegraphics[width=\textwidth]{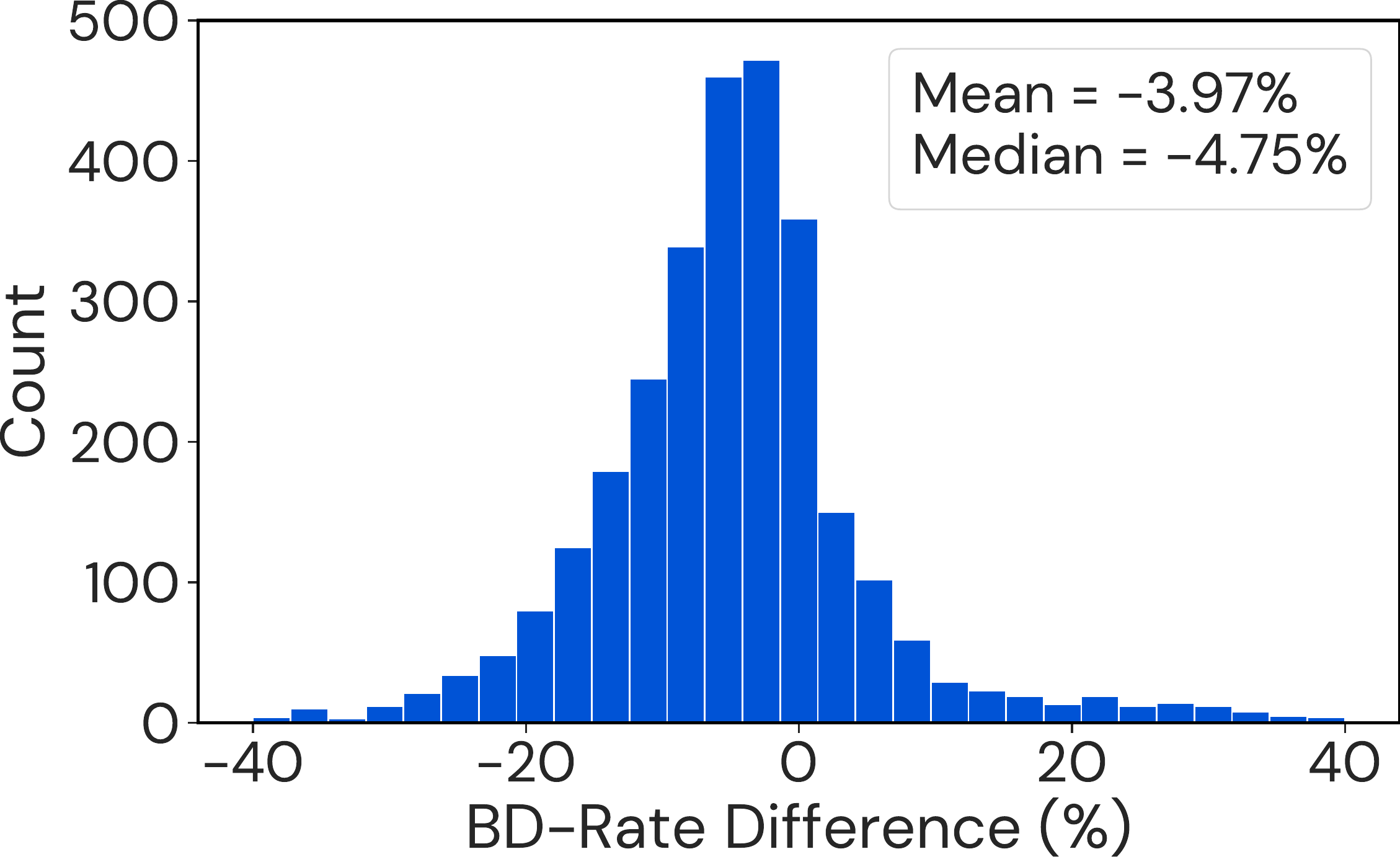}
    	\caption{\muzerorc{}}
    \end{subfigure}
    \begin{subfigure}[c]{0.48\columnwidth}
	   \centering
    	\includegraphics[width=\textwidth]{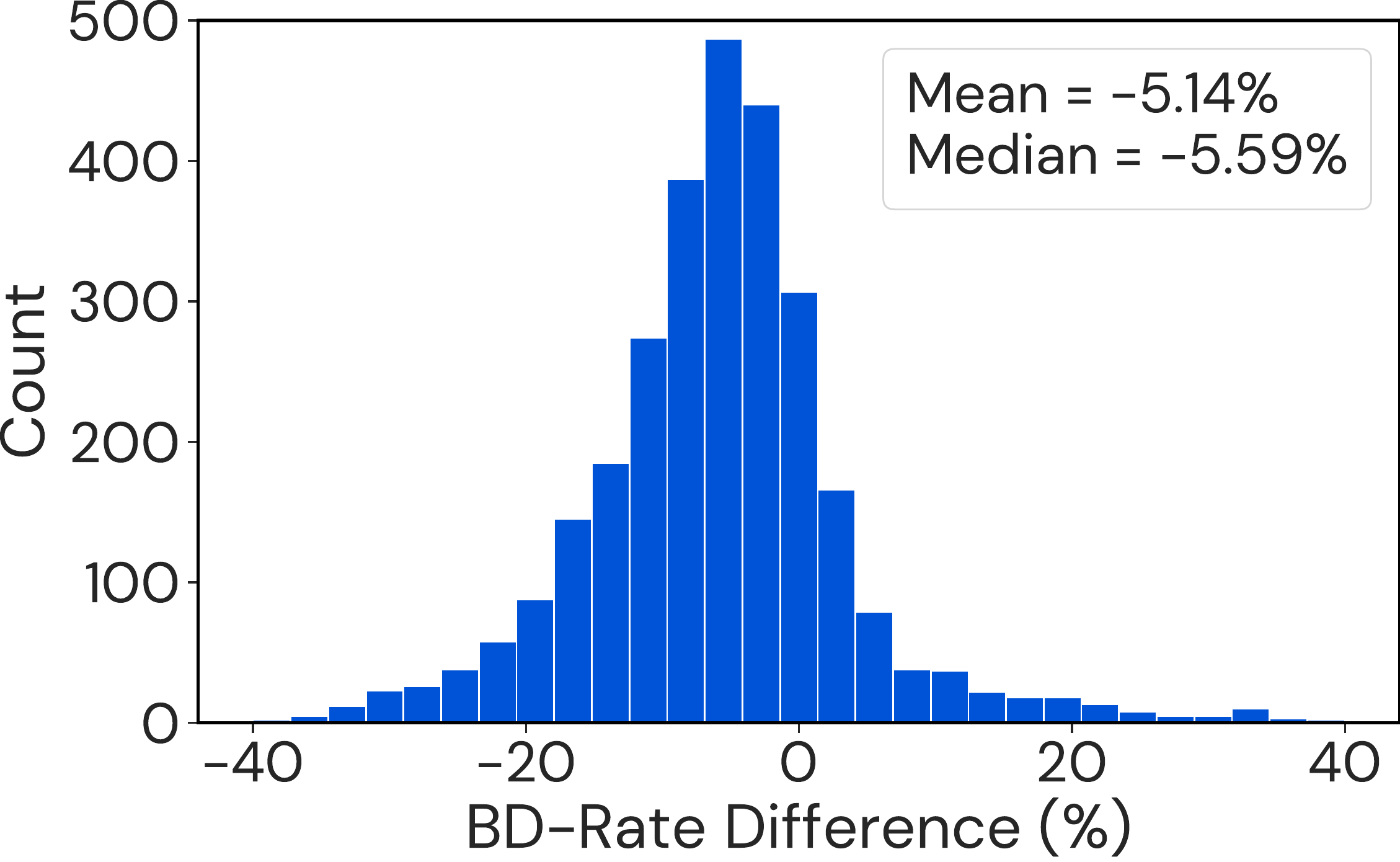}
     	\caption{Augmented \muzerorc{}}
    \end{subfigure}
    \caption{Histogram of SSIM BD-Rate difference of agent compared to \libvpx{} on evaluation set.}
    \label{figure:ssim_hist}
\end{figure}

\begin{figure}[ht]
	\centering
	\begin{subfigure}[c]{0.48\columnwidth}
	   \centering
    	\includegraphics[width=\textwidth]{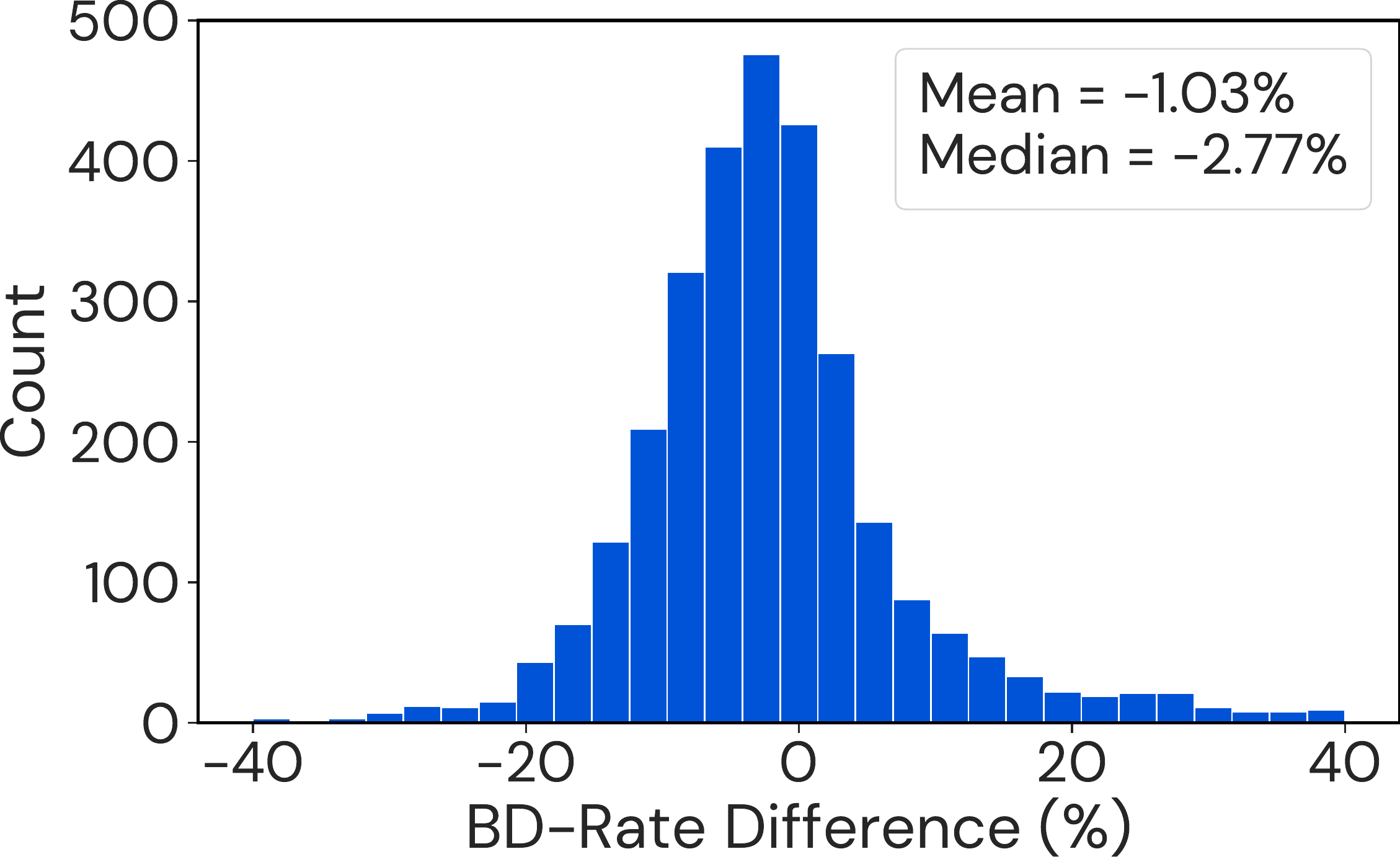}
    	\caption{\muzerorc{}}
    \end{subfigure}
    \begin{subfigure}[c]{0.48\columnwidth}
	   \centering
    	\includegraphics[width=\textwidth]{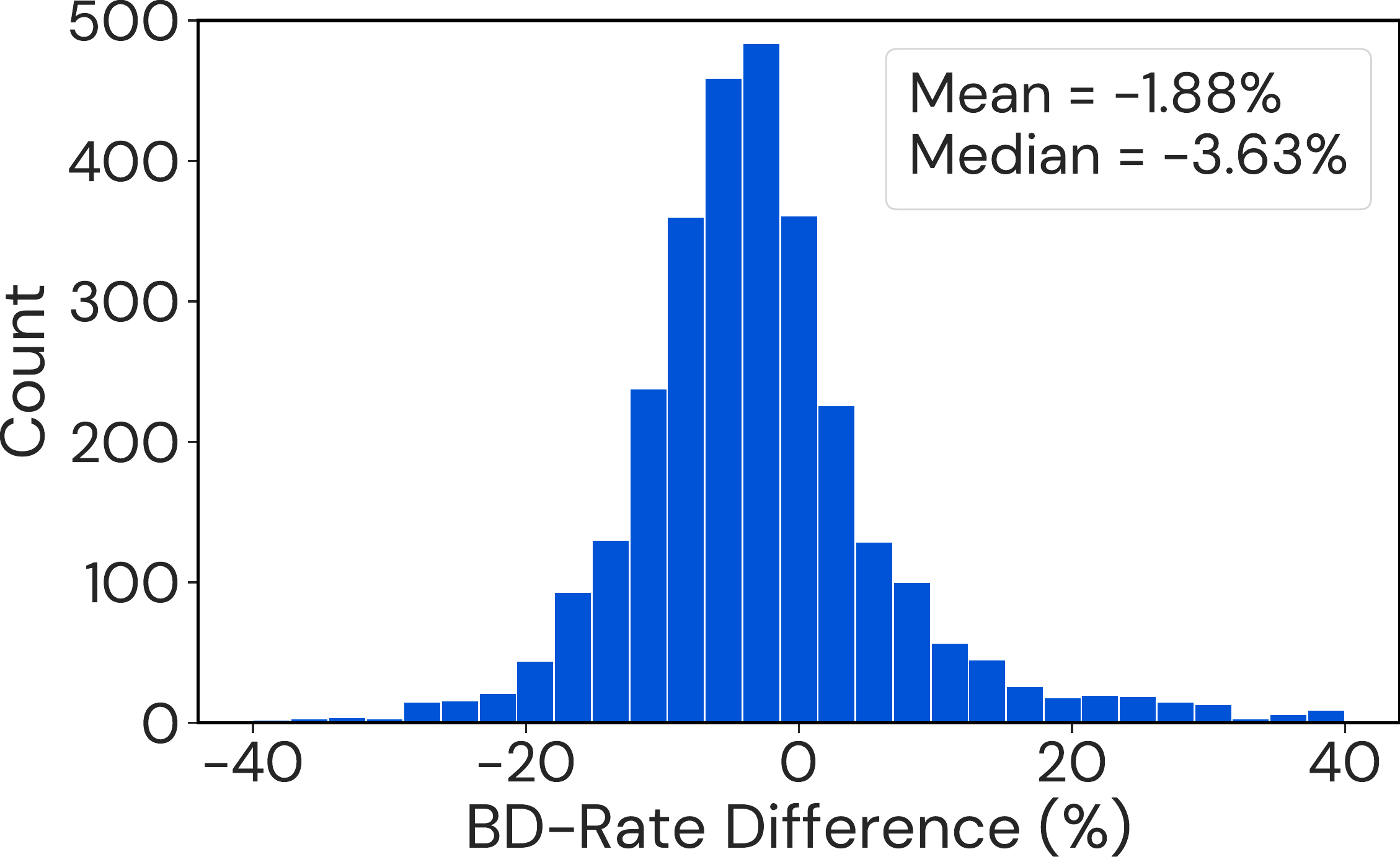}
     	\caption{Augmented \muzerorc{}}
    \end{subfigure}
    \caption{Histogram of VMAF BD-Rate difference of agent compared to \libvpx{} on evaluation set.}
    \label{figure:vmaf_hist}
\end{figure}

This section presents the histograms of BD-rate difference of the behaviors of \muzerorc{} and Augmented \muzerorc{} on 3062 videos from evaluation set compared to \libvpx{}.  See figure \ref{fig:rd-curve-example} for a qualitative example of RD curves and BD-rate difference. The RD curves for all the policies are computed by encoding each video from the evaluation set with nine uniformly spaced target bitrates from the [256, 768] kbps.  The figures \ref{figure:psnr_hist}, \ref{figure:ssim_hist}, and \ref{figure:vmaf_hist} show the histogram of BD-rate differences for PSNR, SSIM, and VMAF respectively for a single run of both agents compared to \libvpx{}.

\newpage
\subsection{Histograms of bitrate constraint satisfaction}
\label{appendix:overshoots}
\begin{figure}[ht]
	\centering
	\begin{subfigure}[c]{0.3\columnwidth}
	   \centering
    	\includegraphics[width=\textwidth]{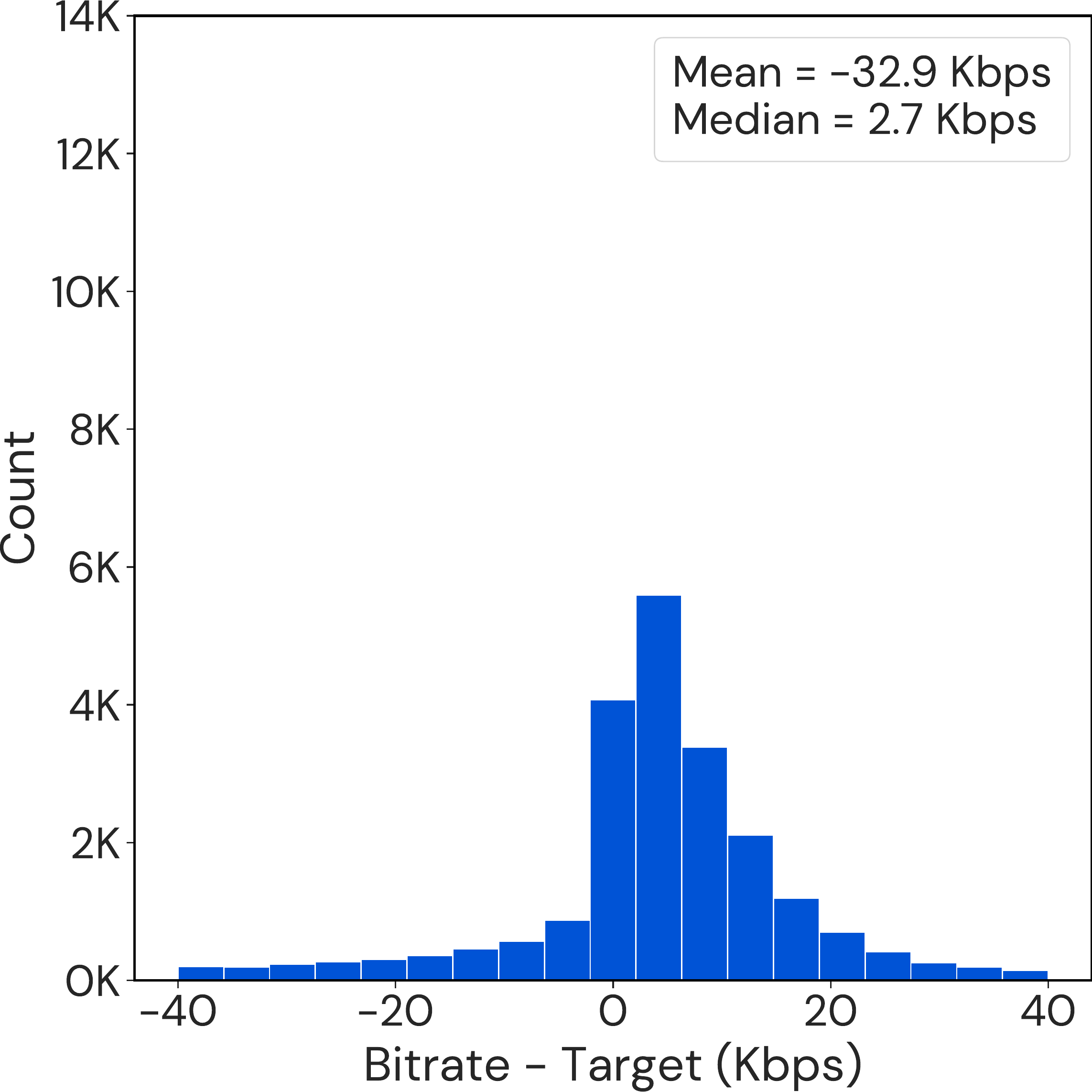}
    	\caption{\libvpx{}}
    \end{subfigure}
    \begin{subfigure}[c]{0.3\columnwidth}
	   \centering
    	\includegraphics[width=\textwidth]{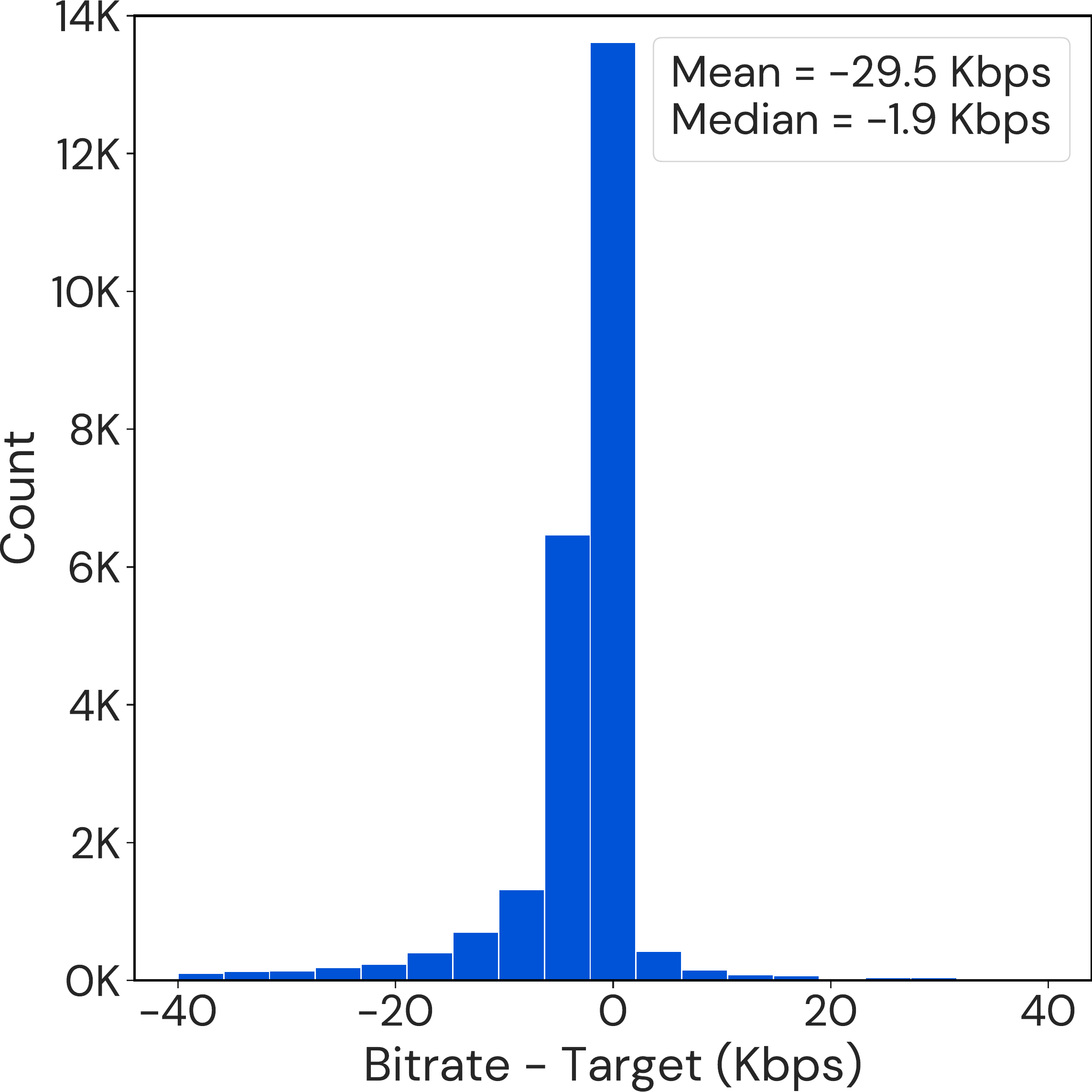}
    	\caption{\muzerorc{}{}}
    \end{subfigure}
    \begin{subfigure}[c]{0.3\columnwidth}
	   \centering
    	\includegraphics[width=\textwidth]{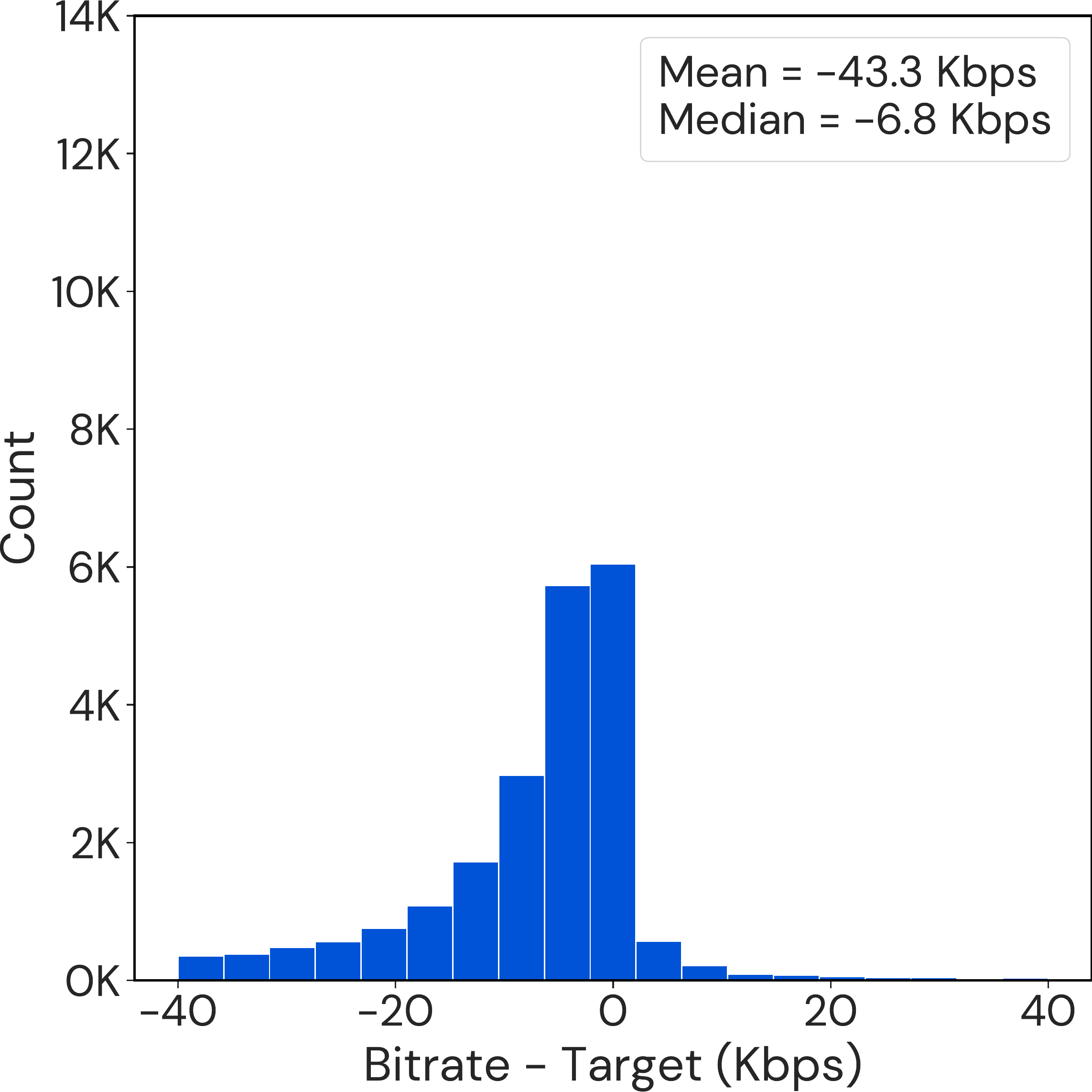}
    	\caption{Augmented \muzerorc{}}
    \end{subfigure}
    \caption{Histogram of overshoots of the agents on the evaluation set for all target bitrates.}
    \label{figure:overshoot_all}
\end{figure}

\begin{figure}[ht]
	\centering
	\begin{subfigure}[c]{0.3\columnwidth}
	   \centering
    	\includegraphics[width=\textwidth]{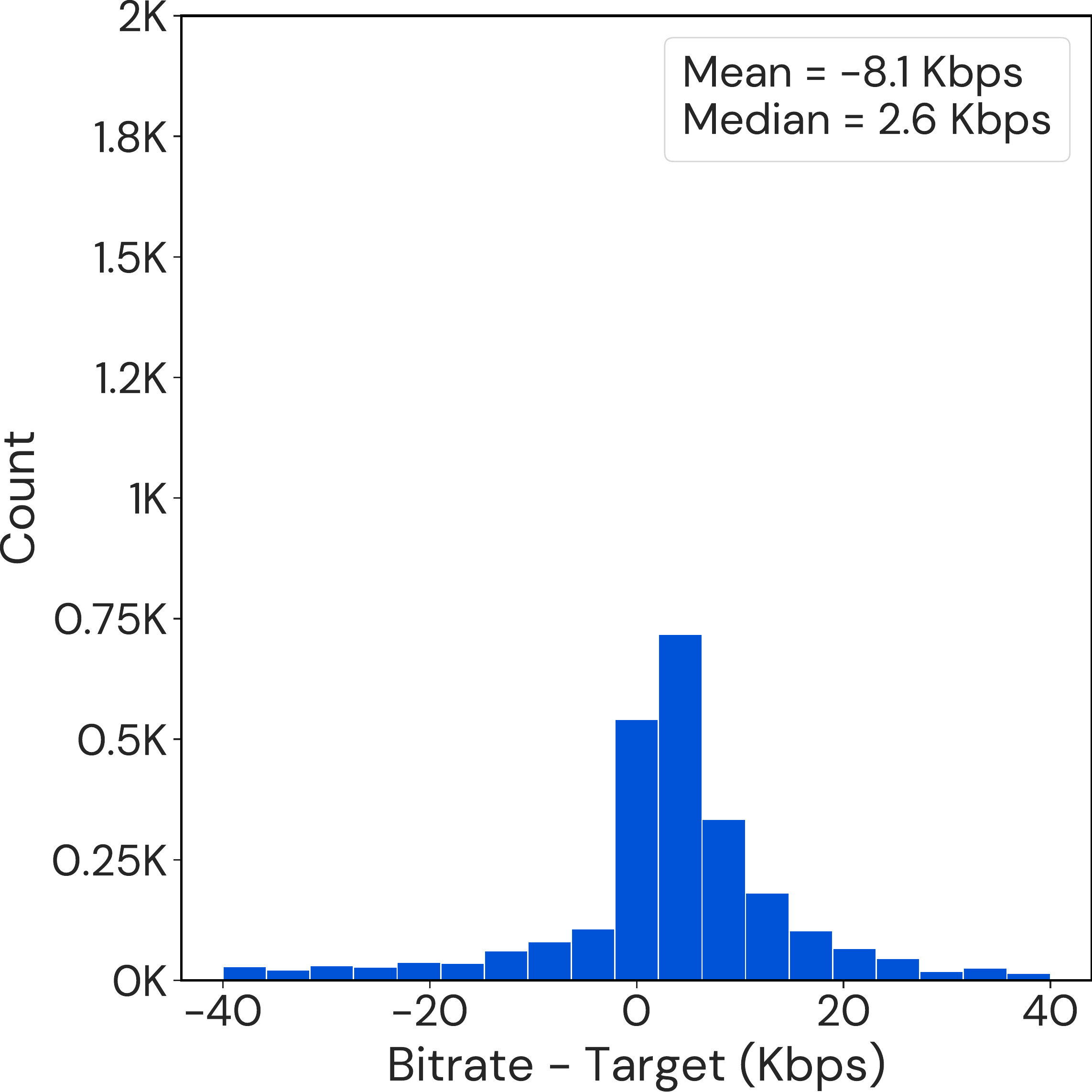}
    	\caption{\libvpx{}}
    \end{subfigure}
    \begin{subfigure}[c]{0.3\columnwidth}
	   \centering
    	\includegraphics[width=\textwidth]{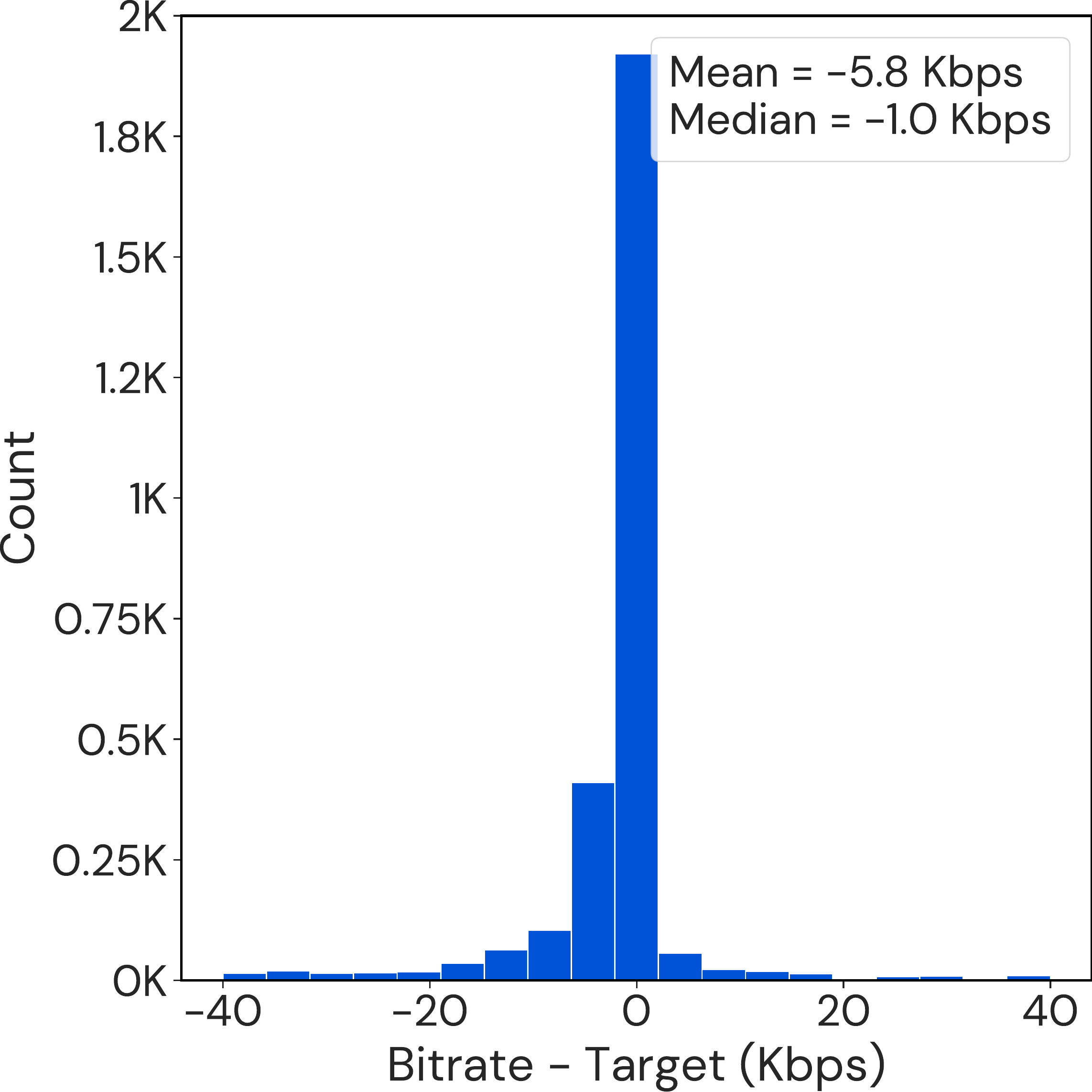}
    	\caption{\muzerorc{}{}}
    \end{subfigure}
    \begin{subfigure}[c]{0.3\columnwidth}
	   \centering
    	\includegraphics[width=\textwidth]{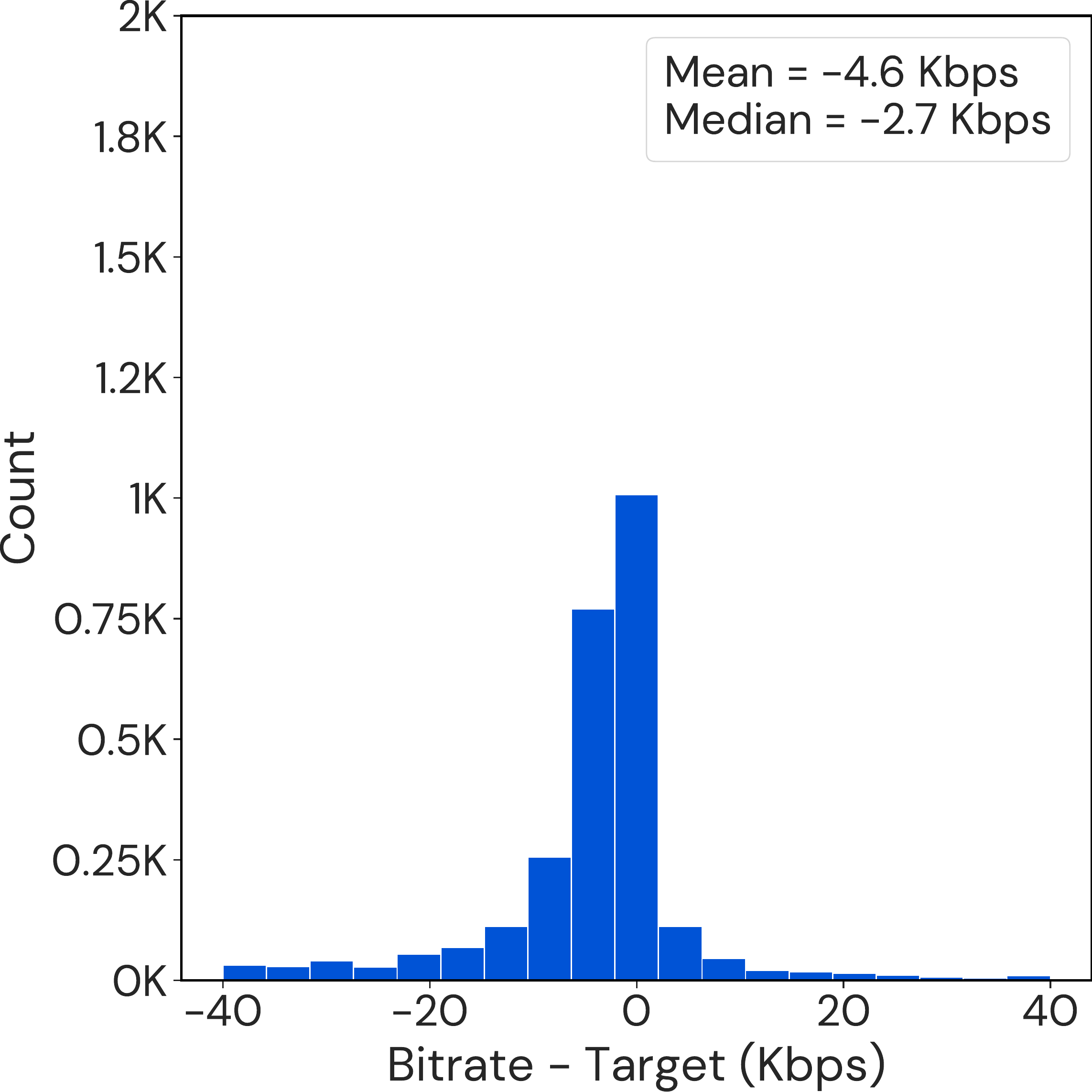}
    	\caption{Augmented \muzerorc{}}
    \end{subfigure}
    \caption{Histogram of overshoots of the agents on the evaluation set for 256 kbps target bitrate.}
    \label{figure:overshoot_256}
\end{figure}

\begin{figure}[ht]
	\centering
	\begin{subfigure}[c]{0.3\columnwidth}
	   \centering
    	\includegraphics[width=\textwidth]{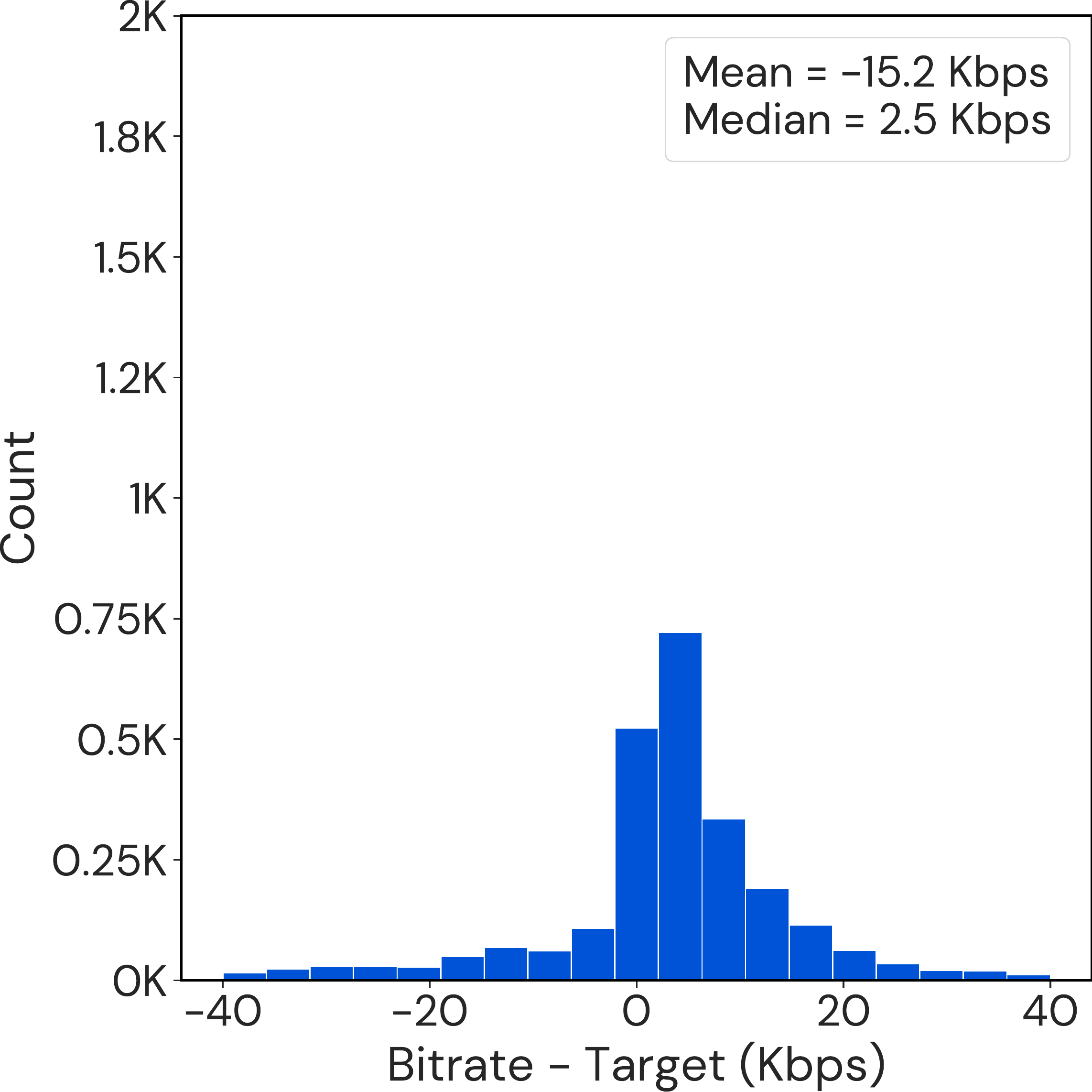}
    	\caption{\libvpx{}}
    \end{subfigure}
    \begin{subfigure}[c]{0.3\columnwidth}
	   \centering
    	\includegraphics[width=\textwidth]{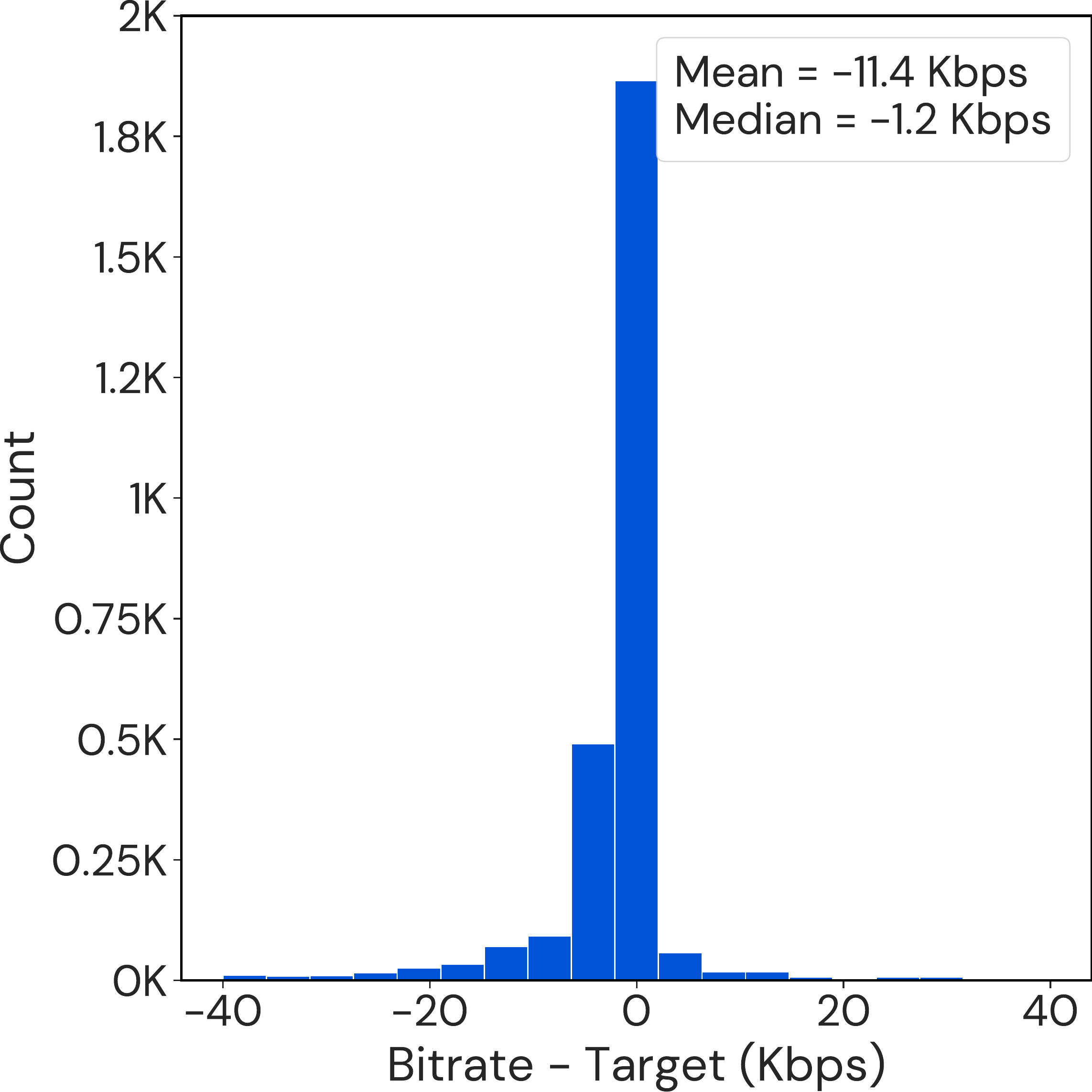}
    	\caption{\muzerorc{}{}}
    \end{subfigure}
    \begin{subfigure}[c]{0.3\columnwidth}
	   \centering
    	\includegraphics[width=\textwidth]{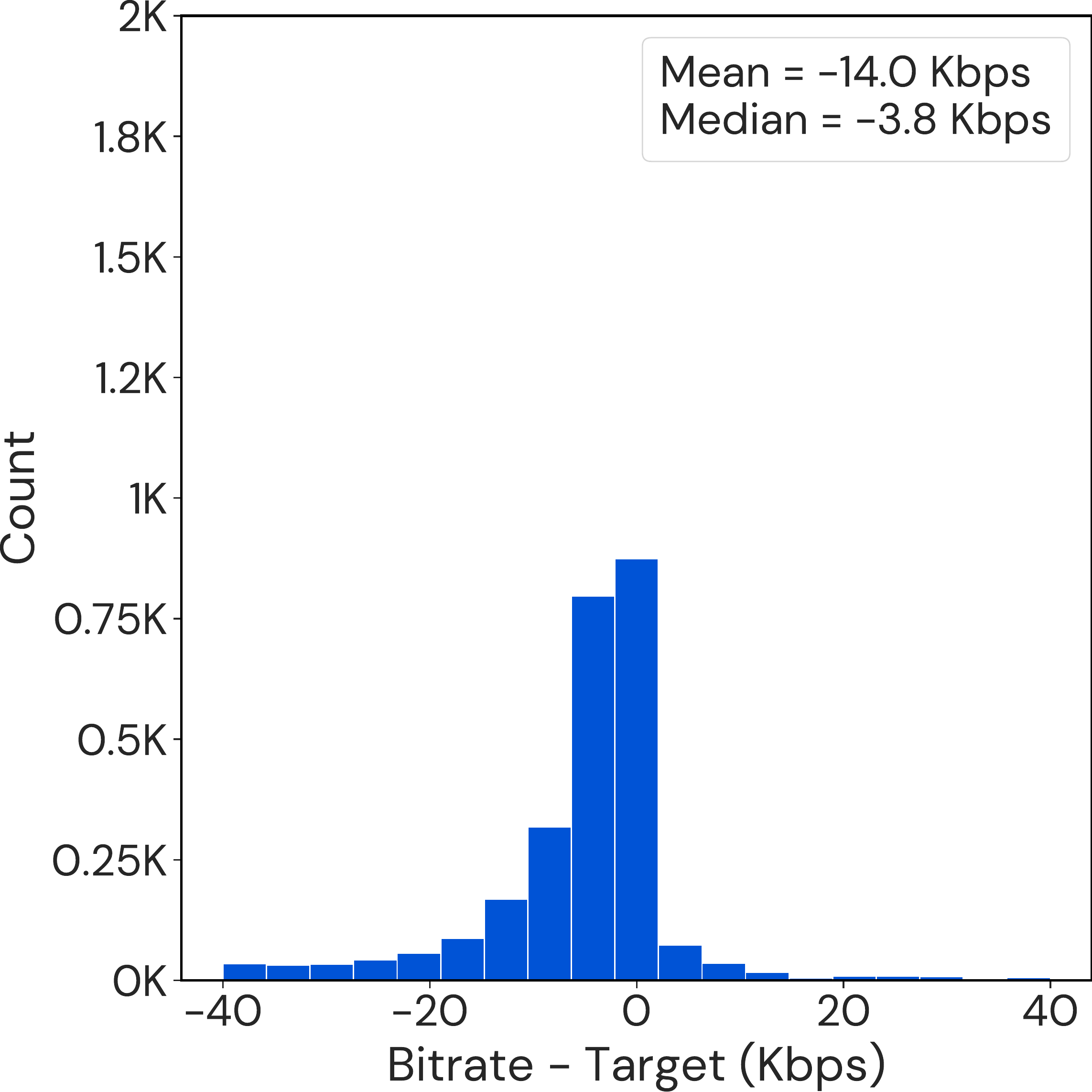}
    	\caption{Augmented \muzerorc{}}
    \end{subfigure}
    \caption{Histogram of overshoots of the agents on the evaluation set for 320 kbps target bitrate.}
    \label{figure:overshoot_320}
\end{figure}

\begin{figure}[ht]
	\centering
	\begin{subfigure}[c]{0.3\columnwidth}
	   \centering
    	\includegraphics[width=\textwidth]{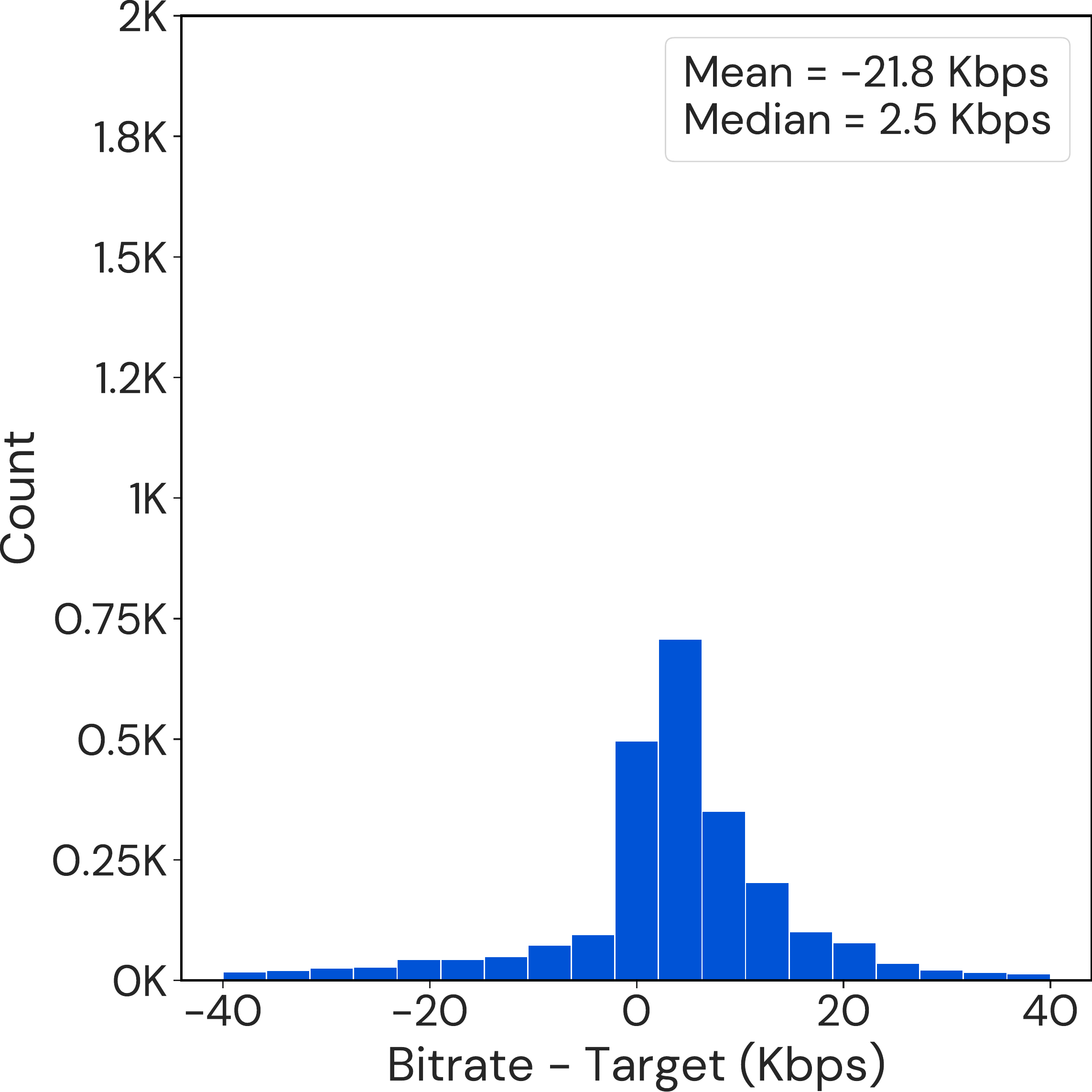}
    	\caption{\libvpx{}}
    \end{subfigure}
    \begin{subfigure}[c]{0.3\columnwidth}
	   \centering
    	\includegraphics[width=\textwidth]{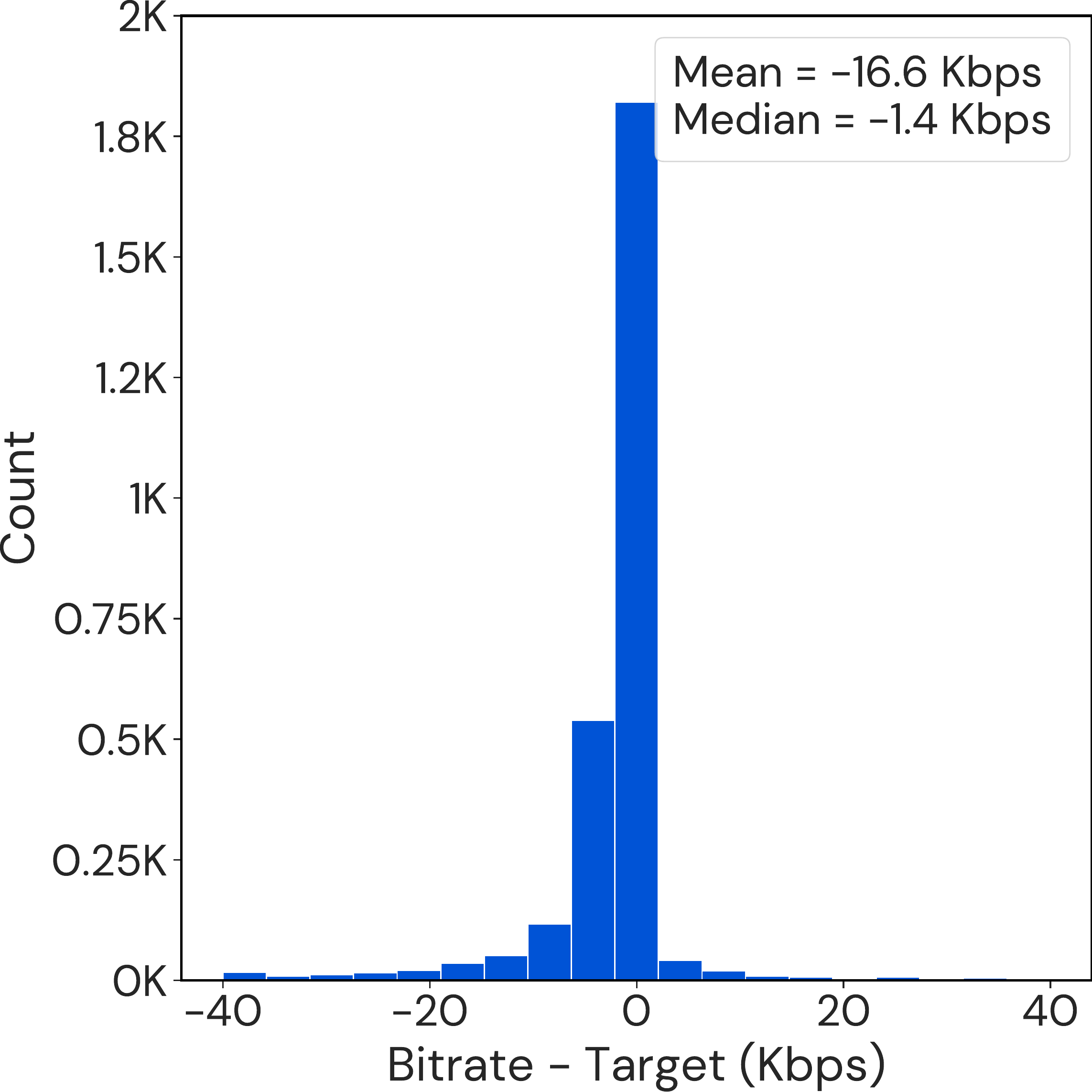}
    	\caption{\muzerorc{}{}}
    \end{subfigure}
    \begin{subfigure}[c]{0.3\columnwidth}
	   \centering
    	\includegraphics[width=\textwidth]{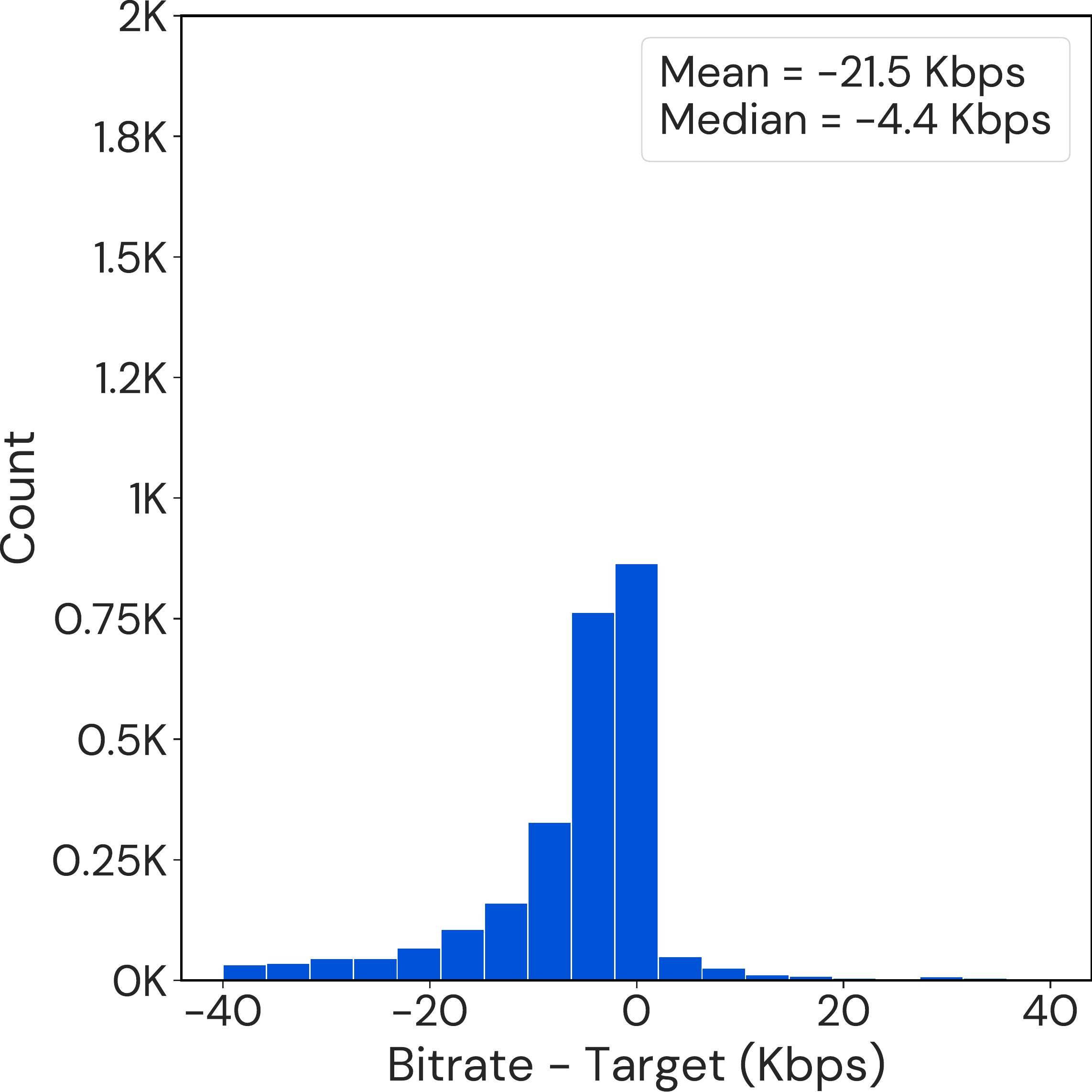}
    	\caption{Augmented \muzerorc{}}
    \end{subfigure}
    \caption{Histogram of overshoots of the agents on the evaluation set for 384 kbps target bitrate.}
    \label{figure:overshoot_384}
\end{figure}

\begin{figure}[ht]
	\centering
	\begin{subfigure}[c]{0.3\columnwidth}
	   \centering
    	\includegraphics[width=\textwidth]{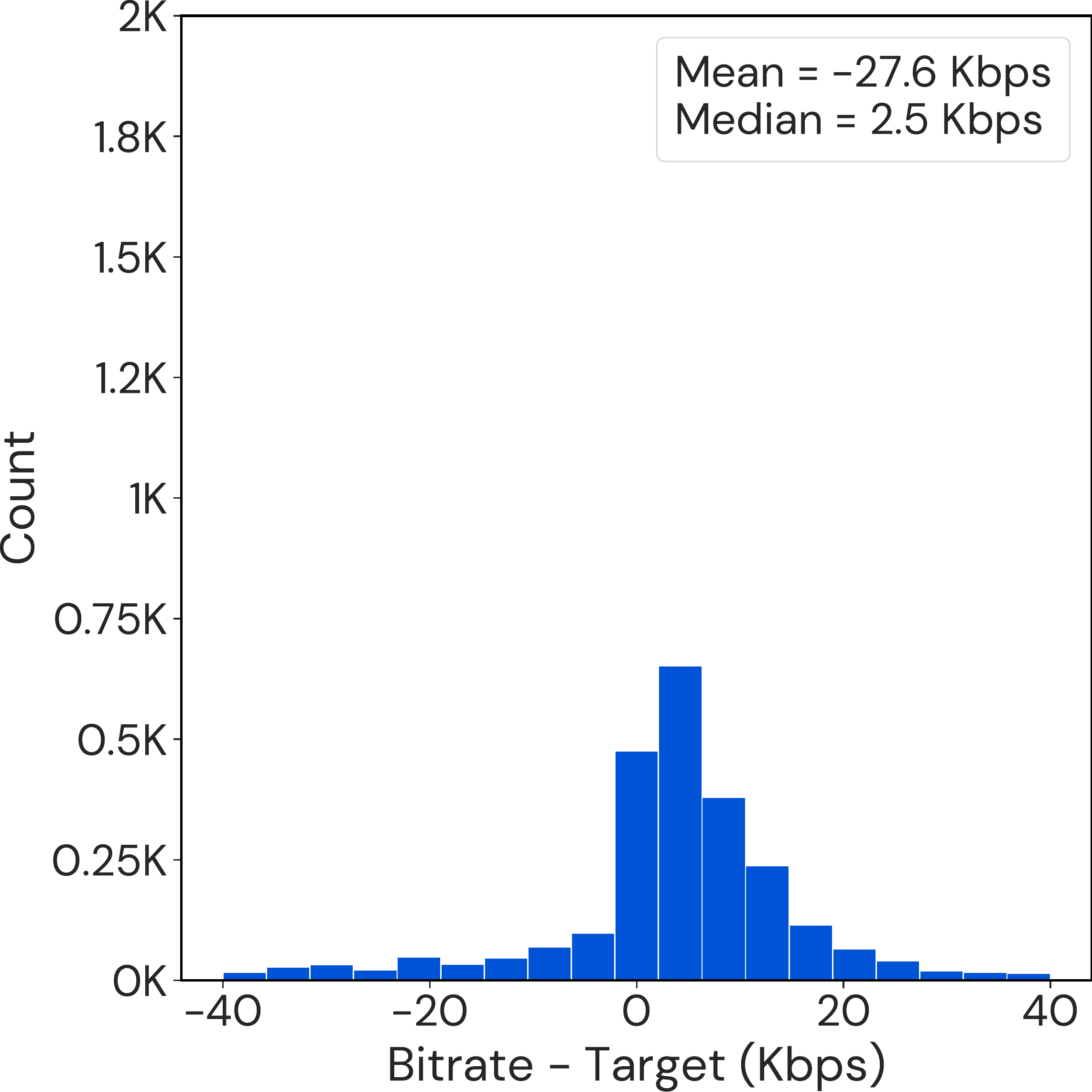}
    	\caption{\libvpx{}}
    \end{subfigure}
    \begin{subfigure}[c]{0.3\columnwidth}
	   \centering
    	\includegraphics[width=\textwidth]{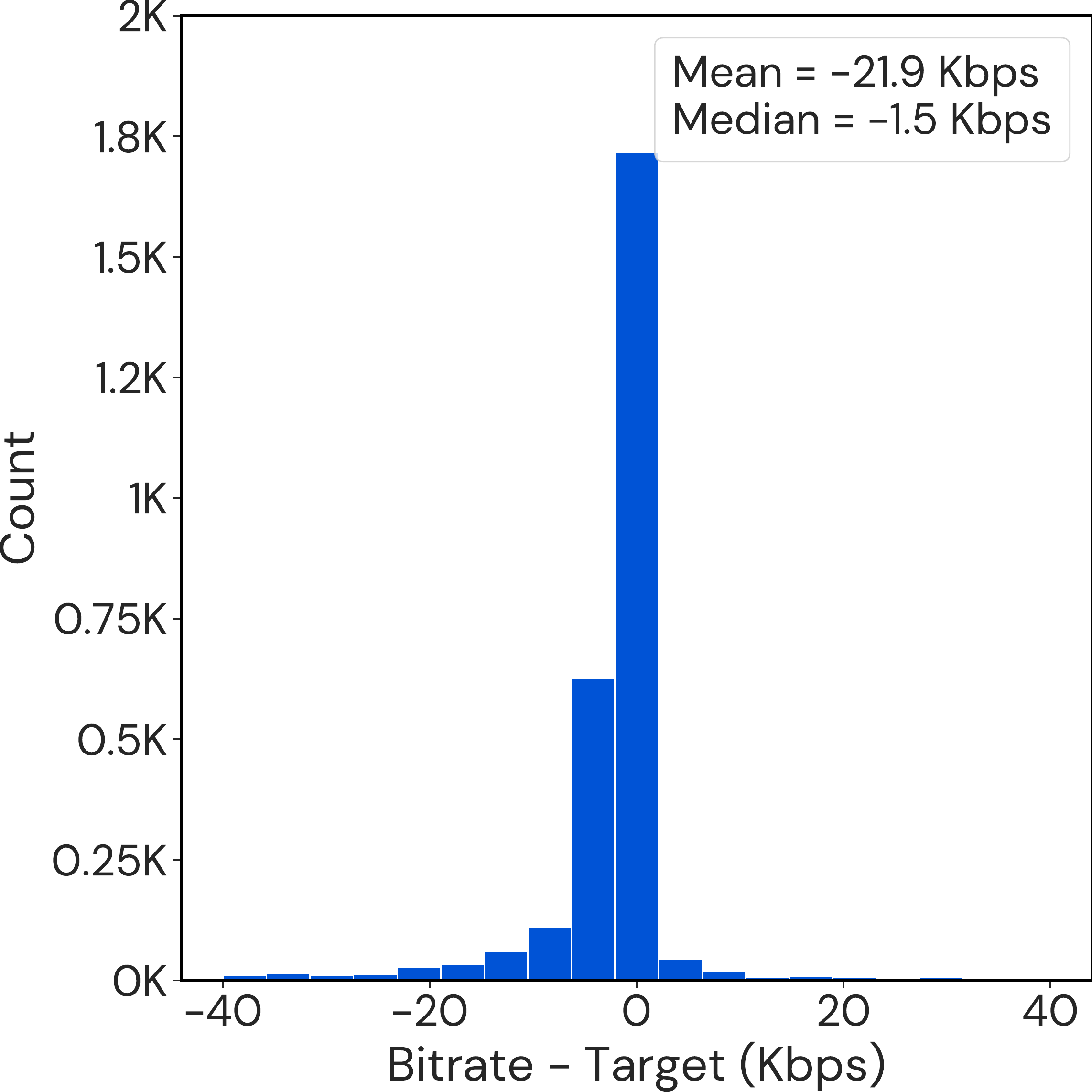}
    	\caption{\muzerorc{}{}}
    \end{subfigure}
    \begin{subfigure}[c]{0.3\columnwidth}
	   \centering
    	\includegraphics[width=\textwidth]{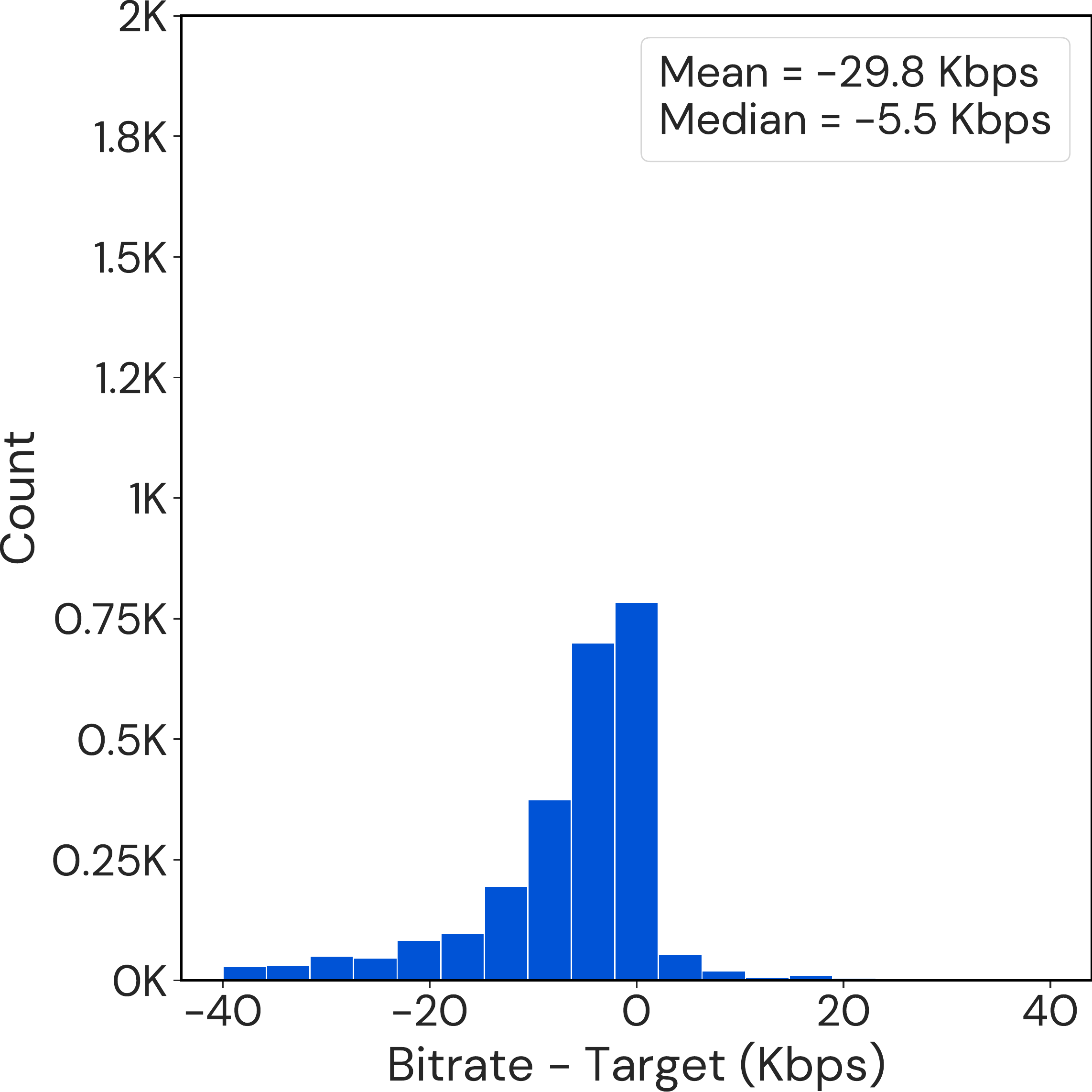}
    	\caption{Augmented \muzerorc{}}
    \end{subfigure}
    \caption{Histogram of overshoots of the agents on the evaluation set for 448 kbps target bitrate.}
    \label{figure:overshoot_448}
\end{figure}

\begin{figure}[ht]
	\centering
	\begin{subfigure}[c]{0.3\columnwidth}
	   \centering
    	\includegraphics[width=\textwidth]{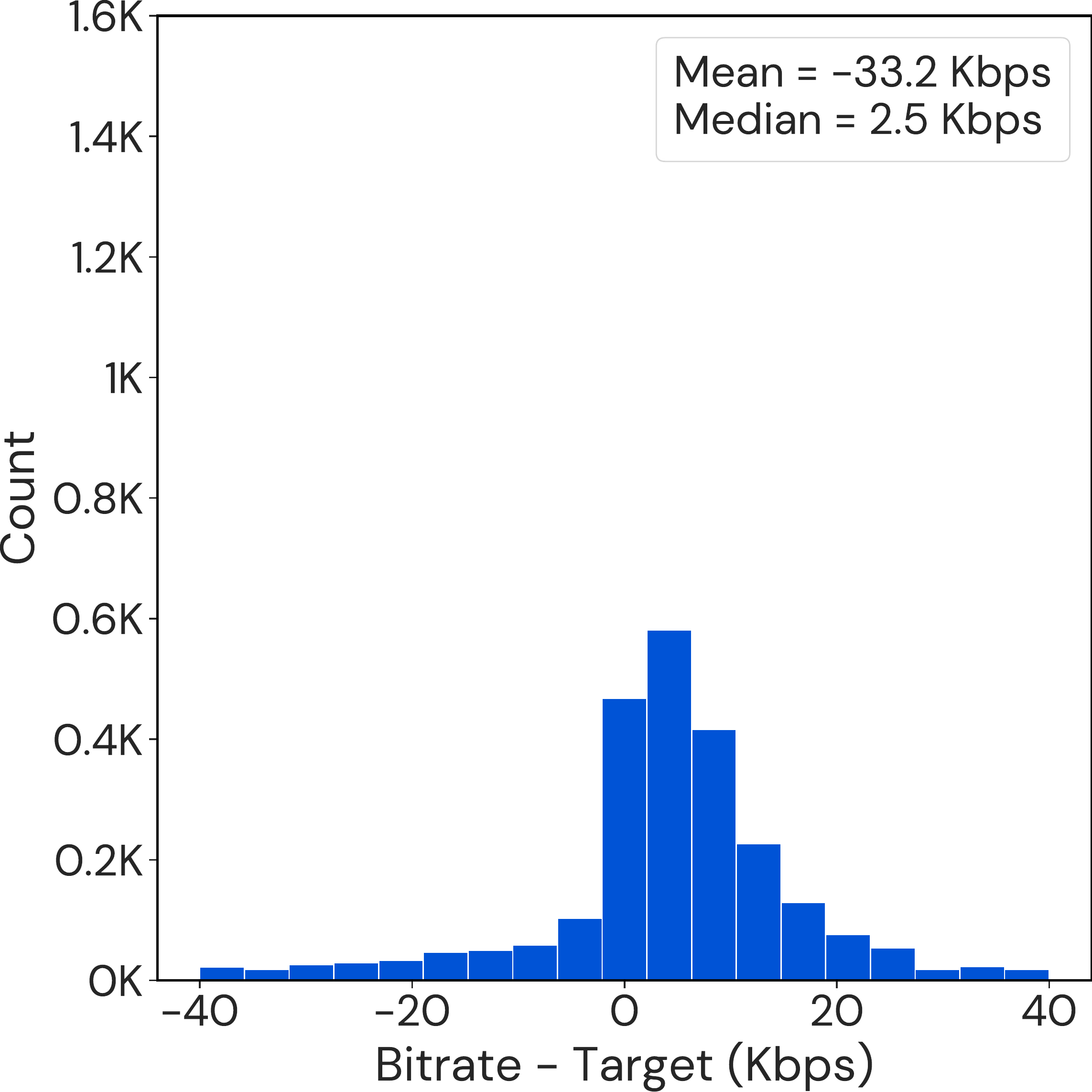}
    	\caption{\libvpx{}}
    \end{subfigure}
    \begin{subfigure}[c]{0.3\columnwidth}
	   \centering
    	\includegraphics[width=\textwidth]{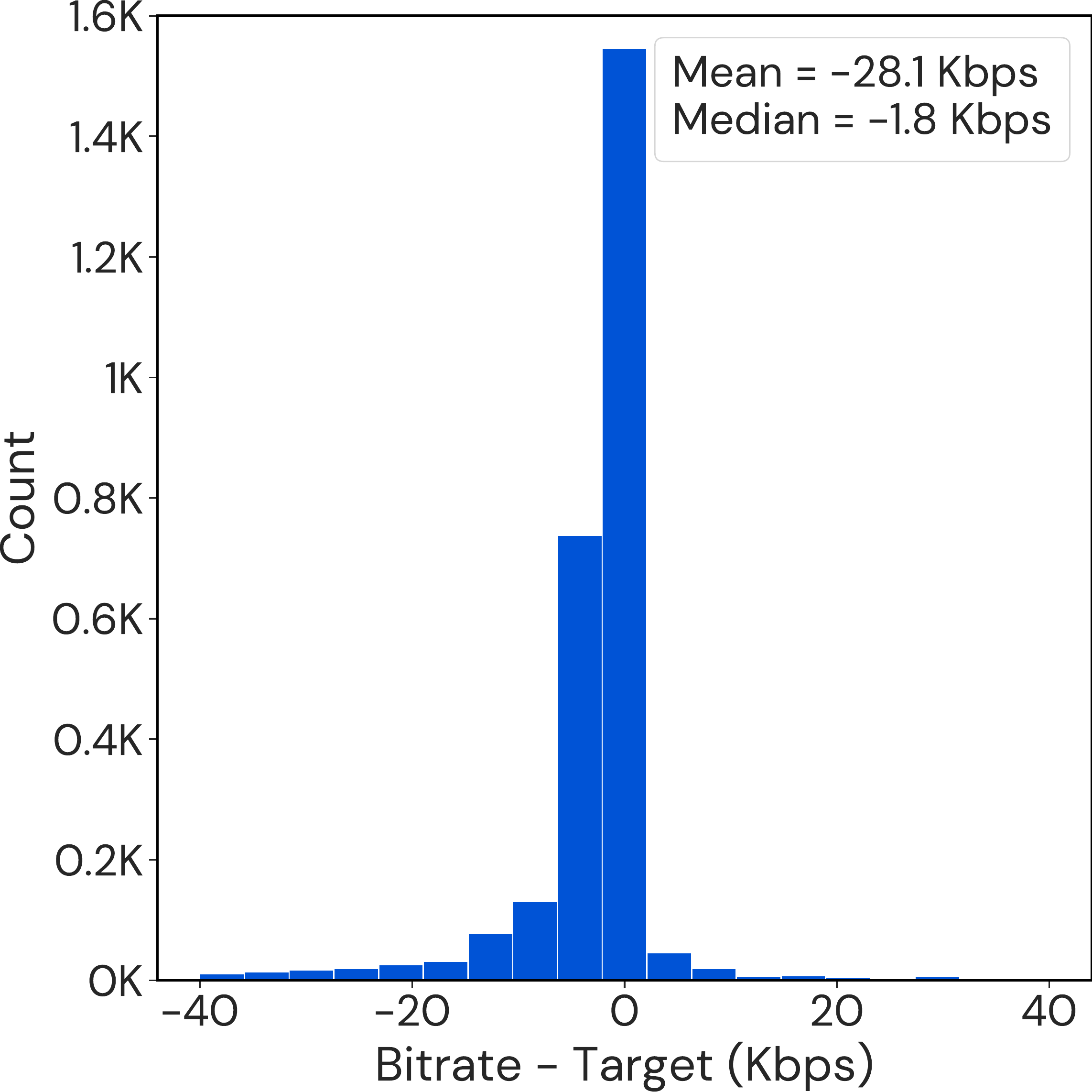}
    	\caption{\muzerorc{}{}}
    \end{subfigure}
    \begin{subfigure}[c]{0.3\columnwidth}
	   \centering
    	\includegraphics[width=\textwidth]{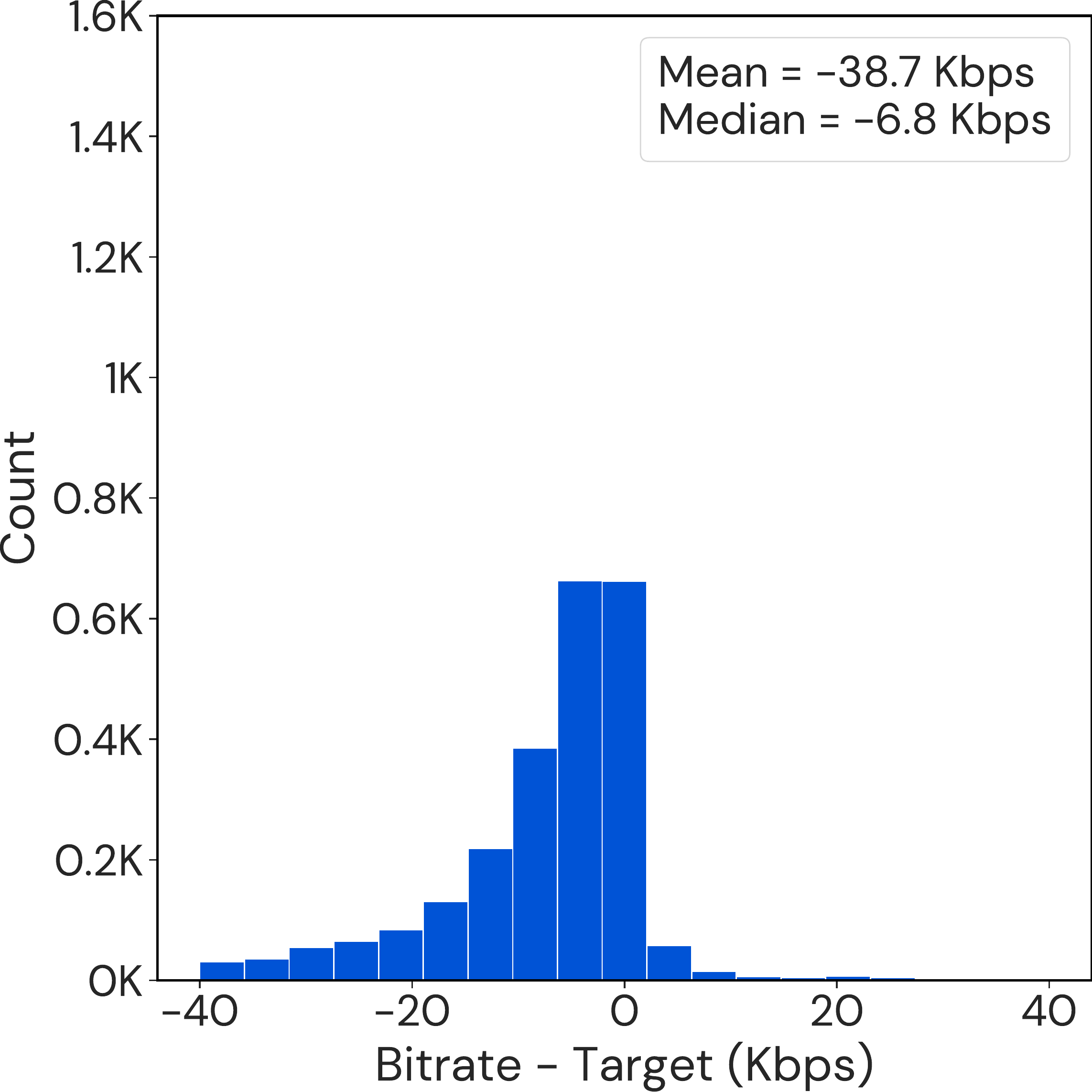}
    	\caption{Augmented \muzerorc{}}
    \end{subfigure}
    \caption{Histogram of overshoots of the agents on the evaluation set for 512 kbps target bitrate.}
    \label{figure:overshoot_512}
\end{figure}

\begin{figure}[ht]
	\centering
	\begin{subfigure}[c]{0.3\columnwidth}
	   \centering
    	\includegraphics[width=\textwidth]{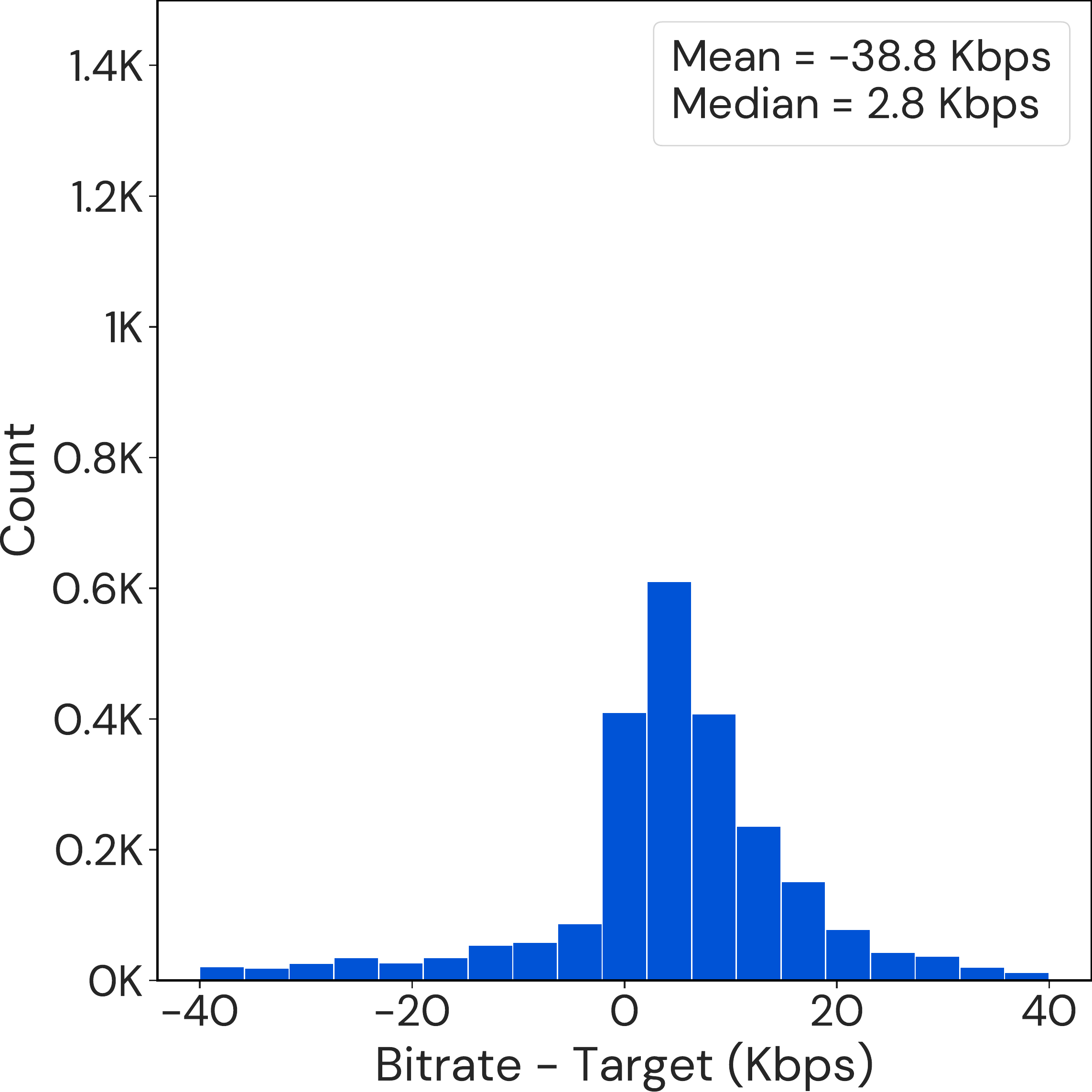}
    	\caption{\libvpx{}}
    \end{subfigure}
    \begin{subfigure}[c]{0.3\columnwidth}
	   \centering
    	\includegraphics[width=\textwidth]{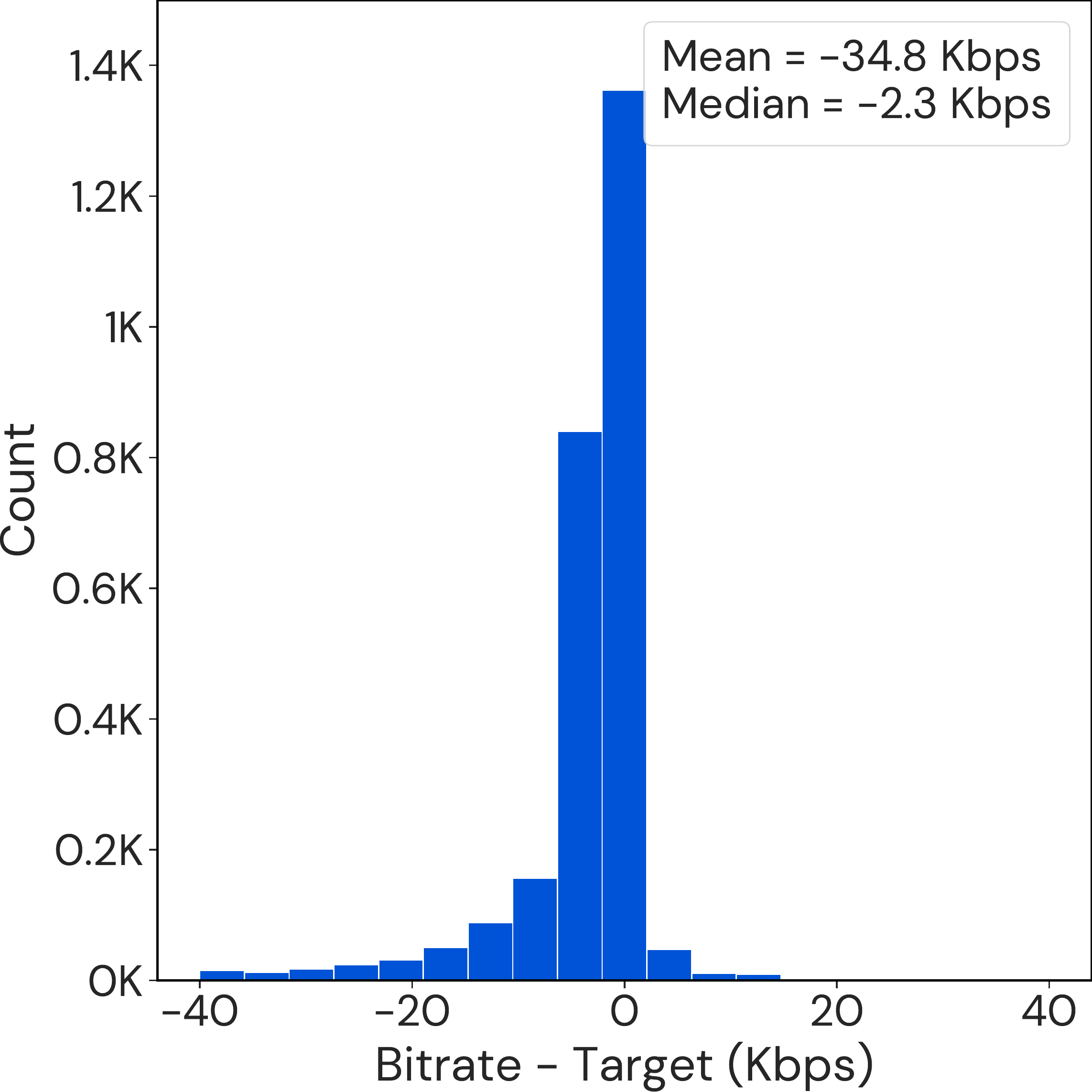}
    	\caption{\muzerorc{}{}}
    \end{subfigure}
    \begin{subfigure}[c]{0.3\columnwidth}
	   \centering
    	\includegraphics[width=\textwidth]{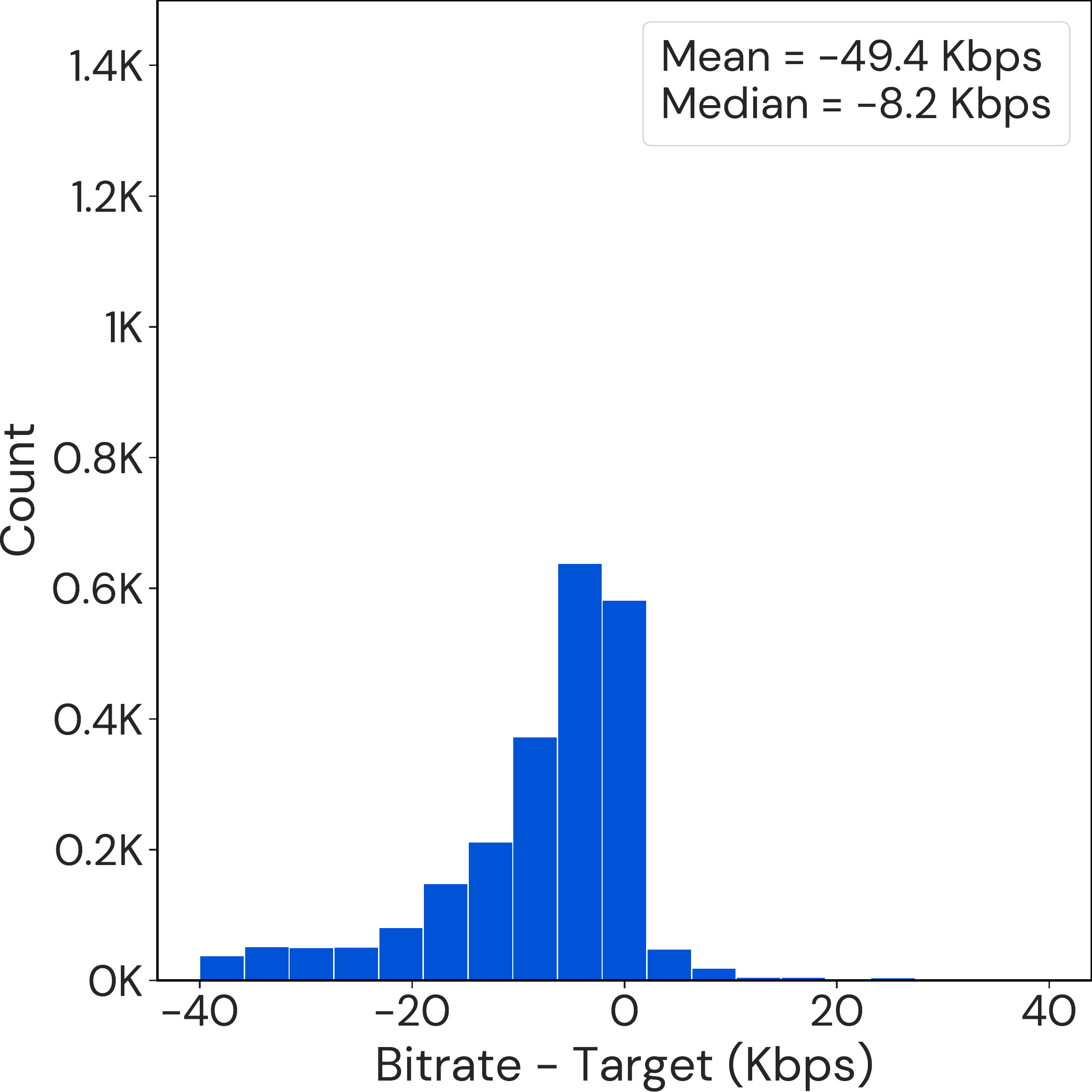}
    	\caption{Augmented \muzerorc{}}
    \end{subfigure}
    \caption{Histogram of overshoots of the agents on the evaluation set for 576 kbps target bitrate.}
    \label{figure:overshoot_576}
\end{figure}

\begin{figure}[ht]
	\centering
	\begin{subfigure}[c]{0.3\columnwidth}
	   \centering
    	\includegraphics[width=\textwidth]{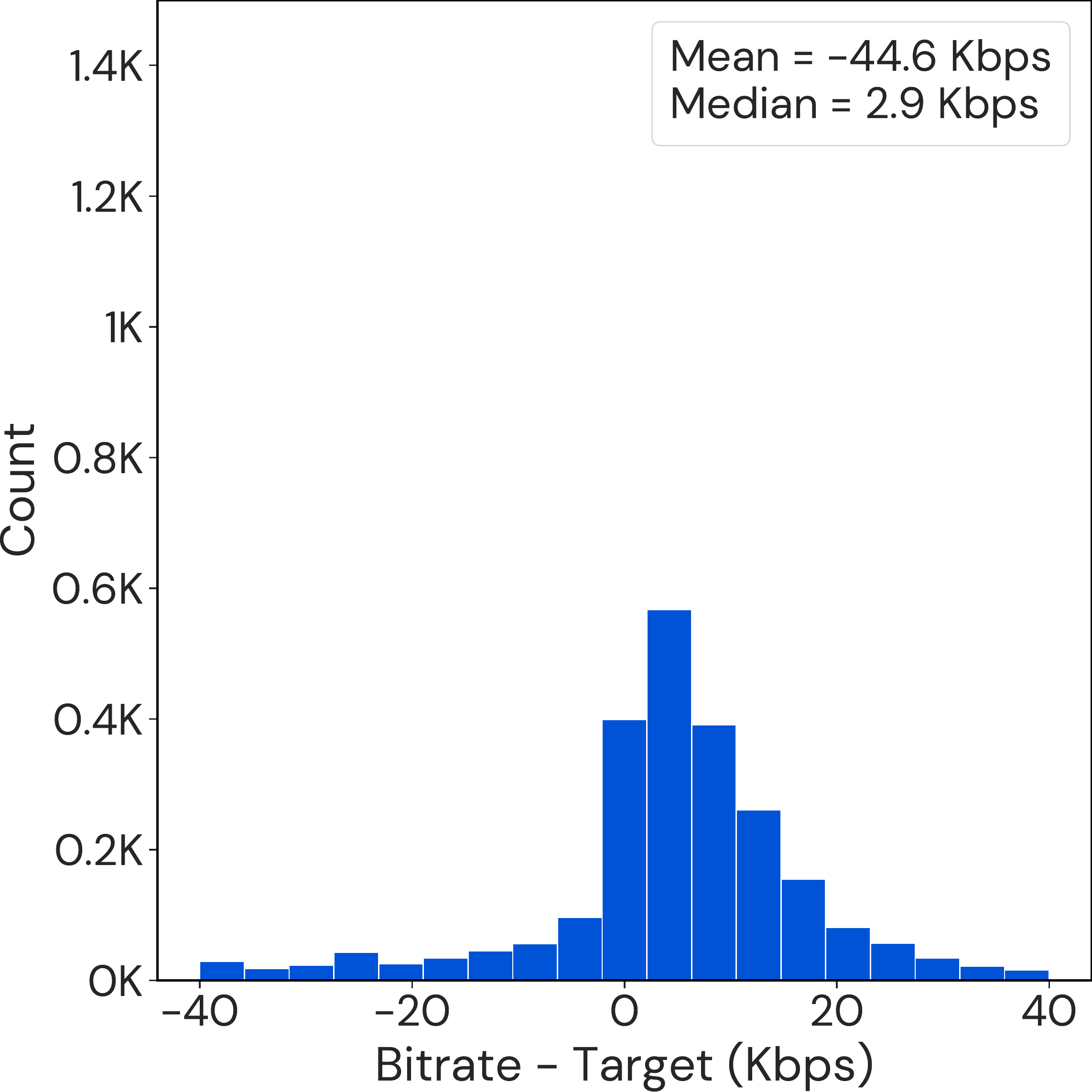}
    	\caption{\libvpx{}}
    \end{subfigure}
    \begin{subfigure}[c]{0.3\columnwidth}
	   \centering
    	\includegraphics[width=\textwidth]{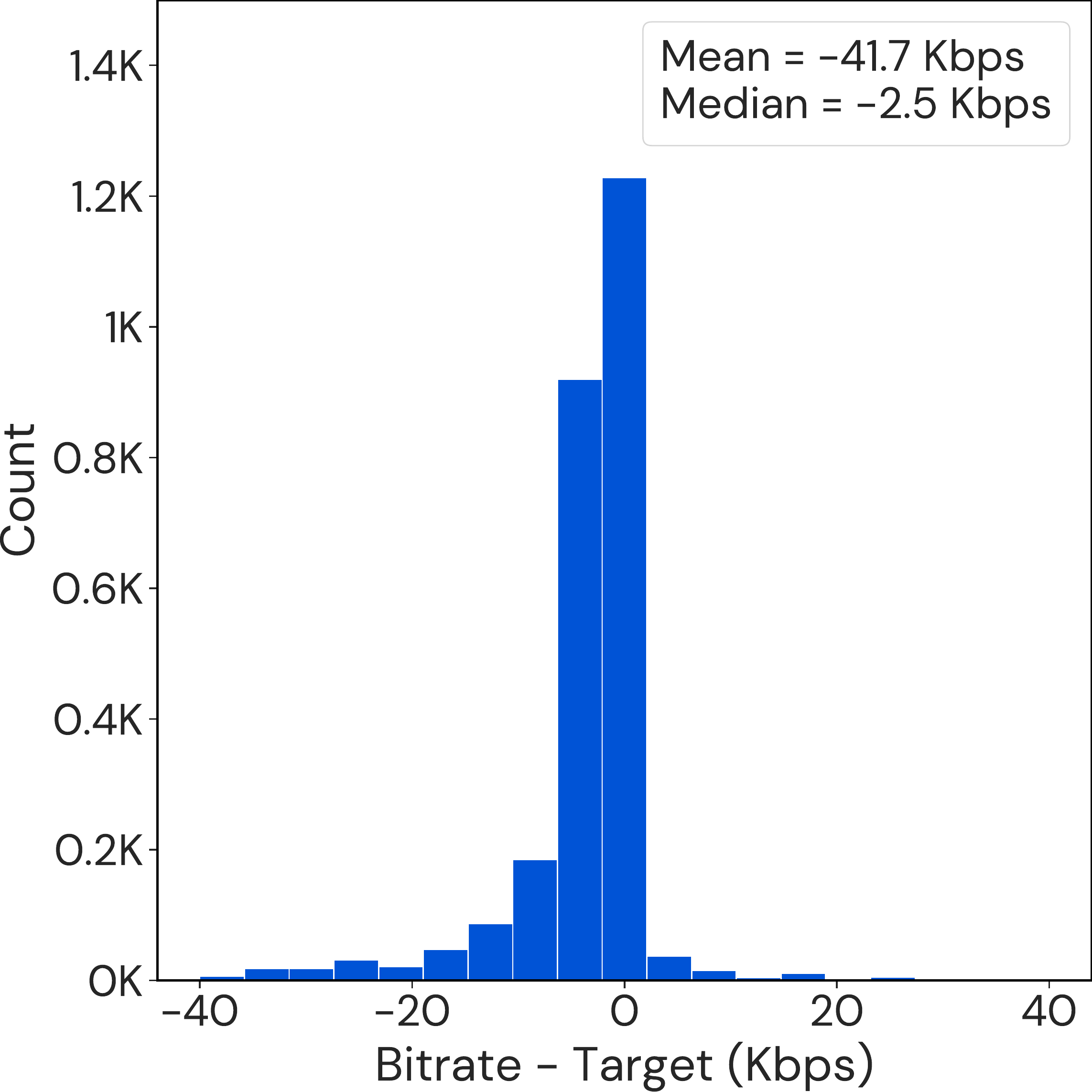}
    	\caption{\muzerorc{}{}}
    \end{subfigure}
    \begin{subfigure}[c]{0.3\columnwidth}
	   \centering
    	\includegraphics[width=\textwidth]{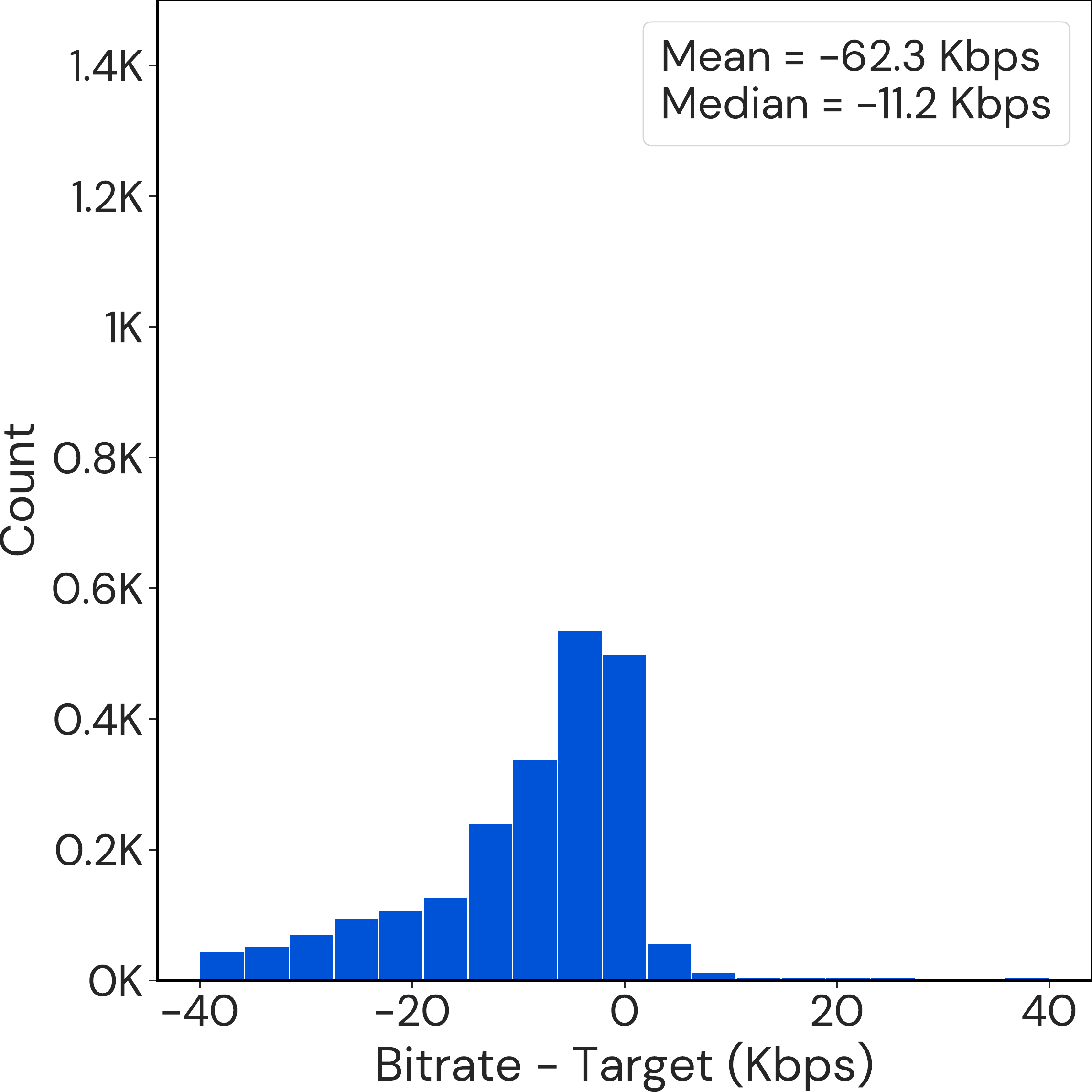}
    	\caption{Augmented \muzerorc{}}
    \end{subfigure}
    \caption{Histogram of overshoots of the agents on the evaluation set for 640 kbps target bitrate.}
    \label{figure:overshoot_640}
\end{figure}

\begin{figure}[ht]
	\centering
	\begin{subfigure}[c]{0.3\columnwidth}
	   \centering
    	\includegraphics[width=\textwidth]{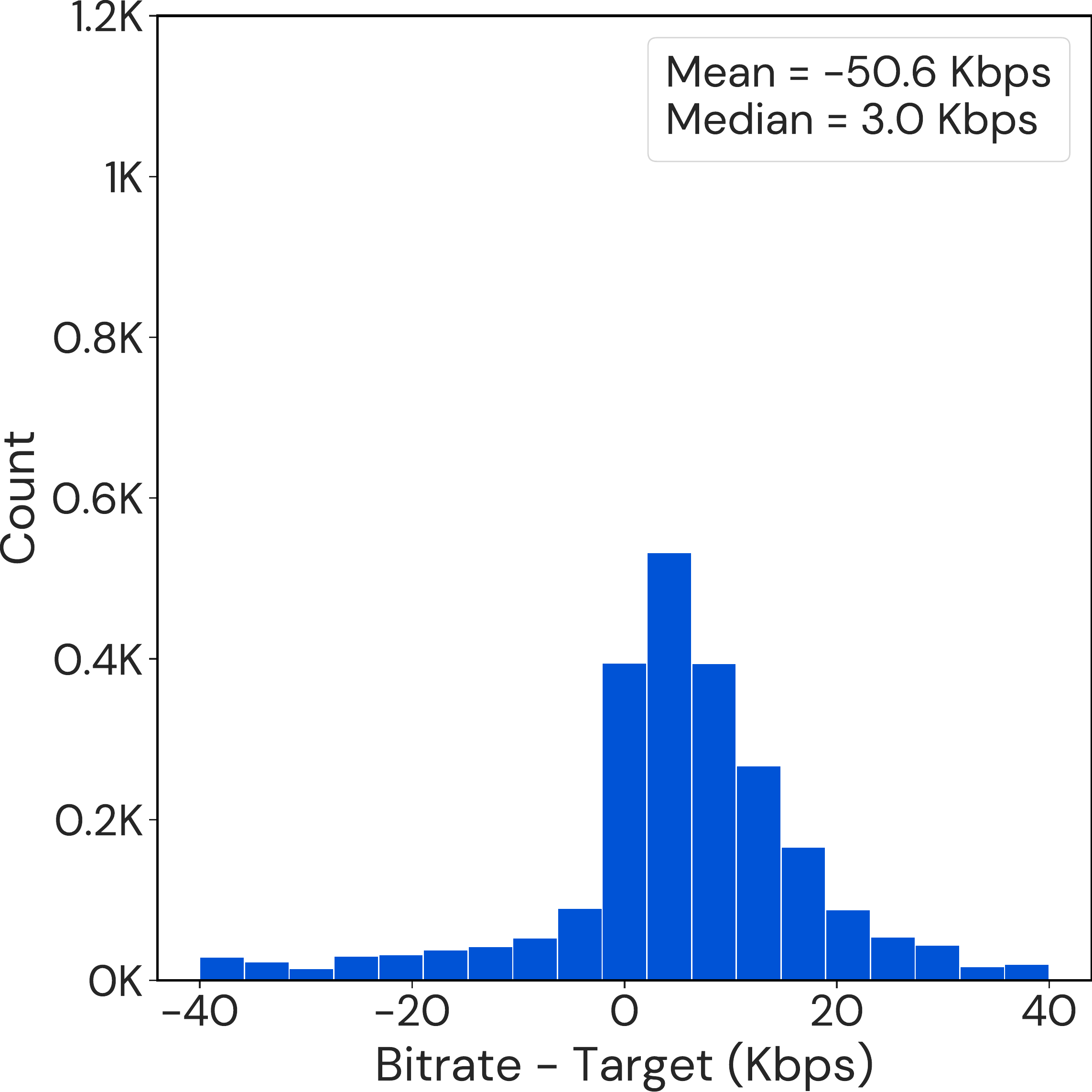}
    	\caption{\libvpx{}}
    \end{subfigure}
    \begin{subfigure}[c]{0.3\columnwidth}
	   \centering
    	\includegraphics[width=\textwidth]{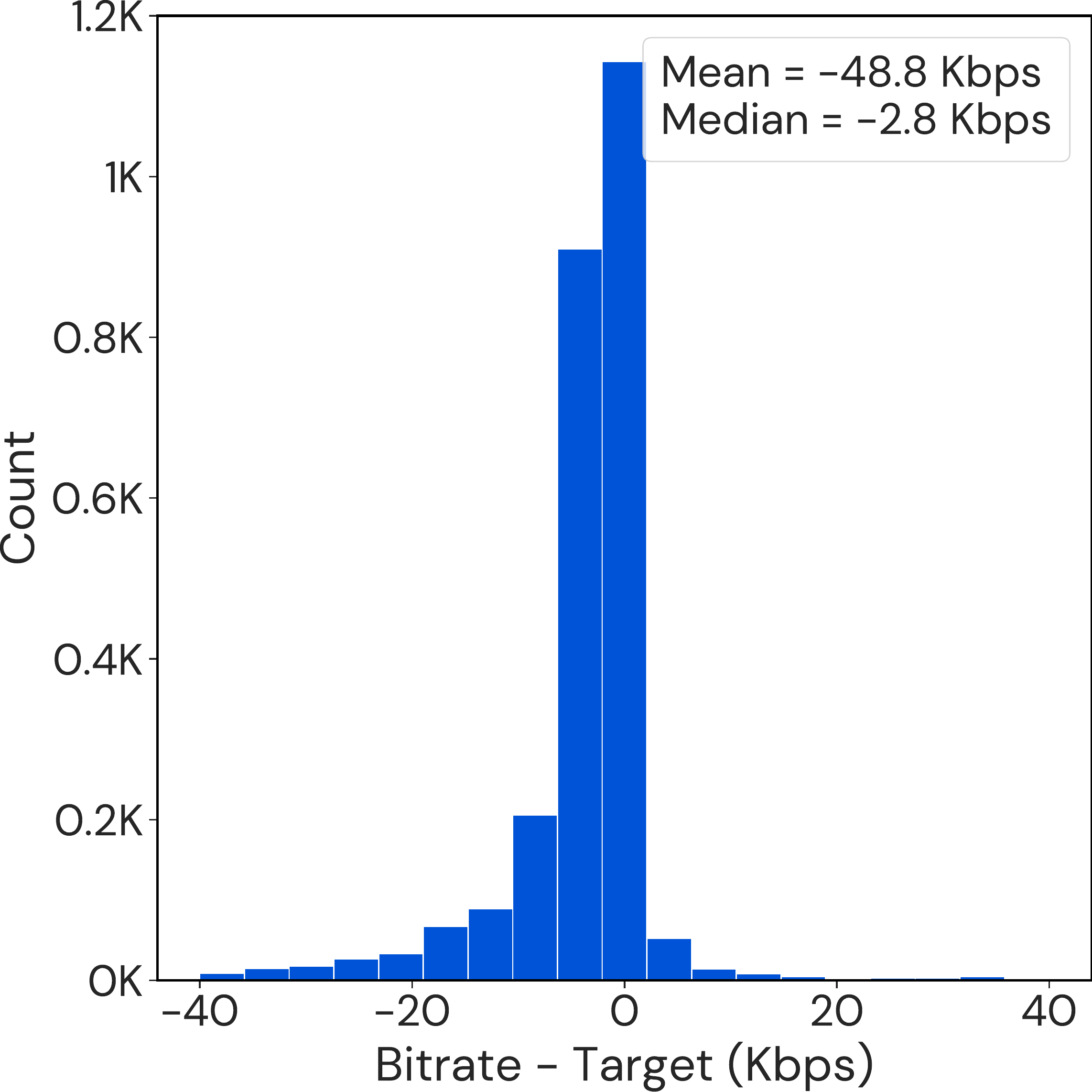}
    	\caption{\muzerorc{}{}}
    \end{subfigure}
    \begin{subfigure}[c]{0.3\columnwidth}
	   \centering
    	\includegraphics[width=\textwidth]{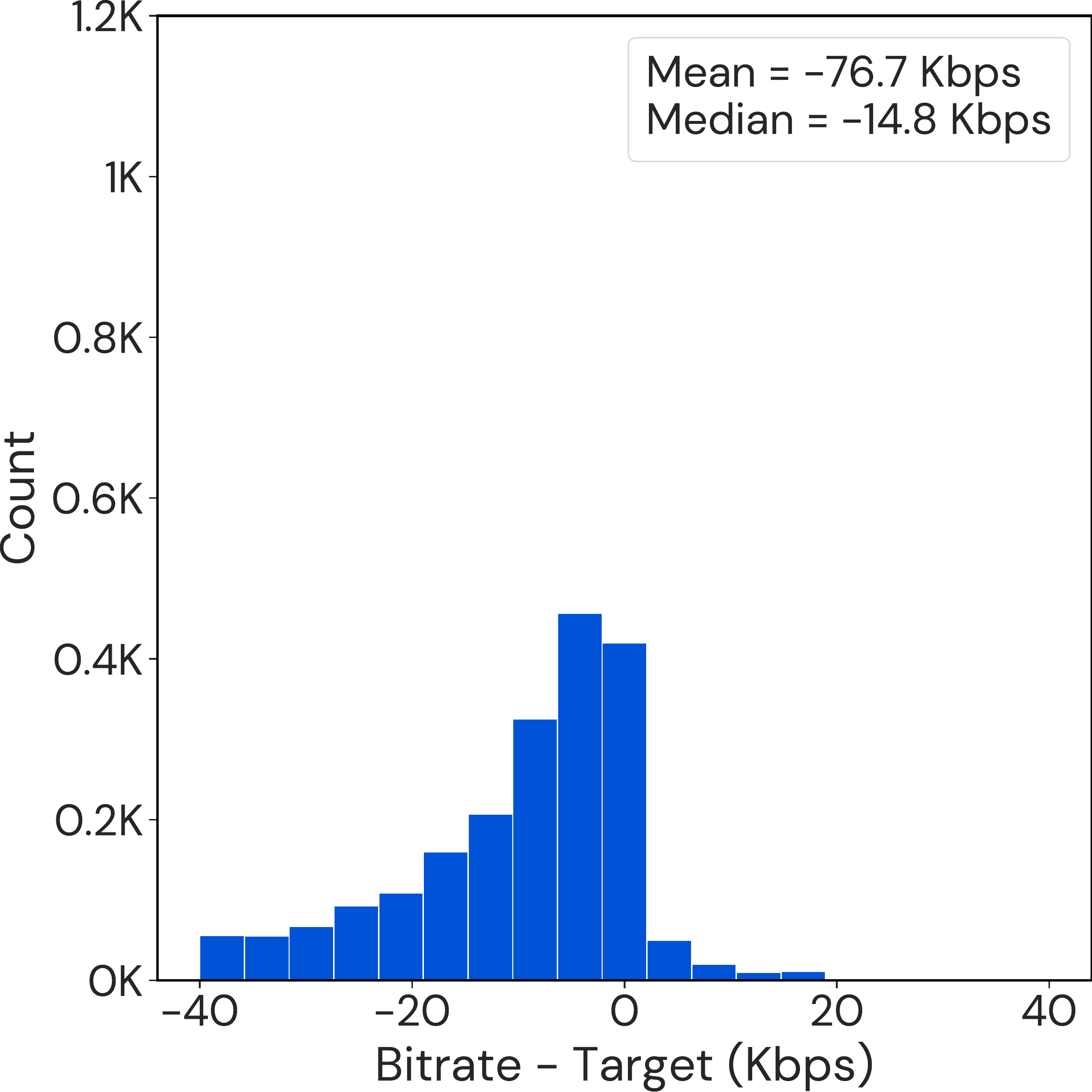}
    	\caption{Augmented \muzerorc{}}
    \end{subfigure}
    \caption{Histogram of overshoots of the agents on the evaluation set for 704 kbps target bitrate.}
    \label{figure:overshoot_704}
\end{figure}

\begin{figure}[ht]
	\centering
	\begin{subfigure}[c]{0.3\columnwidth}
	   \centering
    	\includegraphics[width=\textwidth]{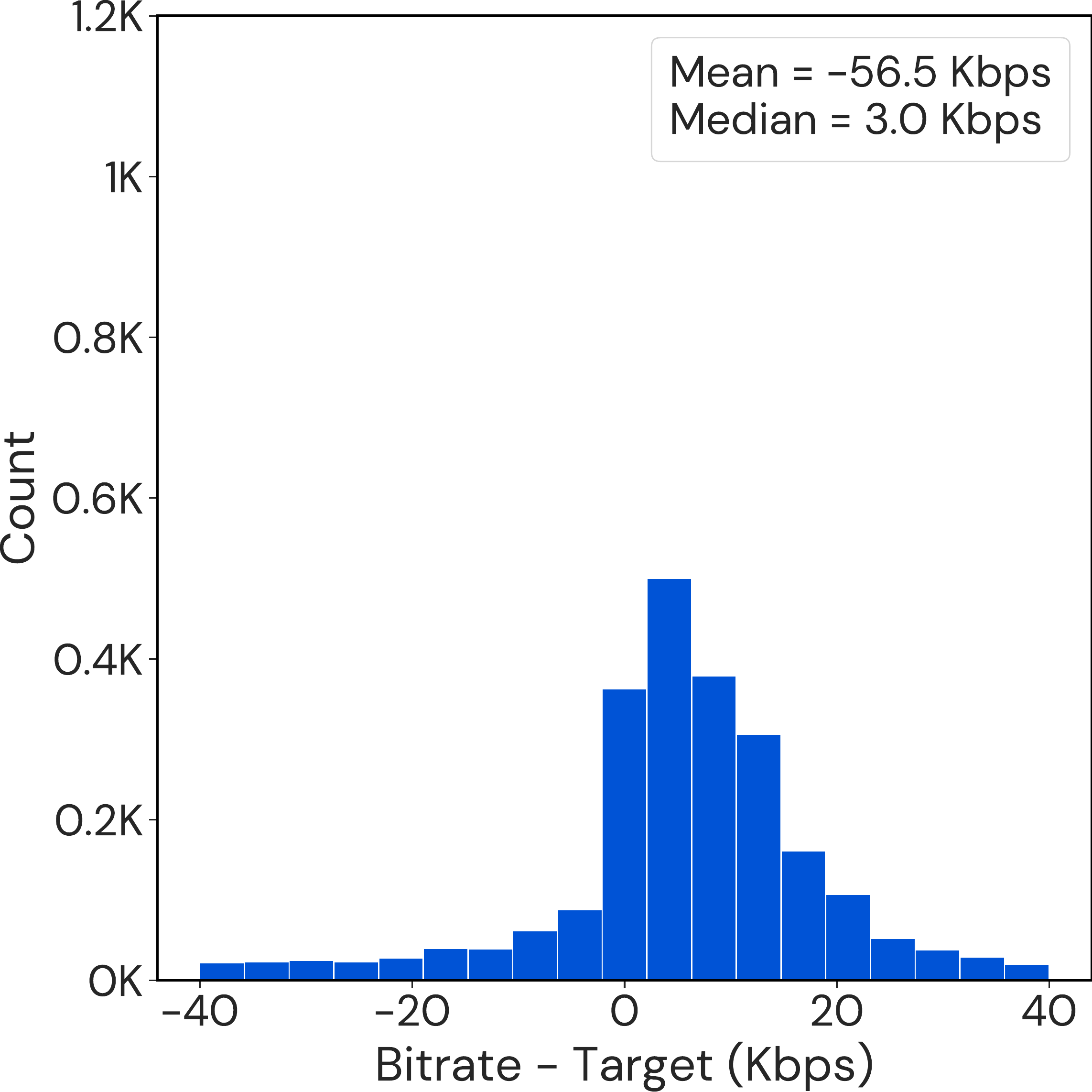}
    	\caption{\libvpx{}}
    \end{subfigure}
    \begin{subfigure}[c]{0.3\columnwidth}
	   \centering
    	\includegraphics[width=\textwidth]{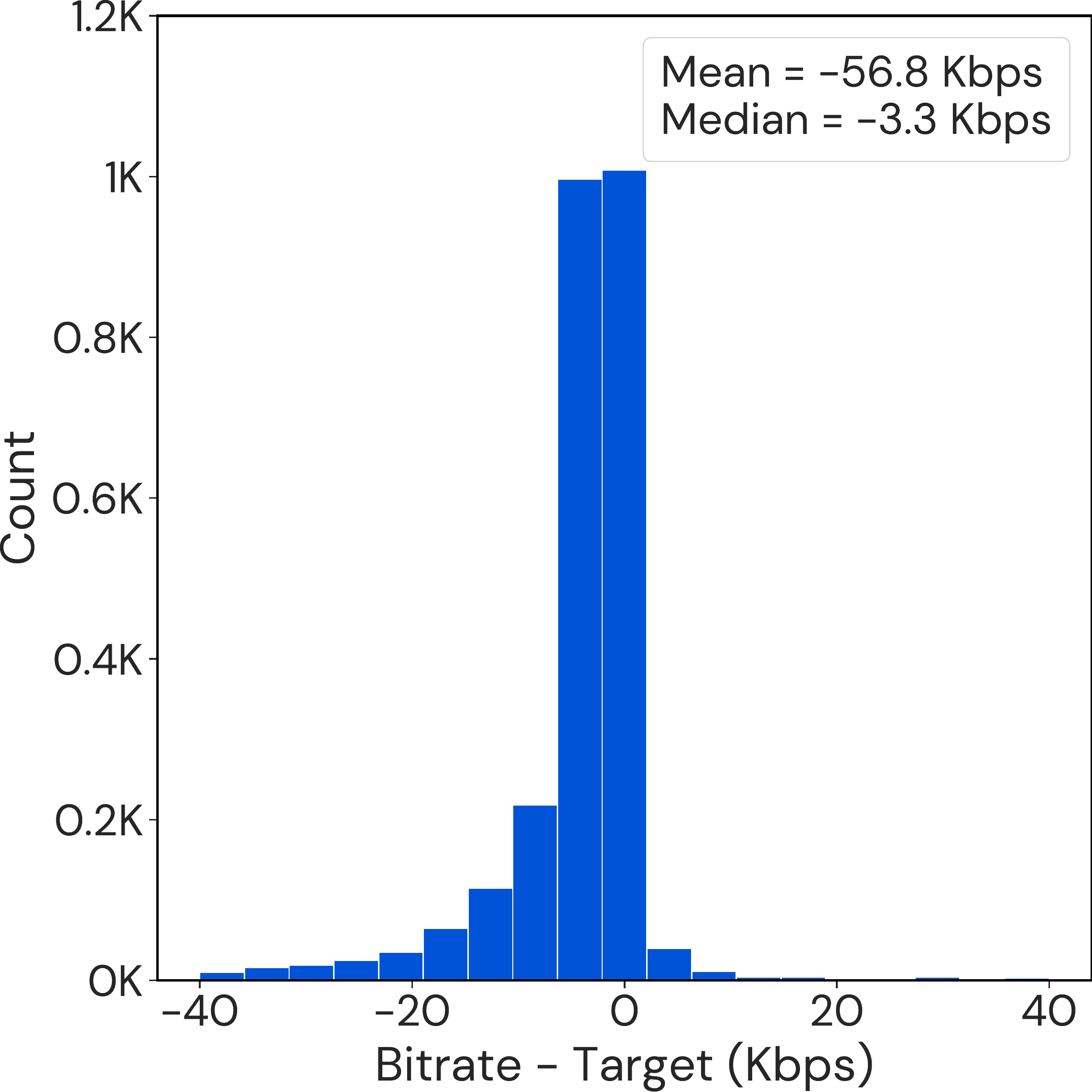}
    	\caption{\muzerorc{}{}}
    \end{subfigure}
    \begin{subfigure}[c]{0.3\columnwidth}
	   \centering
    	\includegraphics[width=\textwidth]{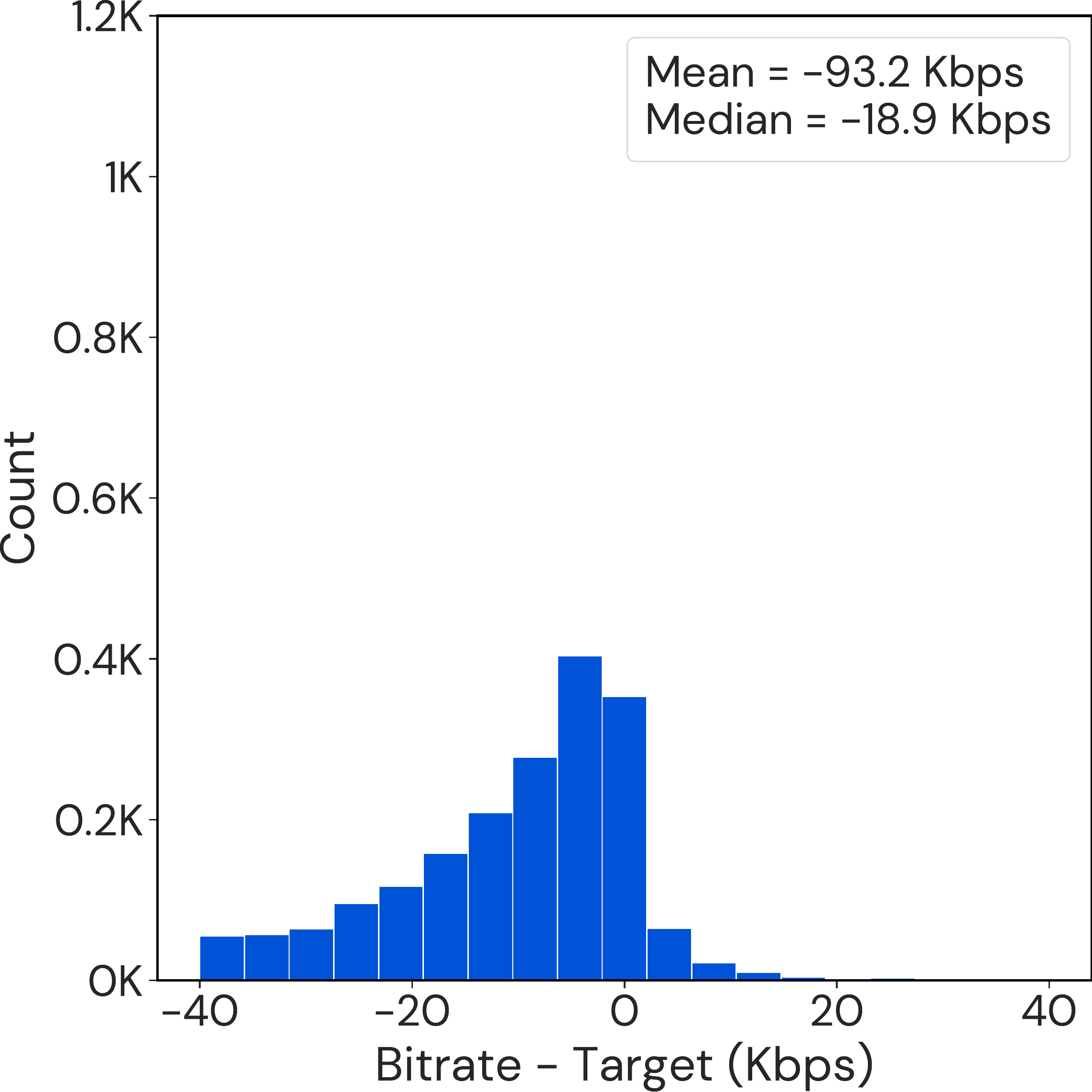}
    	\caption{Augmented \muzerorc{}}
    \end{subfigure}
    \caption{Histogram of overshoots of the agents on the evaluation set for 768 kbps target bitrate.}
    \label{figure:overshoot_768}
\end{figure}

This section presents histograms of constraint satisfaction behaviors of \libvpx{}, \muzerorc{}, and Augmented \muzerorc{} policies on 3062 videos from evaluation set.  We encode the videos using each of these policies with nine uniformly spaced target bitrates from the [256, 768] kbps.  We report the histograms of overshoot (bitrate - target) of the encodings done by each of the policies on aggregate as well as sliced by each of the target bitrate value.  Figure \ref{figure:overshoot_all} shows the histogram of the overshoots for all the policies with all the target bitrates together.  Figures \ref{figure:overshoot_256}-\ref{figure:overshoot_768} show the histograms of the overshoots for all the policies with target bitrate of 256, 320, 384, 448, 512, 576, 640, 704, and 768 kbps respectively.

\end{document}